\documentclass[preprint]{emulateapj}

\usepackage{cancel,soul,ulem,amsmath}
\usepackage{xcolor}
\usepackage{multirow}
\usepackage{rotating}
\usepackage{hyperref}

\def\13co{$^{13}$CO}

\def\c18o{C$^{18}$O}

\def\cm2{cm$^{-2}$}

\def\FCNM{$F_\mathrm{CNM}$}
\def\h2{H$_2$}
\def\hi{HI}
\def\kms{km s$^{-1}$}
\def\k0{$\kappa_{353}$}

\def\n2h{N$_2$H$^+$}
\def\NHI{$N_\mathrm{HI}$}

\def\NCNM{$N_\mathrm{CNM}$}

\def\nhiUnit{$\times\ 10^{20}$ cm$^{-2}$}
\def\NHIthin{$N^{*}_\mathrm{HI}$}

\def\psv{pseudo-Voigt}
\def\s{s$^{-1}$}
\def\s353{$\sigma_{353}$}

\def\taupeak{$\tau_{\mathrm{peak}}$}
\def\TB{$T_{\mathrm{B}}$}
\def\Tbg{$T_{\mathrm{bg}}$}
\def\Tc{$T_{\mathrm{c}}$}

\def\Texp{$T_{\mathrm{exp}}$}
\def\TD{$T_{\mathrm{D}}$}
\def\Tk{$T_{\mathrm{k}}$}

\def\Ts{$T_{\mathrm{s}}$}
\def\t353{$\tau_{353}$}

\newcommand{\Rmnum}[1]{\expandafter\@slowromancap\romannumeral #1@}
\makeatother

\shorttitle{Warm and cold \hi\ gas in the Taurus and Gemini regions}
\shortauthors{Hiep Nguyen et al. (2019)}

\begin{document}

\title{Exploring the properties of warm and cold atomic hydrogen\\ in the Taurus and Gemini regions}

\author{Hiep Nguyen\altaffilmark{1,2}, 
J. R. Dawson\altaffilmark{1},
Min-Young Lee\altaffilmark{3,4},
Claire E. Murray\altaffilmark{5,6},
Sne\v{z}ana Stanimirovi{\'c}\altaffilmark{7},
Carl Heiles\altaffilmark{8},
M.-A. Miville-Desch{\^e}nes\altaffilmark{9},
Anita Petzler\altaffilmark{1}
}

\altaffiltext{1}{Department of Physics and Astronomy and MQ Research Centre in Astronomy, Astrophysics and Astrophotonics, Macquarie University, NSW 2109, Australia. Email: van-hiep.nguyen@hdr.mq.edu.au, nguyenvanhiepiop@gmail.com}
\altaffiltext{2}{Australia Telescope National Facility, CSIRO Astronomy and Space Science, PO Box 76, Epping, NSW 1710, Australia}
\altaffiltext{3}{Max-Planck-Institut fur Radioastronomie, Auf dem H¨ugel 69, D-53121 Bonn, Germany}
\altaffiltext{4}{Korea Astronomy and Space Science Institute, Daedeokdae-ro 776, 34055 Daejeon, Republic of Korea}
\altaffiltext{5}{Department of Physics and Astronomy, Johns Hopkins University, Baltimore, MD 21218, USA}
\altaffiltext{6}{NSF Astronomy and Astrophysics Postdoctoral Fellow}
\altaffiltext{7}{Department of Astronomy, University of Wisconsin-Madison, 475 North Charter Street, Madison, WI 53706, USA}
\altaffiltext{8}{Department of Astronomy, University of California, Berkeley, 601 Campbell Hall 3411, Berkeley, CA 94720-3411}
\altaffiltext{9}{Laboratoire AIM, Paris-Saclay, CEA/IRFU/DAp - CNRS - Universit{\'e} Paris Diderot, 91191, Gif-sur-Yvette Cedex, France}

\begin{abstract}
We report Arecibo 21 cm absorption-emission observations to characterise the physical properties of neutral hydrogen (HI) in the proximity of five giant molecular clouds (GMCs): Taurus, California, Rosette, Mon OB1, NGC 2264. Strong HI absorption was detected toward all 79 background continuum sources in the $\sim$60$\times$20 square degree region. Gaussian decompositions were performed to estimate temperatures, optical depths and column densities of the cold and warm neutral medium (CNM and WNM). The properties of individual CNM components are similar to those previously observed along random Galactic sightlines and in the vicinity of molecular clouds, suggesting a universality of cold HI properties. The CNM spin temperature ($T_\mathrm{s}$) histogram peaks at $\sim$50 K. The turbulent Mach numbers of CNM components vary widely, with a typical value of $\sim$4, indicating that their motions are supersonic. About 60\% of the total HI gas is WNM, and nearly 40\% of the WNM lies in thermally unstable regime 500--5000 K. The observed CNM fraction is higher around GMCs than in diffuse regions, and increases with increasing column density ($N_\mathrm{HI}$) to a maximum of $\sim$75\%. On average, the optically thin approximation ($N^{*}_\mathrm{HI}$) underestimates the total column density by $\sim$21\%, but we find large regional differences in the relationship between $N_\mathrm{HI}$ and the required correction factor, $f$=$N_\mathrm{HI}/N^{*}_\mathrm{HI}$. We examine two different methods (linear fit of $f$ vs $\mathrm{log_{10}}$($N^{*}_\mathrm{HI}$) and uniform $T_\mathrm{s}$) to correct for opacity effects using emission data from the GALFA-HI survey. We prefer the uniform $T_\mathrm{s}$ method, since the linear relationship does not produce convincing fits for all subregions. 
\end{abstract}


\section{Introduction}
Neutral atomic hydrogen (\hi), the most abundant gas in the interstellar medium (ISM), exists in multiple phases basically distinguished by temperature. Theoretical studies (\citealt{Field1969,Silk1969,Watson1972,Silk1975,McKee1977,Wolfire1995,Wolfire2010}) and practical observations (\citealt{Radhakrishnan1972,Davies1975,Garwood1989,Liszt1993,Heiles2003a,Heiles2003b}) have well established a two phase model for interstellar \hi\ gas, in which the phases (characterized by temperature and density) exist because the pressure of the ISM falls in a range where heating and cooling mechanisms permit two thermally stable states to co-exist. This model consists of cold, dense neutral medium (CNM) with kinetic temperatures $T_\mathrm{k} \sim$ 30--200 K and volume densities of $n\sim 5$--120 cm$^{-3}$, and warm, diffuse neutral medium (WNM) typically at $T_\mathrm{k} \sim$ 4100--8800 K, $n\sim 0.03$--1.3 cm$^{-3}$ (\citealt{McKee1977,Liszt2001,Wolfire2003}). The cold gas is readily detected in \hi\ 21 cm absorption and contributes typically 10--50\% to the total \hi\ mass (\citealt{Heiles2003a,Heiles2003b,Dickey2009,Stanimirovic2014,Lee2015,Murray2015}). The variation of the CNM fraction across interstellar environments is poorly constrained; however, it is likely higher in and around molecular clouds \citep[][hereafter S14]{Stanimirovic2014}. The WNM can be easily detected in any direction on the sky via 21 cm emission, and plays an important role in the formation of cold gas, as well as constraining models of the interstellar medium.

The fraction of the neutral ISM that lies in the thermally unstable regime with intermediate kinetic temperatures of 500--5000 K (e.g. \citealt{Dickey1977,Mebold1982,Kanekar2003,Heiles2003a,Heiles2003b}, hereby HT03) has been a topic of intense debate. For instance, HT03 observationally found that 48\% of the total \hi\ column density (or 30\% of total out-of-plane gas) in the Arecibo Millennium survey was thermally unstable; \citet{Roy2013} estimated that at least $\sim$28\% of \hi\ gas toward a sample of 33 compact extragalactic radio sources is in the unstable temperature range. From the highly sensitive 21-SPONGE survey, \citet{Murray2015} and \citet{Murray2018} (hereafter M15 and M18, respectively) find a thermally unstable fraction of $\sim$20\%. While these measurements (all from observations of emission/absorption pairs) are broadly consistent with a recent study by \citet{Kalberla2018}, which decomposed all-sky  \hi\ emission spectra from the HI4PI survey \citep{hi4pi2016}, their estimated $\sim$41\% of thermally unstable HI is on the higher side relative to most recent observational studies. To keep the gas in a thermally unstable state, non-thermal processes are required, since otherwise such gas is expected to settle quickly into one of the two stable phases. Numerical ISM models suggest that dynamical processes (e.g. turbulence, shocks driven by supernova, time-dependent heating processes) may push the gas from stable phases to the thermally unstable phase (e.g. \citealt{Gazol2001,Audit2005,Kim2013,Saury2014}), resulting in significant amounts of unstable neutral gas. However, most recent numerical studies with detailed heating and cooling prescriptions, e.g. \citet{Kim2013,Hill2018}, find a thermally unstable fraction of $<$20\% which is in agreement with most recent observational studies.

The measurement of \hi\ physical properties is most commonly based on observations of the 21 cm line in emission and/or absorption. For warm \hi\ gas at low optical depth, the column density (\NHI) is proportional to the brightness temperature (\TB) and can be directly estimated from the emission profile under the optically thin assumption, \NHIthin. Yet, the relation between \NHI\ and \TB\ is not as simple for cold gas, as opacity effects become significant. In that case, one needs to know both the optical depth and spin temperature to determine the true gas column density. Fortunately these can be estimated by combining the absorption and emission spectra from on-/off-source observations toward strong continuum background sources. Such studies find that in the low column density regime below $\sim$5 \nhiUnit, the total \NHI\ is comparable to \NHIthin; but in denser regions optically thin approximation underestimates the total \hi\ column density by at least (typically) $\sim$10\% (\citealt{Dickey2003}; HT03; \citealt{Liszt2014,Lee2015,Nguyen2018,Murray2018a}). Thus \NHIthin\ is in general a lower limit to the true \NHI.

On-/off-source measurements require good signal-to-noise absorption spectra, and are limited by the population of bright, compact radio continuum background sources, which precludes the creation of high-resolution, large-area Galactic maps of the true \hi\ column density (at least in regions where the optically thin approximation is invalid). Furthermore, the derivation of true \NHI\ from these measurements can remain ambiguous due to the fact that each velocity channel contains a mixture of gas at different temperatures, and it is also possible that pencil-beam absorption measurements may undersample the true sky distribution of dense, compact CNM structures \citep[e.g.][]{Fukui2018}. An alternative is to use the available emission data and correct for the effects of opacity (e.g. by assuming a uniform \Ts; \citealt{Liszt2014b,Remy2017}) or attempt to infer spin temperatures and optical depths indirectly (e.g. from dust optical depth; \citealt{Fukui2015}).

As a follow-up to the \hi\ study of the Perseus molecular cloud (\citealt{Stanimirovic2014,Lee2015}), the main goals of the present study are to (1) explore the properties of the cold and warm atomic \hi\ gas in the proximity of five giant molecular clouds (Taurus, California, Rosette, Mon OB1, NGC 2264) as well as neighbouring diffuse and dense sightlines in and above the Galactic Plane; and (2) to build opacity-corrected \hi\ maps by applying optical depth corrections to emission data from the GALFA-\hi\ survey (\citealt{Stanimirovic2006,Peek2011,Peek2018}). This paper is a part of the GNOMES collaboration (the Galactic Neutral Opacity and Molecular Excitation Survey), which aims to understand the properties of neutral and molecular gas in and around molecular clouds.

There are many reasons why it is crucial to properly take into account \hi\ opacity effects. Estimating the exact amount of each ISM phase is critical to understand the WNM-to-CNM phase transition, and to estimate the contribution made by cold optically thick \hi\ to the so-called ``dark'' ISM \citep[e.g.][]{Grenier2005}. Accurate \hi\ column densities are also essential to test models of \h2\ formation, which predict that there exists a minimum \NHI\ for \h2\ formation in environments where dust shielding is dominant (compared to \h2\ self-shielding); in this case, the distribution of \hi\ column density is expected to be uniform \citep{Krumholz2009}. Interestingly, such a uniform distribution has been observed in Perseus molecular cloud by \citet{Lee2012}.

Our regions of interest cover a wide range of environments, including the Galactic Plane, diffuse off-Plane sightlines and molecule-rich regions. In this work, we first employ the most direct way to estimate the optical depth, spin temperature and \hi\ column density, by using absorption/emission spectra observed with the Arecibo radio telescope toward 79 radio continuum sources. We then derive the ratio of the ``true'' \hi\ column density to the optically thin \hi\ column density, and use this to evaluate two different methods for opacity correction.

Note that in this paper, we will define ``WNM'' as components detected only in emission, and ``CNM'' as components detected in absorption.

We organize this article as follows. In Section \ref{sec:observations}, we briefly summarize the observations and data processing techniques. In Section \ref{sec:data_analysis}, we describe the Gaussian/\psv\ decomposition methods for absorption and emission spectral line data. In Section \ref{sec:results}, we calculate spin temperatures, optical depths and column densities for all CNM and WNM components, and compare our results with previous observational surveys. In Section \ref{sec:opacity_corrections}, we examine two methods for opacity correction and produce opacity-corrected \hi\ maps. Finally, we summarize the results and discuss our future work.

\begin{table*}[htbp]
\begin{center}
\fontsize{6}{5}\selectfont
\caption{79 sightlines}
\centering
\label{table:source_list} 
\begin{tabular}{lcccccc}
\noalign{\smallskip} \hline \hline \noalign{\smallskip}
\shortstack{Sources\\ (NVSS name)} & \shortstack{R.A (J2000)\\(hh:mm:ss)} & \shortstack{DEC (J2000)\\(dd:mm:ss)} & \shortstack{$l$\\$(^{o})$} & \shortstack{$b$\\$(^{o})$} &
\shortstack{S$_\mathrm{1.4\ GHz}$\\(Jy)\tablenotemark{*}} & \shortstack{T$_\mathrm{sky}$\\(K)} \\
\hline

J034053+073525 (4C+07.13) & 03:40:53.73 & 07:35:25.40 & 178.87 & -36.27 & 1.01 & 4.07\\
J032153+122114 (PKS0319+12) & 03:21:53.11 & 12:21:14.00 & 170.59 & -36.24 & 1.91 & 4.51\\
J032723+120835 (4C+11.15) & 03:27:23.11 & 12:08:35.80 & 171.98 & -35.48 & 1.21 & 4.17\\
J031857+162833 (4C+16.09) & 03:18:57.77 & 16:28:33.10 & 166.64 & -33.6 & 8.03 & 6.93\\
J033626+130233 (3C090) & 03:36:26.56 & 13:02:33.20 & 173.15 & -33.29 & 1.99 & 4.67\\
J035613+130535 & 03:56:13.81 & 13:05:35.80 & 177.02 & -29.78 & 0.89 & 4.14\\
J035900+143622 (3C096) & 03:59:00.91 & 14:36:22.50 & 176.27 & -28.26 & 1.2 & 4.37\\
J042725+085330 (4C+08.15) & 04:27:25.05 & 08:53:30.30 & 186.21 & -26.51 & 0.94 & 4.08\\
J032504+244445 (4C+24.06) & 03:25:04.35 & 24:44:45.60 & 161.92 & -26.26 & 0.81 & 4.13\\
J035633+190034 (4C+18.11) & 03:56:33.46 & 19:00:34.60 & 172.23 & -25.66 & 1.05 & 4.15\\
J041140+171405 (4C+17.23) & 04:11:40.77 & 17:14:05.10 & 176.36 & -24.24 & 1.03 & 4.26\\
J042022+175355 (3C114) & 04:20:22.17 & 17:53:55.20 & 177.3 & -22.24 & 1.11 & 4.23\\
J042524+175525 (4C+17.25) & 04:25:24.43 & 17:55:25.30 & 178.11 & -21.31 & 0.88 & 4.16\\
J042756+175242 (4C+17.26) & 04:27:56.98 & 17:52:42.80 & 178.56 & -20.88 & 1.01 & 4.22\\
J044907+112128 (PKS0446+11) & 04:49:07.65 & 11:21:28.20 & 187.43 & -20.74 & 0.85 & 4.16\\
J034008+320901 (3C092) & 03:40:08.54 & 32:09:01.30 & 159.74 & -18.41 & 1.61 & 3.95\\
J042846+213331 (4C+21.17) & 04:28:46.64 & 21:33:31.40 & 175.7 & -18.36 & 1.34 & 4.35\\
J040442+290215 (4C+28.11) & 04:04:42.82 & 29:02:15.90 & 166.06 & -17.22 & 0.95 & 3.69\\
J052424+074957 (4C+07.16) & 05:24:24.04 & 07:49:57.10 & 195.51 & -15.35 & 0.82 & 4.25\\
J051240+151723 (PKS0509+152) & 05:12:40.99 & 15:17:23.80 & 187.41 & -13.79 & 0.97 & 4.11\\
J053239+073243 & 05:32:39.01 & 07:32:43.50 & 196.84 & -13.74 & 2.73 & 4.96\\
J051930+142829 (4C+14.14) & 05:19:30.95 & 14:28:29.00 & 189.04 & -12.85 & 0.86 & 4.15\\
J053450+100430 (4C+09.21) & 05:34:50.82 & 10:04:30.30 & 194.89 & -11.98 & 1.06 & 4.62\\
J041236+353543 (4C+35.07) & 04:12:36.28 & 35:35:43.20 & 162.58 & -11.36 & 0.86 & 3.93\\
J052109+163822 (3C138) & 05:21:09.93 & 16:38:22.20 & 187.41 & -11.34 & 8.6 & 7.59\\
J053056+133155 (PKS0528+134) & 05:30:56.44 & 13:31:55.30 & 191.37 & -11.01 & 1.56 & 4.64\\
J042353+345144 (3C115) & 04:23:53.25 & 34:51:44.80 & 164.76 & -10.24 & 1.3 & 3.88\\
J060536+014512 (4C+01.17) & 06:05:36.56 & 01:45:12.70 & 206.08 & -9.37 & 0.61 & 4.07\\
J045956+270602 (4C+27.14) & 04:59:56.09 & 27:06:02.90 & 175.83 & -9.36 & 0.93 & 3.9\\
J051740+235110 (4C+23.14) & 05:17:40.81 & 23:51:10.20 & 180.86 & -8.01 & 0.97 & 4.32\\
J045323+312924 (3C131) & 04:53:23.34 & 31:29:24.20 & 171.44 & -7.8 & 2.87 & 4.04\\
J053557+175600 (4C+17.33) & 05:35:57.42 & 17:56:00.70 & 188.22 & -7.67 & 0.83 & 4.23\\
J053444+192721 (PKS0531+19) & 05:34:44.51 & 19:27:21.70 & 186.76 & -7.11 & 7.02 & 6.48\\
J054046+172839 (4C+17.34) & 05:40:46.05 & 17:28:39.20 & 189.21 & -6.93 & 1.47 & 4.5\\
J050929+295755 (4C+29.16) & 05:09:29.51 & 29:57:55.80 & 174.77 & -5.97 & 1.06 & 4.03\\
J061900+050630 & 06:19:00.21 & 05:06:30.80 & 204.66 & -4.84 & 0.69 & 4.19\\
J062812+010926 (4C+01.19) & 06:28:12.55 & 01:09:26.00 & 209.24 & -4.64 & 0.94 & 4.27\\
J062152+043834 (4C+04.22) & 06:21:52.90 & 04:38:34.50 & 205.41 & -4.43 & 1.06 & 4.35\\
J062551+043540 (4C+04.24) & 06:25:51.89 & 04:35:40.20 & 205.92 & -3.57 & 0.81 & 4.28\\
J053425+273223 (B0531+2730) & 05:34:25.73 & 27:32:23.90 & 179.87 & -2.83 & 1.04 & 4.25\\
J061622+115553 & 06:16:22.32 & 11:55:53.90 & 198.33 & -2.2 & 0.83 & 4.41\\
J055955+190852 (4C+19.18) & 05:59:55.70 & 19:08:52.50 & 190.09 & -2.17 & 1.03 & 4.45\\
J060157+192216 (4C+19.19) & 06:01:57.97 & 19:22:16.30 & 190.13 & -1.64 & 0.8 & 4.43\\
J060933+162207 (4C+16.15)\tablenotemark{**} & 06:09:33.51 & 16:22:07.00 & 193.64 & -1.53 & 0.73 & 4.58\\
J061857+133631 (4C+13.32) & 06:18:57.50 & 13:36:31.50 & 197.15 & -0.85 & 1.96 & 4.86\\
J054844+263600 (4C+26.18)\tablenotemark{**} & 05:48:44.28 & 26:36:00.30 & 182.36 & -0.62 & 1.16 & 3.98\\
J054211+290147 (4C+29.19) & 05:42:11.81 & 29:01:47.80 & 179.53 & -0.59 & 0.98 & 4.13\\
J063313+081318 (4C+08.21) & 06:33:13.01 & 08:13:18.90 & 203.54 & -0.27 & 1.1 & 4.46\\
J060351+215937 (4C+22.12) & 06:03:51.56 & 21:59:37.50 & 188.07 & 0.04 & 2.77 & 5.15\\
J062141+143211 (3C158) & 06:21:41.09 & 14:32:11.90 & 196.64 & 0.17 & 2.24 & 4.86\\
J063215+102201 (4C+10.20) & 06:32:15.34 & 10:22:01.10 & 201.53 & 0.51 & 2.38 & 5.0\\
J062545+144019 (4C+14.18) & 06:25:45.97 & 14:40:19.70 & 196.98 & 1.1 & 2.44 & 5.01\\
J064516+053132 (3C167) & 06:45:16.84 & 05:31:32.00 & 207.31 & 1.15 & 1.33 & 4.76\\
J053752+361112 (4C+36.10) & 05:37:52.39 & 36:11:12.90 & 172.98 & 2.44 & 1.04 & 4.34\\
J062019+210229 & 06:20:19.54 & 21:02:29.70 & 190.74 & 2.94 & 0.9 & 4.33\\
J062250+220025 (4C+22.16) & 06:22:50.64 & 22:00:25.60 & 190.16 & 3.91 & 1.06 & 4.34\\
J065601+083407 (4C+08.23) & 06:56:01.03 & 08:34:07.10 & 205.81 & 4.91 & 0.68 & 4.12\\
J063451+190940 & 06:34:51.38 & 19:09:40.90 & 193.99 & 5.1 & 0.84 & 4.31\\
J065917+081331 & 06:59:17.97 & 08:13:31.80 & 206.48 & 5.48 & 0.91 & 4.12\\
J065548+104258 (4C+10.21) & 06:55:48.36 & 10:42:58.80 & 203.85 & 5.82 & 0.91 & 4.15\\
J064841+144020 (4C+14.20) & 06:48:41.15 & 14:40:20.60 & 199.52 & 6.04 & 1.03 & 4.17\\
J071523+031105 (4C+03.12) & 07:15:23.75 & 03:11:05.00 & 212.82 & 6.78 & 0.67 & 3.97\\
J071004+061705 (4C+06.28) & 07:10:04.83 & 06:17:05.40 & 209.43 & 7.0 & 0.58 & 4.05\\
J071924+021035 & 07:19:24.88 & 02:10:35.80 & 214.18 & 7.22 & 0.66 & 3.87\\
J064524+212145 (3C166) & 06:45:24.08 & 21:21:45.70 & 193.12 & 8.3 & 2.59 & 4.8\\
J071028+091954 (4C+09.27) & 07:10:28.52 & 09:19:54.30 & 206.72 & 8.44 & 1.02 & 4.09\\
J072140+053050 (PKS0719+056) & 07:21:40.87 & 05:30:50.90 & 211.43 & 9.23 & 0.83 & 4.01\\
J070001+170922 & 07:00:01.52 & 17:09:22.30 & 198.47 & 9.58 & 0.65 & 4.09\\
J065937+211742 (4C+21.22) & 06:59:37.81 & 21:17:42.20 & 194.63 & 11.26 & 0.57 & 3.98\\
J071404+143620 (3C175.1) & 07:14:04.70 & 14:36:20.90 & 202.29 & 11.53 & 1.93 & 4.38\\
J072551+103202 (4C+10.22) & 07:25:51.27 & 10:32:02.30 & 207.31 & 12.37 & 1.23 & 4.09\\
J070937+195443 (4C+19.26) & 07:09:37.27 & 19:54:43.00 & 196.91 & 12.8 & 0.69 & 4.03\\
J072525+123624 (PKS0722+12) & 07:25:25.60 & 12:36:24.60 & 205.35 & 13.17 & 0.61 & 3.77\\
J072832+121010 (4C+12.30) & 07:28:32.88 & 12:10:10.50 & 206.09 & 13.67 & 1.07 & 3.92\\
J072516+142513 (4C+14.23) & 07:25:16.80 & 14:25:13.50 & 203.64 & 13.91 & 1.07 & 3.87\\
J072810+143736 (3C181) & 07:28:10.26 & 14:37:36.00 & 203.75 & 14.63 & 2.3 & 4.53\\
J071810+201002 (PKS0715+20) & 07:18:10.64 & 20:10:02.70 & 197.52 & 14.74 & 0.73 & 3.98\\
J073035+151512 (4C+15.20) & 07:30:35.41 & 15:15:12.60 & 203.42 & 15.42 & 1.54 & 4.23\\
J073037+173951 (4C+17.41) & 07:30:37.09 & 17:39:51.50 & 201.13 & 16.42 & 0.69 & 3.94\\

\noalign{\smallskip} \hline \noalign{\smallskip}

\multicolumn{7}{l}{\textsuperscript{*}\footnotesize{\citet{Condon1998}}} \\
{\textsuperscript{**}\footnotesize{Excluded from the analysis}}

\end{tabular}
\end{center}
\end{table*}

\section{Observations}
\label{sec:observations}

\begin{figure*}
 \center
  \includegraphics[width=\textwidth]{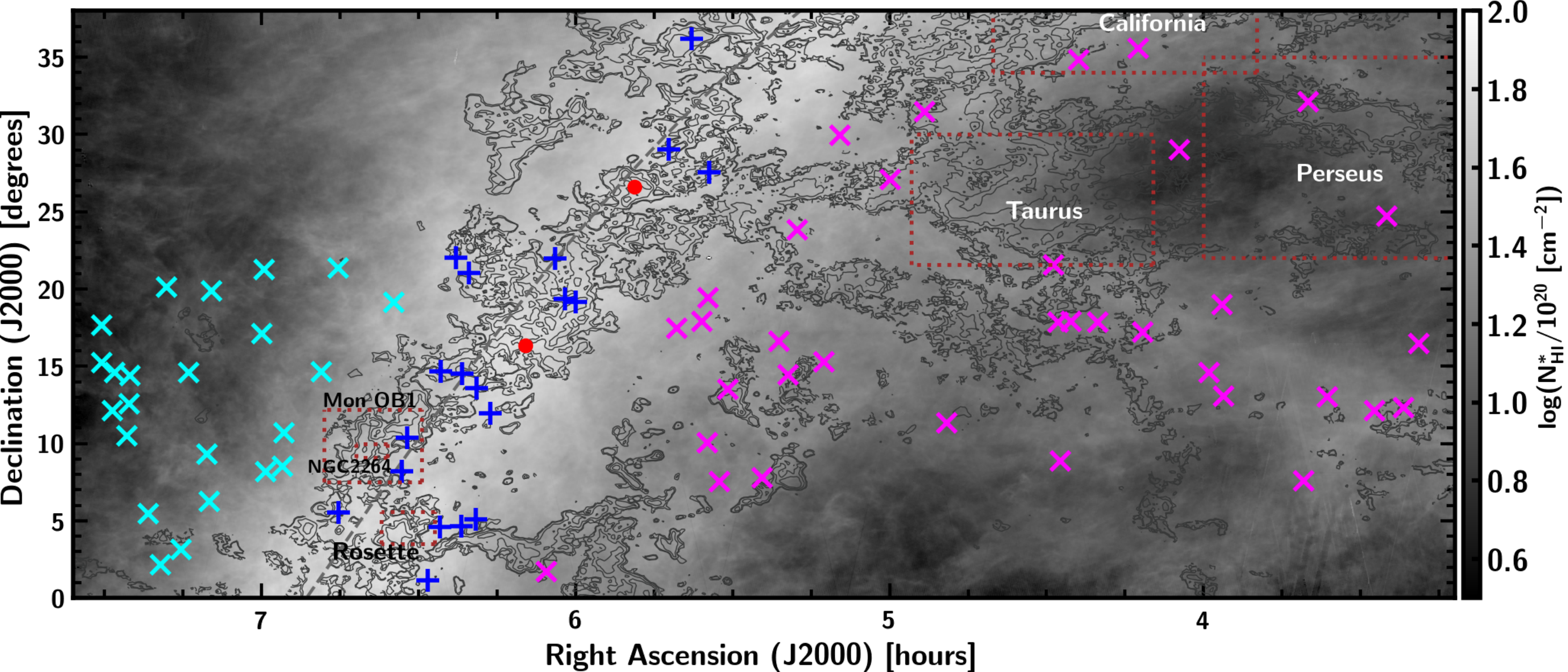}
  \caption{Locations of all 79 background radio continuum sources considered in this study, overlaid on the map of \hi\ column density \NHIthin\ from the Arecibo GALFA-\hi\ survey \citep[DR2;][]{Peek2018}. The \NHIthin\ range is (1--80) \nhiUnit. The blue plus markers show the sightlines toward 19 radio sources near the Galactic plane ($|b|<5^{\circ}$), cyan crosses show the 23 sightlines above the Galactic plane, magenta crosses show the 35 sightlines below the Galactic plane, and the red dots show the two saturated sightlines excluded from the analysis (4C+16.15 and 4C+26.18). The dashed line shows the Galactic plane (Galactic latitude $b = 0^{\circ}$). Dotted boxes show roughly the locations of five molecular clouds. The contours represent the logarithm of integrated intensity W$_{\mathrm{CO}(1-0)}$ from the all-sky extension to the maps of \citet{Dame2001} (T. Dame, private communication). The levels are [0.01, 0.25, 0.75, 1., 1.75] in units of log$_{10}(\mathrm{K\ km\ s^{-1}})$.}\vspace{0.4cm}
  \label{fig:src_locations}
\end{figure*}

We have conducted \hi\ and OH absorption/emission measurements toward 79 extragalactic radio continuum sources in the vicinity of five giant molecular clouds (Taurus, California, Rosette, Mon OB1 and NGC 2264) using the Arecibo 305-m radio telescope (project a2769, PI Stanimirovi{\'c}). The continuum sources used in this work were selected from the NVSS catalogue \citep{Condon1998} and have typical flux densities of S $\gtrsim$ 0.6 Jy at 1.4 GHz. The source positions in equatorial coordinates (J2000) are presented in Figure \ref{fig:src_locations}. In Table \ref{table:source_list}, we list the basic information of these sources: right ascension (R.A.), declination (DEC), Galactic longitude/latitude, flux density at 1.4 GHz, and the diffuse background radio continuum emission.

The observations were conducted with Arecibo using the L-wide receiver, which permits simultaneous recording of \hi\ (centered at 1420.406 MHz), and the two OH ground-state main lines at 1665.402 and 1667.359 MHz, with a velocity resolution of ~0.16 km s$^{-1}$ and an angular resolution of 3$'$.5 at these frequencies.

We use the same observing procedure described in detail by \citet{Heiles2003a}. Briefly, we make a 17 data point observation (so-called Z16), consisting of one on-source absorption spectrum at the position of the background radio source, and 16 off-source emission spectra with innermost positions located at 1.0 HPBW and the outermost positions at $\sqrt{2}$ HPBW from the central source. This particular observation pattern was designed to measure the fluctuations of 21 cm intensity within an area with an angular diameter of $\sim$13$'$.5 around the source, and take these spatial variations into account to derive an accurate off-source ``expected'' emission spectrum (\Texp, the emission profile we would observe in the absence of the radio source; see Section \ref{sec:data_analysis} for details of the \Texp\ derivation). Our integration time, on average, is $\sim$1 hour per source (including all on- and off-source integrations); leading to a characteristic noise level in optical depth of $\sim$5 $\times\ 10^{-3}$ per 1 km s$^{-1}$ velocity channel.

In the 21-SPONGE survey, M15 apply a main beam efficiency of 0.94 to Arecibo emission spectra taken with the L-wide receiver. To verify that this factor is also appropriate for our dataset, we compare our \Texp\ profiles with emission data from the Arecibo GALFA-\hi\ survey \citep[Data Release 2, DR2;][]{Peek2018}. We extract the GALFA-\hi\ emission spectra around the location of each source within the outermost angular radius of the Z16 observation pattern, excluding the central pixels inside the Arecibo main beam (since these may be affected by absorption), and average the remaining pixels. We then compare the resultant spectra with our \Texp\ profiles on a channel-by-channel basis, for all channels above our 5$\sigma$ noise level, and find that the \Texp\ values from the present work are systematically lower by a factor of $\sim$1.06 (standard deviation of 0.08). Similarly, our \Texp\ integrated intensities are found to be systematically lower than GALFA-\hi\ by the same factor of $\sim$1.06 (standard deviation of 0.04). We also make the same comparison using the Leiden-Argentine-Bonn \citep[LAB,][]{Hartmann1997,Kalberla2005} and HI4PI \citep{hi4pi2016} surveys. While these have lower resolutions than Arecibo ($36'$ and $16'$ respectively), they are both stray-radiation-corrected and have excellent absolute calibration. We find a consistent factor of $\sim$1.07 (standard deviation of 0.07). We therefore apply a beam efficiency factor of $\eta_b = 1/1.07 = 0.93$ to our data.

Note that while we do not attempt to correct for the effects of stray radiation in our data, visual inspection of our spectra reveal no strong signs of the broad, wide wings that typically characterise such contamination. \citet{Peek2011} note that contamination from stray radiation in GALFA-\hi\ is small -- typically within their $1\sigma$ uncertainties. Once the main beam efficiency has been applied, the absolute differences between our spectra and those of GALFA-\hi\ are also small; specifically, on a velocity channel basis, the difference spectra have broad features, but these mostly fall within the 2$\sigma$ uncertainty profiles of our $T_\mathrm{exp}(v)$. This results in a relative difference of at most 12\% between the optically thin column densities of the two surveys along our 79 sightlines, with a median difference of (2.0$\pm$1.1)\% (the uncertainty of the median is estimated by bootstrap re-sampling). Thus, our visual inspection suggests that while stray radiation is likely present at a low level, its contribution to our profiles is not significant enough to affect our scientific conclusions.

\section{Data Analysis: \\ \hi\ absorption \& emission fitting}
\label{sec:data_analysis}

The root of the analysis is to solve the two equations of radiative transfer for on-/off-source measurements, under the assumption that the on and off positions sample the same gas:

\begin{equation}
T_\mathrm{B}^\mathrm{ON} (v) = (T_\mathrm{bg}+T_\mathrm{c})e^{-\tau_{v}} + T_\mathrm{s}(1-e^{-\tau_{v}}),
\label{eq:on_simp}
\end{equation}
\begin{equation}
T_\mathrm{B}^\mathrm{OFF} (v) = T_\mathrm{bg}e^{-\tau_{v}} + T_\mathrm{s}(1-e^{-\tau_{v}}),
\label{eq:off_simp}
\end{equation}
\begin{equation}
T_\mathrm{exp} (v) = T_\mathrm{B}^\mathrm{OFF} (v) - T_\mathrm{bg}.
\label{eq:texp}
\end{equation}

\noindent Hence:

\begin{equation}
e^{-\tau_{v}} = \frac{T_\mathrm{B}^\mathrm{ON}(v) - T_\mathrm{B}^\mathrm{OFF}(v)}{T_\mathrm{c}}.
\label{eq:emt}
\end{equation}\\

\noindent In the above, $T_\mathrm{B}^\mathrm{ON} (v)$ and $T_\mathrm{B}^\mathrm{OFF} (v)$ are the brightness temperatures of the on-source and off-source profiles respectively, \Ts\ is the spin temperature, $\tau_{v}$ is the optical depth, \Tc\ is the brightness temperature of the continuum source, and \Tbg\ is the background brightness temperature including the 2.725 K isotropic radiation from the CMB and the Galactic synchrotron background at the source position. We obtain \Tbg\ from the 1.4 GHz radio continuum maps of CHIPASS \citep[in the Southern sky][]{Calabretta2014} and from the Stockert and Villa-Elisa telescopes \citep[for the Northern sky][]{Reich2001}, with typical values of around 4.4 K.

Following HT03, we derive \Texp, the ``expected'' emission profile at the location of the background source. We express the 16 off-source spectra as both first and second order Taylor series expansions of \Texp, plus a small contribution from the source intensity attenuated by optical depth, and perform a least-squares fit for the 17-point measurements to derive simultaneously: (1) the on-source optical depth profile ($e^{-\tau_{v}}$), (2) the off-source expected emission spectrum as well as its spatial derivatives, (3) their uncertainty profiles: $\sigma_\mathrm{\tau}$ (also $\sigma_\mathrm{e^{-\tau}}$) and $\sigma_{T_\mathrm{exp}}$. Here, we use a slightly simpler approach than HT03, by not including their fine-tuning for gain variation, under the assumption that an accurate knowledge of the spatially-varying telescope gain and beam-shape properties is required to properly account for the off-source gain. As pointed out by \citet{Stanimirovic2014}, we also find that the second-order Taylor expansion is noisier (because the higher order variation in \hi\ emission requires more fitting parameters) but likely better models the actual complex variations in \hi\ emission and thus is the more accurate approach. We therefore use the $T_\mathrm{exp}(v)$ and $\tau(v)$ profiles from the second-order expansion for all sources.

In this paper, to estimate the physical properties of individual \hi\ clouds along a sightline, we follow the methodology applied by \citet{Heiles2003a} to the Millennium Survey \hi\ profiles. Essentially we assume cold gas (`CNM') components are seen in both absorption and emission (i.e. in both opacity and brightness temperature profiles), and warm gas (`WNM') is seen only in emission, giving only a brightness temperature (see \citealt{Heiles2003a} for further detail). A key difference in the present work, however, is that besides Gaussian profile fitting, we also carry out \psv\ fitting, to examine whether the derived properties are affected by the choice of profile shape. Each Gaussian has three free parameters, and each \psv\ has four (see Section \ref{subsec:psv} for details).

\subsection{Initial Component Generation}
\label{sg-bic}
The opacity and expected emission profiles consist of multiple components at different velocities. A crucial question is how many components should be added to fit each spectrum. The noise level of the spectrum and residuals of the fit can be used as a measure of the goodness of the fit. However, while adding more components will naturally reduce the residuals, it may result in overfitting, and may not necessarily reflect physical reality. To minimise subjective judgment as much as possible, we generate a common set of initial guesses for both Gaussian and pseudo-Voigt fittings using a Savitzky-Golay filter for peak/trough detections. Note that the numbers of CNM and WNM components from these common initial guesses are not optimized for both types of fittings. We then employ the Bayesian Information Criterion ($BIC$) to automatically obtain the number of fit components, $N_C$, for the absorption ($e^{-\tau (v)}$) and emission spectra ($T_\mathrm{exp} (v)$):

\begin{equation}
BIC = \chi^{2} + d\ ln(N)
\label{eq:bic} 
\end{equation}

\noindent where $\chi^{2}$ is the chi-squared from the fit, $d$ is the number of free parameters (which is proportional to $N_C$, e.g. $d=3N_C$ for Gaussians, and $d=4N_C$ for pseudo-Voigts) and $N$ is the sample size. Clearly, the $BIC$ adds a new term to $\chi^{2}$ to penalize the number of parameters in the model. For two given models, the model with the lower value of $BIC$ is preferred. We calculate the $BIC$ value for each value of $d$, and adopt the number of components from the most preferred $d$ (see also \citealt{Allison2012, Gordon2017}). After this step, we have two sets of initial guesses for the Gaussian and \psv\ fittings, each with typically a different number of components. In the few cases where the spectra are either noisy or marginally saturated, we slightly adjust the initial guesses (central velocities and number of components) to allow the fits to converge.\\

\begin{figure*}[htbp]
\centering
\includegraphics[width=1.\linewidth]{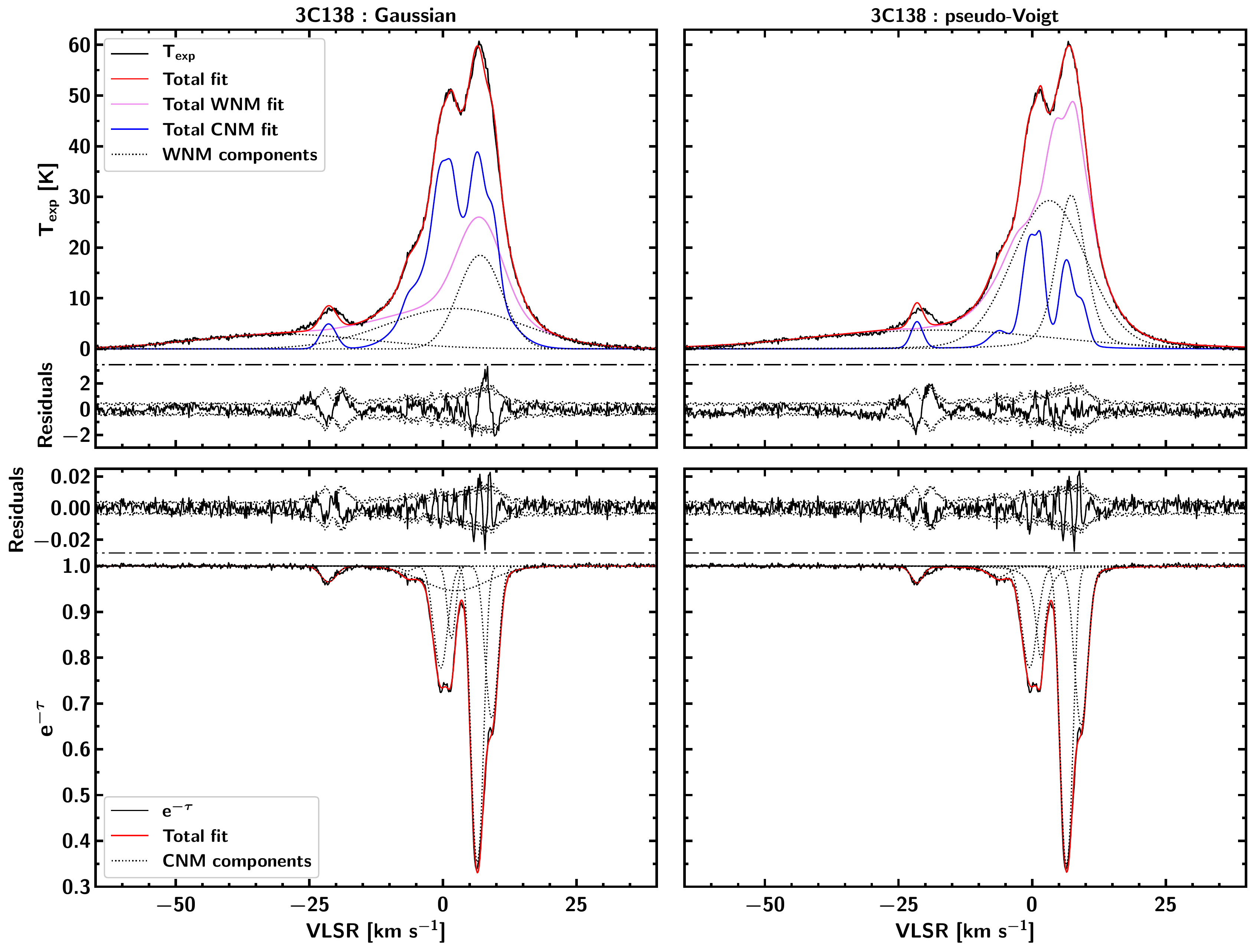}
\caption{Example of Gaussian (left) and \psv\ (right) fits to a pair of emission and absorption spectra. In the upper panels, the black solid line is the expected emission profile \Texp; the black dotted lines are the WNM Gaussian/\psv\ components; the pink line shows the contribution of the WNM components to the \Texp\ profile; the blue line displays the contribution to the \Texp\ profile by the CNM components from the absorption profile; the red solid line is the total WNM and CNM fit. The residuals from the fit are plotted below zero, with $\pm\sigma_{T_\mathrm{exp}}$ superimposed. In the lower panels, the black line shows the optical depth profile (e$^{-\tau_{v}}$); the dotted lines are the CNM Gaussian/\psv\ components, the red line represents the fit to the optical depth profile. The residuals from the absorption fit are plotted above one, with $\pm\sigma_{e^{-\tau}}$ superimposed. The derivation of uncertainty profiles ($\sigma_{T_\mathrm{exp}}$ and $\sigma_{e^{-\tau}}$) is described in Section \ref{sec:data_analysis}.}
\label{fig:eg_plot1}
\end{figure*}

\subsection{Gaussian spectral decomposition}
In this section, we decompose both the absorption and emission spectra into sets of Gaussian components. The expected emission profile $T_\mathrm{exp}(v)$ consists of both warm and cold components:

\begin{equation}
T_\mathrm{exp}(v) = T_\mathrm{B,CNM}(v) + T_\mathrm{B,WNM}(v)
\label{eq_texp} 
\end{equation}

\noindent where $T_\mathrm{B,CNM}(v)$ is the brightness temperature of the cold gas and $T_\mathrm{B,WNM}(v)$ is the brightness temperature of the the warm gas. The on-source spectrum (Equation \ref{eq:emt}) contains only cold gas seen in absorption.

First, we fit the opacity spectrum $\tau(v)$ with a set of $N$ Gaussian components using a least-squares method with the assumption that each component is independent and isothermal with a spin temperature $T_\mathrm{s,n}$:

\begin{equation}
\tau(v) = \sum\limits_{n=0}^{N-1} \tau_{0,n}e^{-4ln2\ [(v-v_{0,n})/\delta v_{n}]^2}.
\label{eqtau} 
\end{equation}

\noindent Here the parameters used to fit the absorption spectrum, $\tau_{0,n}$, $v_{0,n}$, $\delta v_{n}$ are respectively the peak optical depth, central velocity and FWHM of $n^{th}$ component. These parameters are assumed to be independent, and the number of components and the initial guesses for the fit parameters are generated as described in Section \ref{sg-bic}. This least-squares fit gives all the values of $\tau_{0,n}$, $v_{0,n}$ and $\delta v_{n}$. 

We now fix all the parameters derived for these CNM components (namely $\tau_{0,n}$, $v_{0,n}$ and $\delta v_{n}$). Next, we also use the least squares technique to fit the expected emission profile, \Texp($v$), which is assumed to consist of the $N$ cold components seen in the absorption spectrum, plus any warm components seen only in emission. The contribution of the cold components is given by
\begin{equation}
T_\mathrm{B,CNM}(v) = \sum\limits_{n=0}^{N-1} T_{s,n}(1-e^{-\tau_{n}(v)}) e^{-\sum\limits_{m=0}^{M-1} \tau_{m}(v)}
\label{eqtbc} 
\end{equation}

\noindent in which the subscript $m$ describes $M$ absorption clouds lying in front of the $n^{th}$ cloud. For WNM, we use a set of $K$ Gaussian functions to represent the contribution to the expected profile $T_\mathrm{exp}(v)$, (where the number of components and their initial guesses are also generated as described in Section \ref{sg-bic}). For each $k^{th}$ component, we assume that a fraction $F_{k}$ of WNM lies in front of all CNM components, so the remaining ($1-F_{k}$) lies behind and is absorbed by CNM components. Therefore, the brightness temperature from these $K$ WNM components is:
\begin{equation}
\begin{split}
T_\mathrm{B,WNM}(v) =\ & \sum\limits_{k=0}^{K-1} [F_\mathrm{k} + (1-F_\mathrm{k})e^{-\tau_{v}}]\\ & \times T_\mathrm{0,k}e^{-4ln2\ [(v-v_\mathrm{0,k})/\delta v_\mathrm{k}]^2}
\label{eqtbw}
\end{split}
\end{equation}

\begin{figure*}[ht]
\centering
\includegraphics[width=1.\linewidth]{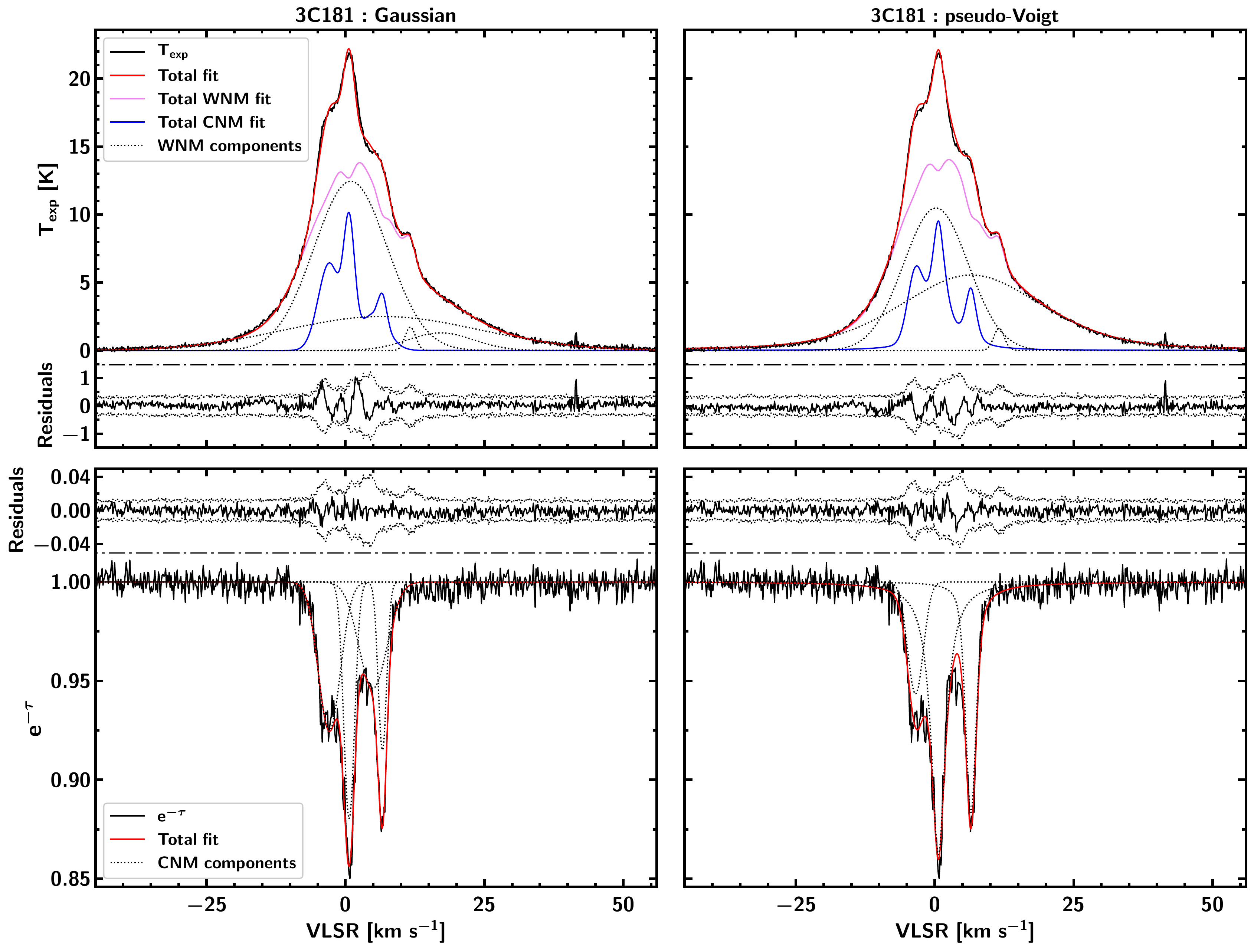}
\caption{See Figure \ref{fig:eg_plot1} for details.}
\label{fig:eg_plot2}
\end{figure*}

\noindent Here the independent fit parameters $T_\mathrm{0,k}$, $v_\mathrm{0,k}$, $\delta v_\mathrm{k}$ are respectively the peak brightness temperature, central velocity and FWHM of the $k^{th}$ emission component. This least-squares fit to the emission spectrum gives $T_\mathrm{0,k}$, $v_\mathrm{0,k}$, $\delta v_\mathrm{k}$, $F_\mathrm{k}$ for warm components and $T_\mathrm{s,n}$ for cold components. For each sightline, we perform the \Texp$(v)$ fit with all possible orderings of the $N$ absorption components ($N!$), and choose the one that gives the smallest residuals. As seen in Equation \ref{eqtbw}, the fraction $F_\mathrm{k}$ determines the contribution of $T_\mathrm{B,WNM}$ to \Texp\ and so it has an important effect on the derived CNM spin temperatures ($T_\mathrm{s,n}$). However, the results of the $F_\mathrm{k}$ fractions from different CNM ordering fits are difficult to distinguish statistically by the difference in the fit residuals. Hence, following the suggestion in HT03, after including $K$ WNM emission-only components, we calculate the spin temperature of CNM by assigning three values (0, 0.5, 1) to each $F_\mathrm{k}$. We then experiment with all possible combinations of $K$ WNM components ($3^K$) along the line of sight. We obtain the final spin temperatures $T_\mathrm{s,n}$ for CNM components as a weighted average over all trials, and the weight of each trial is the inverse of the variance estimated from the residuals the $T_\mathrm{exp} (v)$ fit. The uncertainty of the final spin temperature for each CNM component is estimated from the variations in \Ts\ with $F_\mathrm{k}$ in all trials (as described in Section 3.5 of \citealt{Heiles2003a}). Apart from the dependence of CNM \Ts\ on the ordering of the $N$ CNM components and the fraction $F_\mathrm{k}$, we do not observe strong correlations between other independent CNM and WNM Gaussian parameters. Example spectra and fits are shown in Figures \ref{fig:eg_plot1} and \ref{fig:eg_plot2}.

\subsection{Pseudo$-$Voigt spectral decomposition}
\label{subsec:psv}
Gaussian fitting has been widely used in ISM radio astronomy because the velocity profiles are assumed to be truly Gaussian. This assumption holds when internal motions are characterized by thermal processes or small-scale turbulence, and the damping wings in the spectra are insignificant. However, some authors have chosen to fit \psv\ profiles instead (\citealt{PLC2015,Remy2017}), with the reasoning that \psv\ function fits broad pedestals better. Therefore, we have tested both options in this work and assess the impact (if any) on the astrophysical conclusions (see \hyperref[apd:compare-GnV]{Appendix A} for more details).

We repeat the above fitting procedures using \psv\ functions instead of pure Gaussians. A \psv\ profile, a simple approximation for the Voigt function (\citealt{Wertheim1974}), is  a linear combination of a Gaussian curve $G(x)$ and a Lorentzian curve $L(x)$ sharing the same full width at half-maximum values. The normalized \psv\ profile by definition is given by:
\begin{equation}
V(x) = \eta L(x) + (1-\eta) G(x)
\label{eq:psvoigt}
\end{equation}
\noindent where $L(x)$ and $G(x)$ are the normalized Gaussian
and Lorentzian functions, with FWHM $\delta_V$ =  $\sqrt{2(ln2)}\delta_G = 2\delta_L$:
\begin{equation}
G(x) = \frac{1}{\sqrt{\pi}\delta_G}e^{-x^2/\delta^2_G}
\label{eq:normGaussian}
\end{equation}
\begin{equation}
L(x) = \frac{1}{\pi\delta_L}\left(\frac{1}{1+x^2/\delta^2_L}\right)
\label{eq:normLorentzian}
\end{equation}
\noindent and $\eta$ is a parameter which mixes the two functions. The results for the \psv\ fits are shown in Table \ref{table:parameters0}; example spectra and fits are shown in Figures \ref{fig:eg_plot1} and \ref{fig:eg_plot2}.

\section{Results}
\label{sec:results}
In Table \ref{table:parameters0} we list the parameters obtained from both the \psv\ and Gaussian fits for all CNM and WNM components along each sightline. Column 1 shows the Galactic longitude and latitude for each sightline; Columns 2 and 10 list the peak brightness temperature for each component. For the WNM components, this is the unabsorbed height $T_\mathrm{B}$, quoted with its error from the fits. For the CNM components, this is the spin temperature obtained from the fits multiplied by $(1-e^{-\tau})$ (see Equation \ref{eqtbc}), and is quoted without uncertainty. In Columns 3 and 11, we list the optical depth of each component: for CNM components this is the peak optical depth and its associated uncertainty, for WNM components it is the upper limit to peak opacity (set to be equal to the 1$\sigma$ noise level in absorption at the central velocity of the emission component) and its error is not quoted. Columns 4 and 12 show the central velocities for each CNM and WNM component; Columns 5 and 13 are the linewidth (FWHM) in VLSR of each CNM and WNM component; Columns 7 and 14 list the spin temperatures: for CNM components the spin temperature is derived from the fit to the \Texp\ profile and is quoted with its error, for WNM components \Ts\ is a lower limit calculated by the upper limit on optical depth in Columns (3, 11) $T_\mathrm{s,k} = T_\mathrm{0,k}/\tau_\mathrm{0,k}$ and has a very large error. In Columns 8 and 15, we quote the \hi\ column densities and their errors for each CNM and WNM component; Columns 9 and 16 are the fraction ($F$) of each WNM component lying in front of all CNM components or the order ($O$) of each CNM component along the sightline (integer numbers). The $\eta$ values in Column 6 are the fractions of the Lorentzian function in the \psv\ profile. Note that while we detect \hi\ absorption along all 79 sightlines, we exclude two sightlines (4C+16.15 and 4C+26.18) from further analysis because their opacity spectra ($e^{-\tau}$) are saturated and result in huge optical depths with enormous uncertainties.

\begin{figure*}[htbp]
 \center
\includegraphics[width=1.0\linewidth]{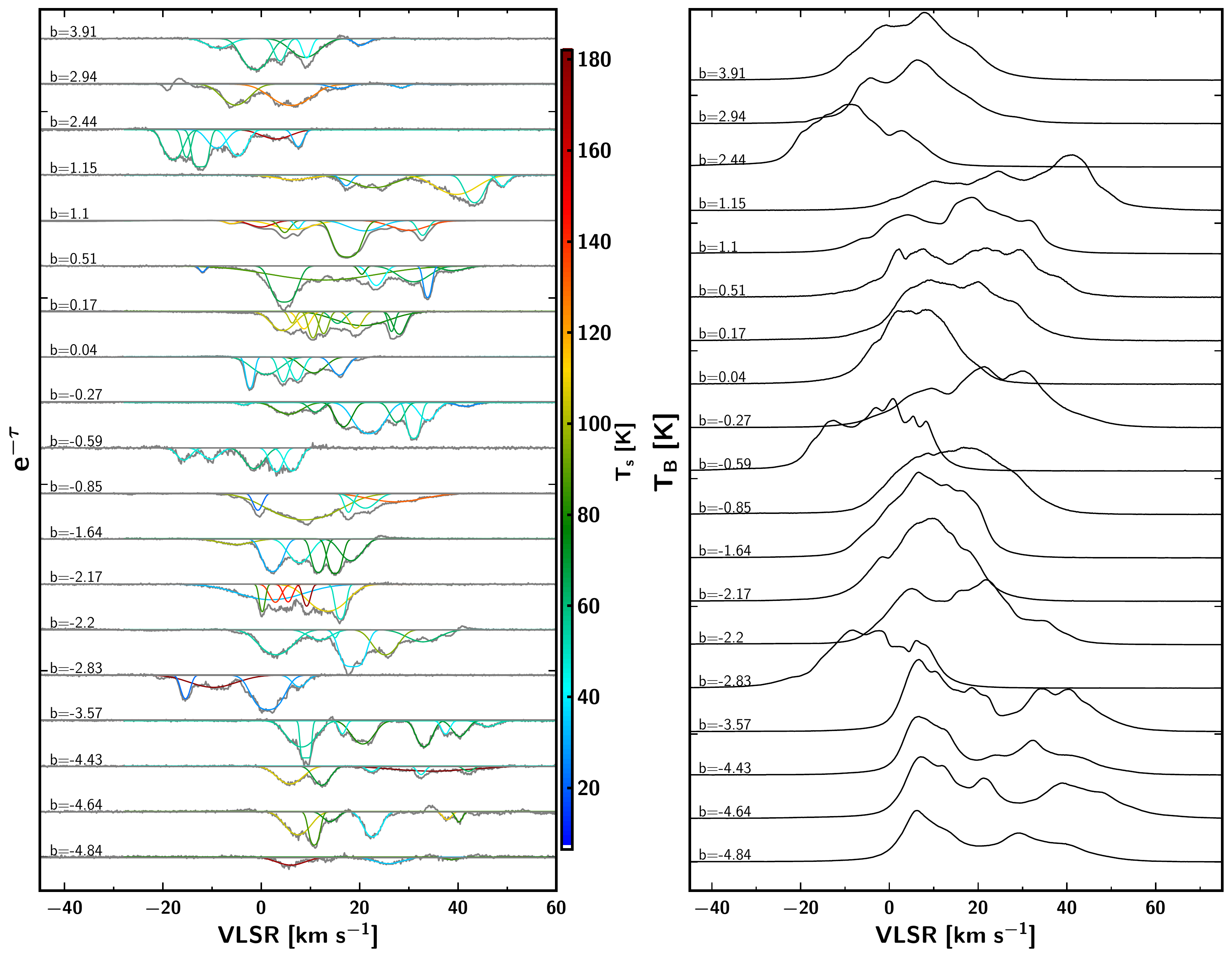}
\caption{Comparisons of absorption profiles (left) and emission expected profiles (right) for 19 sightlines in near Galactic plane. The colorbar on the left panel shows the \hi\ spin temperatures of CNM components. The sightlines are sorted in order of decreasing Galactic latitude $b$ and are offset along the y-axis for comparison.}
\label{fig:GP_all_spectra}
\end{figure*}

Along the remaining 77 sightlines, the total numbers of CNM components from the Gaussian and \psv\ fits are (349, 323), with column densities of (956.1, 1013.8) \nhiUnit. The numbers of WNM components are (327, 284), with column densities of (1609.7, 1712.5) \nhiUnit. Thus the total (WNM+CNM) column densities derived from Gaussian and \psv\ fits in this study are (2565.8, 2726.3) \nhiUnit, which equates to cold and warm gas fractions of 40\% and 60\%. 

The distributions of properties derived from Gaussian and \psv\ fits are almost indistinguishable. For consistency with past studies, we will present the results of the Gaussian fits for the rest of this paper, however, we present a detailed comparison between the Gaussian and \psv\ fits in the \hyperref[apd:compare-GnV]{Appendix A}.

\begin{figure*}
 \center
\includegraphics[width=1.0\linewidth]{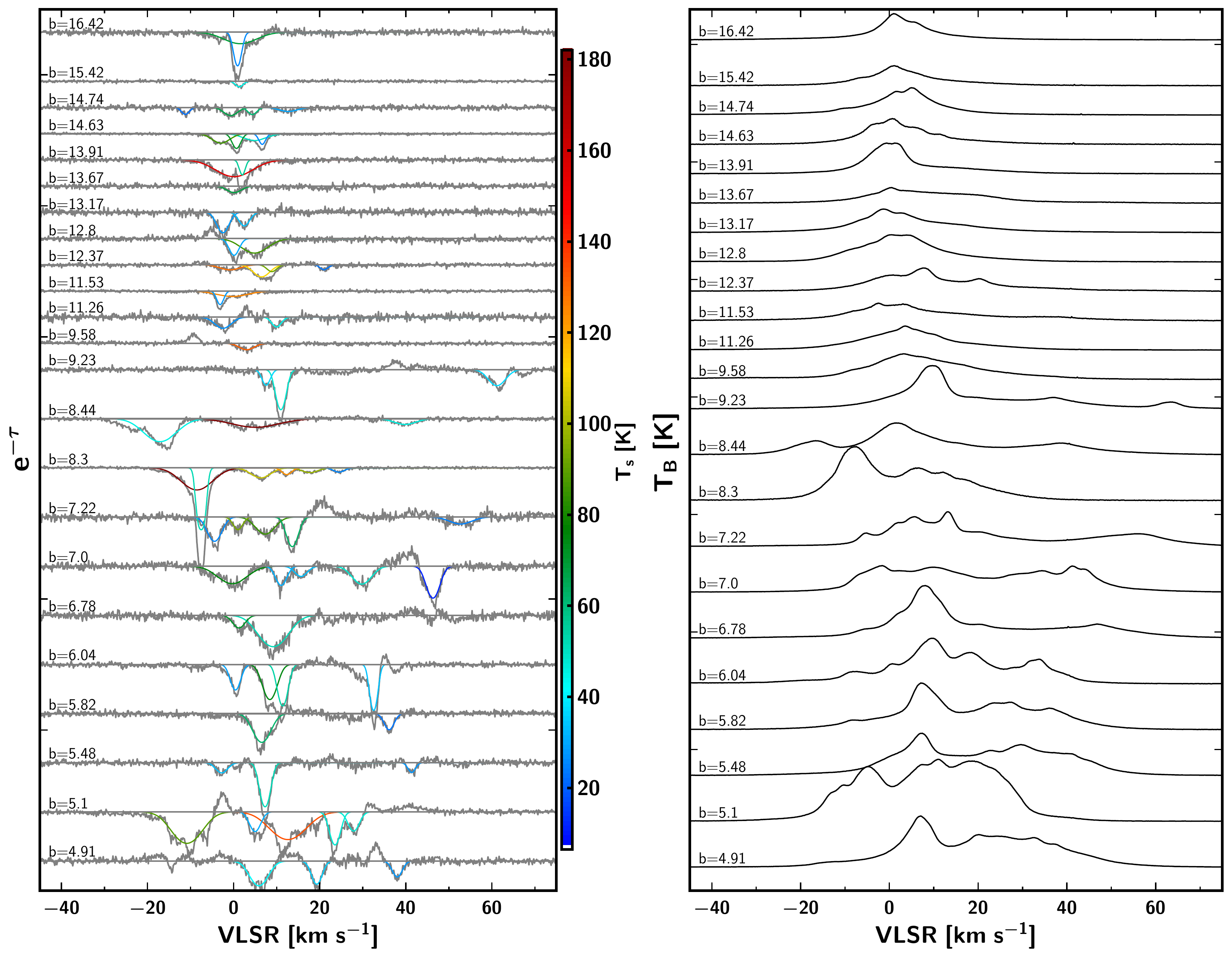}
\caption{The same as Figure \ref{fig:GP_all_spectra} but for 23 sightlines toward Gemini region.}
\label{fig:HGL_all_spectra}
\end{figure*}

\begin{figure*}
 \center
\includegraphics[width=1.0\linewidth]{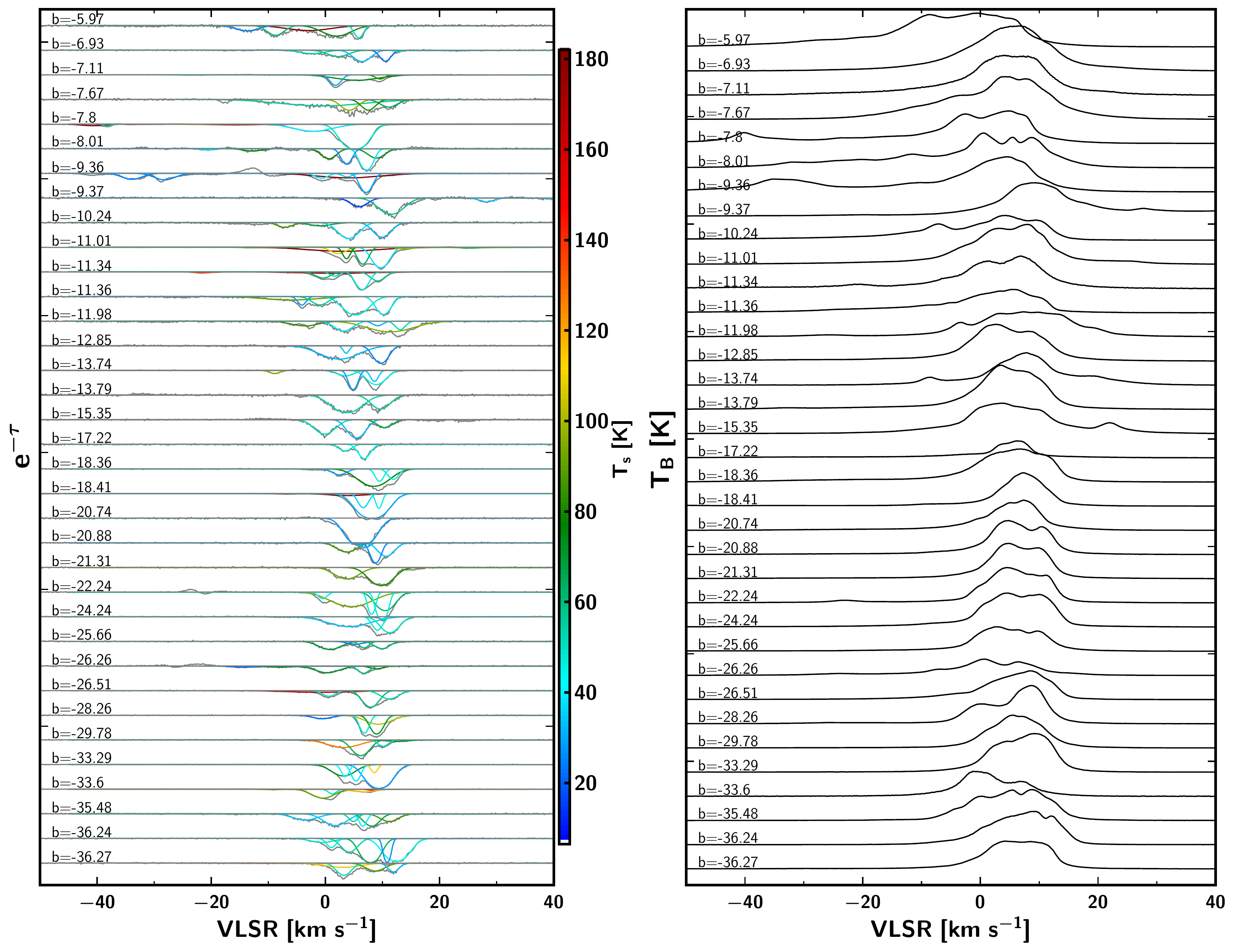}
\caption{The same as Figure \ref{fig:GP_all_spectra} but for 35 sightlines toward Taurus region.}
\label{fig:all_spectra_taurus}
\end{figure*}

Figures \ref{fig:GP_all_spectra}, \ref{fig:HGL_all_spectra} and \ref{fig:all_spectra_taurus} show the absorption and emission spectra for all sightlines, in the Galactic Plane, Gemini and Taurus regions respectively. In a few cases, we see the presence of ``emission'' features in absorption spectra ($e^{-\tau} > 1$) toward the weakest background sources. These sources are probably quite faint to measure the \hi\ absorption with Arecibo telescope, thus their derived absorption spectra are essentially noisier. During the spectral decompositions, we do not take these ``emission'' features into the fit.

Here (and from hereon) we define the `Galactic Plane' as $|b| < 5^{\circ}$ (19 sightlines), including Rosette, Mon OB1 and NGC 2264; the `Gemini region' as $b >$ 5$^{\circ}$ (23 sightlines), including minimal molecular gas; and the `Taurus region', including both Taurus and California, as $b <$ $-$5$^{\circ}$ (35 sightlines) (as illustrated in Figure \ref{fig:src_locations}). These latitude divisions (basically based on the distribution of $W_\mathrm{CO(1-0)}$ from \citealt{Dame2001}) roughly divide our sample into three characteristic physical regimes -- in-Plane, out-of-Plane with mostly diffuse sightlines (Gemini) and out-of-Plane with many dense/molecular sightlines (Taurus). The colors in the absorption spectra in the plots indicate the \hi\ spin temperature of each CNM component. In general, we detect strong absorption and emission in a wide range of VLSR from $-$10 to 50 \kms\ for sightlines in Gemini, from $-$20 to 20 \kms\ for those in Taurus and $-$50 to 50 \kms\ near the Galactic plane. The strongest absorption/emission is located at $\sim$0 \kms, and the closer to the Galactic plane, the more complex the profile shapes.

Besides the spectral decompositions, we also applied two alternative methods (one \Ts\ per velocity channel and single harmonic-mean \Ts\ along full sightline) to estimate the \hi\ spin temperature and column density from the absorption/emission pairs. We compare the results obtained from these different approaches in \hyperref[apd:ts-nhi-from-other-mehods]{Appendix B}.

\subsection{Optical depth}
\label{subsec: optical_depth}
\begin{figure}
\includegraphics[width=1.0\linewidth]{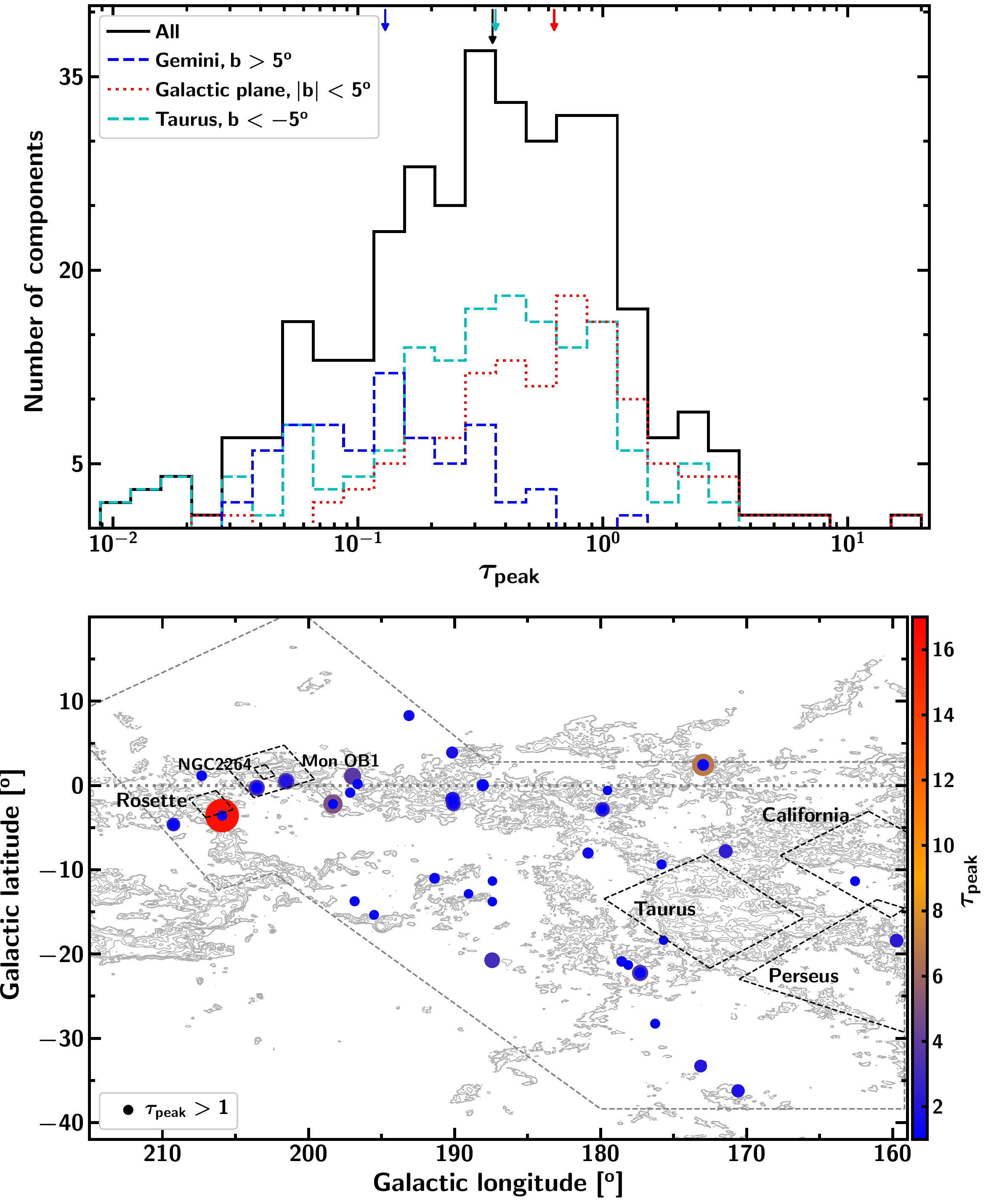}
\caption{Top panel: Histograms of \taupeak\ for CNM components along 77 sightlines: black for all CNM components, blue, red and cyan for the subsamples in Gemini, Galactic plane and Taurus respectively. The arrows show the medians. Bottom panel: The locations of components having \taupeak\ $>$ 1; both the colors and sizes of the circles represent the magnitudes of \taupeak; the horizontal dotted line marks the Galactic plane; the dashed rectangles roughly show the areas of molecular clouds; and the gray dashed lines roughly show the boundaries of the regions of interest.}
\label{fig:tau_hist}
\end{figure}

The peak optical depths, \taupeak, derived from Gaussian fitting range from $\sim$0.01 to 16.2. The values are summarized in histograms in Figure \ref{fig:tau_hist} (top panel), where we show the \taupeak\ distribution for all CNM components, as well as for each individual region. As expected, the optical depths in the Galactic plane sample are the highest, with a mean and median of (1.0, 0.63), and also relatively high toward Taurus, with mean and median values (0.54, 0.37) --  triple those toward the Gemini region (0.18, 0.13). The mean and median values for the total sample are $\sim$0.64 and $\sim$0.35. About 17\% of the CNM components have \taupeak\ $ > 1$, namely 60/349 Gaussian components; corresponding to dense gas regions. As shown in bottom panel of Figure \ref{fig:tau_hist}, the locations of these high optical depths are all either coincident with or close to sightlines with CO detections.

\subsection{Temperatures of cold and warm gas}
\label{subsec:temperatures}

The cold gas is characterized by spin temperature \Ts, defined from the populations of the upper and lower hyperfine levels. The spin temperature is influenced by many factors from the surrounding environment: the ambient radiation field, scattering by Lyman-$\alpha$ radiation, or collisions of \hi\ atoms with electrons, protons and other H atoms. In high density CNM regions ($n \gtrsim 100$ cm$^{-3}$, e.g. \citealt{Shaw2017}), collisions dominate the excitation of \hi\ atoms, so that \Ts\ is approximately equal to the kinetic temperature, \Tk\ (e.g. \citealt{Kulkarni1988}). In contrast, in more diffuse WNM gas, collisions are insufficient to thermalise the 21 cm transition, hence \Ts\ is generally expected to be lower than \Tk\ \citep[e.g.][]{Field1958PIRE,Deguchi1985,Liszt2001}. However, the Lyman-$\alpha$ radiation field from Galactic and extragalactic sources can couple the \hi\ spin temperature in the WNM with local gas motions. This mechanism, known as the Wouthuysen-Field effect \citep{Wouthuysen1952,Field1958PIRE,Field1959}, requires a large number of resonance scatterings of Lyman-$\alpha$ photons (i.e. large optical depth) and a recoil effect of the scattering atom (i.e. momentum transfer between \hi\ atom and Lyman-$\alpha$ photon). The Wouthuysen-Field effect is likely responsible for the high WNM spin temperatures inferred from sensitive absorption studies \citep[$\sim$10$^4$ K;][]{Murray2014,Murray2017,Murray2018}, which cannot be explained by steady-state collisional excitation.

As expected, our Gaussian fitting shows that the FWHM of WNM components is obviously broader than that of CNM components, with a mean value of 16 \kms\ vs 4 \kms\ respectively. For the warm gas, we calculate Doppler temperature (\citealt{Payne1982}) from the line-width ($\Delta V = $ FWHM) of each WNM component:
\begin{equation}
T_\mathrm{D}=21.86\times \Delta V^{2}
\label{eq:Doppler_temp}
\end{equation}
\noindent $T_\mathrm{D}$ is an upper limit on the kinetic temperature because the observed FWHM results from the combination of thermal and turbulent broadening. We also estimate $lower$ limits on \Ts\ for our WNM component, imposed by the upper limits on optical depth (see Table \ref{table:parameters0}).

\begin{figure}[htbp]
\includegraphics[width=1.0\linewidth]{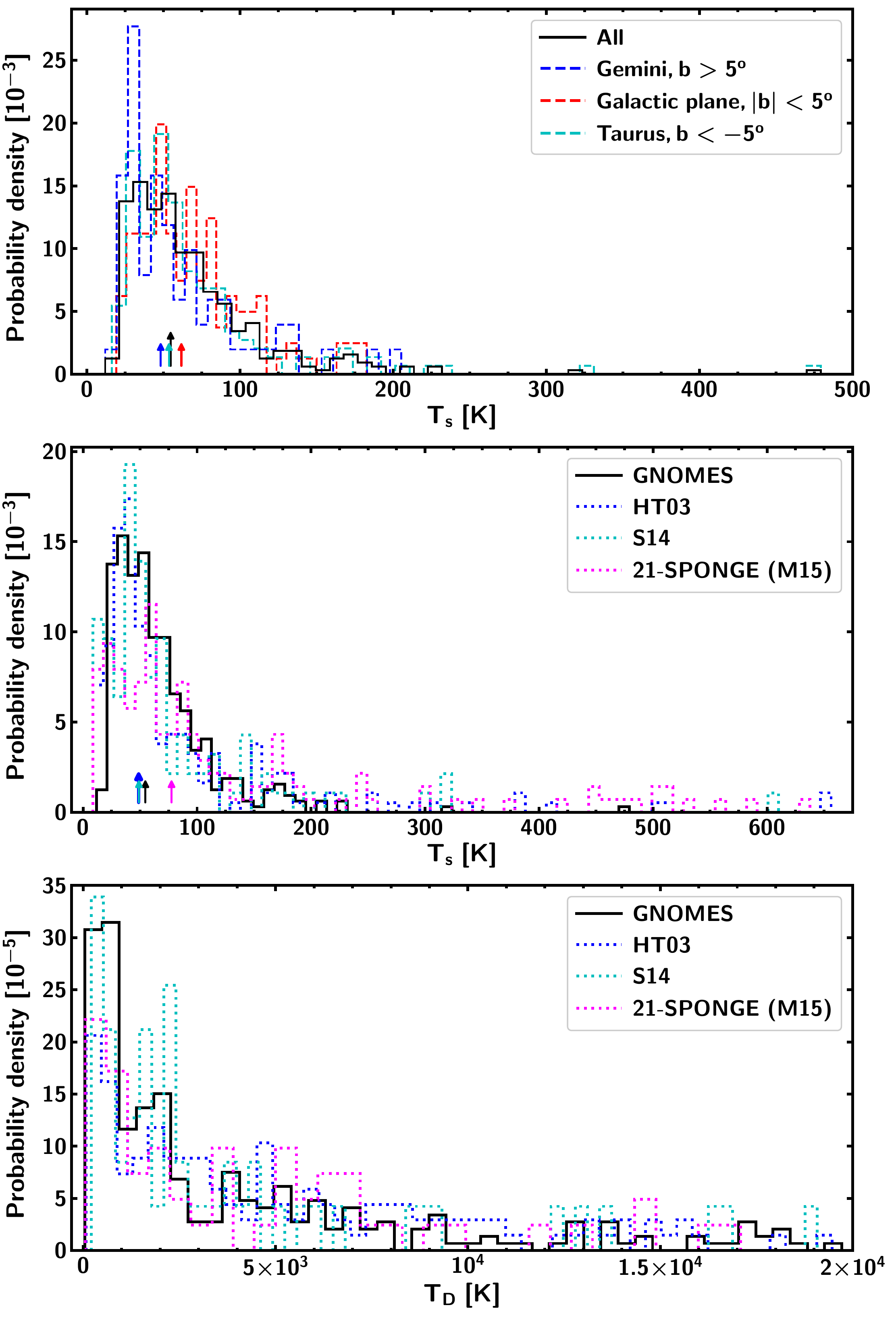}
\caption{Normalized histograms of \Ts\ for CNM components (top and middle), $T_\mathrm{D}$ for WNM components (bottom). In the top panel, the solid black line shows all CNM components; blue, red and cyan dashed lines show the subsamples in Gemini, the Galactic plane and Taurus respectively. In the middle and bottom panels, GNOMES in shown in solid black, HT03 in dotted blue, S14 in dotted cyan, and 21-SPONGE (M15) in dotted pink. The arrows show the medians.}
\label{fig:ts_tkmax}
\end{figure}

In Figure \ref{fig:ts_tkmax}, we show the distributions of temperatures derived from the Gaussian fits to all 77 sightlines, with the top and middle panels showing CNM \Ts; and the bottom panel WNM \TD. In the top panel, we also show the \Ts\ distributions of our three different regions; in the others, we plot the temperature histograms obtained from other studies (HT03, S14, M15) for comparison. Here is important to note that our optical depth sensitivity ($\sim$5 $\times$ 10$^{-3}$ per 1 \kms, obtained from the median of the 1$\sigma$ noise level in optical depth spectra) is lower than that of previous studies: $\sim$2 $\times$ 10$^{-3}$ per 1 \kms\ for both HT03 and S14, and $\sim$9 $\times$ 10$^{-4}$ per 0.42 \kms\ for M15.

\subsubsection{Temperatures of cold gas}
\label{subsec:CNM_temperatures}
For the CNM components, the spin temperatures fall in the range $\sim$10--480 K, with our \Ts\ distribution spreading mostly from 25--150 K with a peak at $\sim$50 K and a median at 54.6$\pm$4.6 K, as well as a column-density-weighted median of 74.8$\pm$9.6 K (the uncertainties of the medians are obtained from bootstrap re-sampling). Indeed, we see no obvious tail at \Ts\ $>$ 200 K. In order to examine how \Ts\ medians vary with the sensitivity in optical depth, we group the \Ts\ of CNM components into four different ranges of optical depth sensitivity with similar sample sizes. From high to low sensitivities, the spin temperature medians are 61.7$\pm$16.5 K, 61.2$\pm$10.4 K, 54.3$\pm$6.6 K and 48.3$\pm$8.3 K respectively; and the highest \Ts\ values belong to the highest sensitivity range. Hence, while the uncertainty is large, this (unsurprisingly) suggests a dependence of the median measured \Ts\ on the opacity sensitivity. This is also consistent with the fact that the highest-sensitivity survey plotted in Figure \ref{fig:ts_tkmax} (21-SPONGE; M15) measured a median \Ts\ 25 K higher than the three others. This simply indicates that observations with improved sensitivity in \hi\ absorption are able to directly measure the \Ts\ of warmer \hi\ gas (in the temperature range of UNM or WNM), resulting in a higher \Ts\ median.

From our results, $\sim$98\% of CNM components (equivalent to 97\% by column density of the cold gas) lie below 200 K. In fact, most of the CNM gas (80\% by number of components and 74\% by column density) has \Ts\ = 25--100 K. About 14\% of the gas has 100 $<$ \Ts\ $<$ 200 K, and it contributes $\sim$22\% to CNM column density. The remaining of 4\% of the gas is very cold with \Ts\ $<$ 25 K but contributes only $<$1\% in CNM column density. Such extremely cold gas has been observed by several studies: HT03, S14, M15, \citet{Meyer2006,Meyer2012}. In particular, \citet{Knapp1972} claimed to have observed a cold \hi\ cloud with spin temperature of 7--20 K; and \citet{Wannier1999} also measured a \Ts\ of 20--60 K in the Perseus B5 molecular cloud. While some fraction of very low \Ts\ components may be spurious fits, the minimum CNM \Ts\ from analytical models is $\sim$20 K (\citealt{Wolfire2003}), and such low temperatures might also occur in regions where photoelectric heating by dust grains is reduced \citep[e.g.][]{Wolfire1995, Heiles2003b}.

We find no evidence for differences in the distribution of CNM spin temperatures in our three regions, despite their different characteristic environments (as shown in top panel of Figure \ref{fig:ts_tkmax}). The CNM \Ts\ distributions for the Gemini, Galactic plane and Taurus regions, have relatively close medians of 48.1$\pm$8.8 K, 61.6$\pm$8.7 K, 53.5$\pm$4.9 K respectively, and the differences are broadly consistent with the background continuum source flux density distributions -- Gemini has the highest proportion of weak sources; the Galactic plane has the lowest (as listed in Table \ref{table:source_list}). Conversely, if there were significant differences in the underlying CNM \Ts\ distribution, then we might have observed different medians that do not scale with the background source intensity.

These \Ts\ medians are also in good general agreement with the results from HT03 (48.8 K, along random sightlines in the Arecibo sky), S14 (49.0 K, around the Perseus molecular cloud), and \citet{Denes2018} (48.0 K, in the Riegel-Crutcher cloud). Our values are lower than those of the 21-SPONGE survey (M15, M18 at 77.6 K, 73.6 K respectively), reflecting its higher optical depth sensitivity. More importantly, the general shape of the spin temperature histograms of the various surveys are similar, with the exception of the obvious higher-\Ts\ tail in 21-SPONGE (see the middle panel of Figure \ref{fig:ts_tkmax}). Together, this is consistent with a picture in which the spin temperature distribution of the CNM remains relatively consistent throughout the Galactic ISM; in particular, measurements along random Galactic sightlines agree well with those in fields focused around molecular clouds. Any measured differences are largely a result of higher-sensitivity observations recovering larger portions of the higher-\Ts\ tail. This may suggest a universality of cold \hi\ properties, which will be explored more in the following sections. The same conclusion was drawn by S14.

\subsubsection{Temperatures of warm gas}
\label{subsec:WNM_temperatures}
For the WNM components on the bottom panel of Figure \ref{fig:ts_tkmax}, we show the histogram of \TD. Most of the WNM lies below 5000 K, but the \TD\ distribution exhibits a long tail well beyond 20,000 K. About 45\% of the WNM components (equivalent to 40\% by column density) is in the thermally unstable regime with \TD\ = 500--5000 K \citep[see e.g. HT03,][M15, M18]{Roy2013}, implying that $\sim$24\% of all of the atomic gas does not lie in thermal equilibrium. This fraction is consistent with previous studies: HT03 found that at least 48\% of the WNM (equivalent to 30\% of the total gas) is thermally unstable, and \citet{Roy2013} also found at least $\sim$28\% of the gas in the unstable range. The fractions in M15 and M18 are slightly lower, at $\sim$20\%, but agree well with our finding within their uncertainties of $\gtrsim$10\%. Nearly 14\% of our WNM components have \TD\ $<$ 500 K, corresponding to 4\% in WNM column density, with the lowest limit at $\sim$108 K (only two out of 349 WNM components have \TD\ $<$ 200 K, one of which is shown in the upper panels of Figure \ref{fig:eg_plot2}). In this temperature range the WNM gas is too cold to classify as thermally unstable. From examination of the spectra, we believe these components are real, and not artifacts of poor fitting. They are not seen in absorption likely because they are thin, cold \hi\ clouds in the diffuse medium with very low column density (as reported by \citealt{StanimirovicHeiles2005} and \citealt{Stanimirovic2007}). Typical 1$\sigma$ upper limits on $\tau$ for these ``cold'' WNM components (\TD\ $<$ 500 K) are at a median of $\sim$0.07, and thus their lower limits on \Ts\ are at a median of 148 K. Most of the rest of the WNM (25\% in component number, 37\% in column density) lies between 5000 K and 20,000 K. The fraction of the gas with \TD\ $>$ 20,000 K (not shown in the histogram) is 16\% in components and 19\% in column density. In reality, the gas would be ionized at temperatures above $\sim$10,000 K, so WNM components with \TD\ $>$ 10,000 K either must either include multiple narrower components or have highly supersonic motions (e.g. HT03).

\subsection{Turbulent Mach Number Distribution}
From \hi\ observations, it is almost impossible to constrain the sonic Mach number for the WNM. However, the Mach number of non-thermal CNM motions can be estimated from the line-width ($\Delta V$= FWHM) and spin temperature. Assuming that the FWHM of each CNM component results from both thermal broadening and turbulent motions (\citealt{Leung1976,Liszt2001}), $\Delta V$ can be expressed as:

\begin{equation}
\frac{(\Delta V)^2}{8~ln2} = \frac{k_\mathrm{B} T_\mathrm{k}}{m_\mathrm{H}} + V^2_\mathrm{turb,1D}
\label{eq:tkmax_turb}
\end{equation}

\noindent where $k_\mathrm{B}$ is Boltzmann's constant, $T_\mathrm{k}$ is the kinetic temperature, $m_\mathrm{H}$ is the mass of a hydrogen atom and $V^2_\mathrm{turb,1D}$ is the one-dimensional mean square turbulent velocity which relates to the mean square three-dimensional turbulence velocity of the CNM as $V^2_\mathrm{turb,3D}=3V^2_\mathrm{turb,1D}$. For the CNM, since thermal equilibrium is established quickly, we may assume $T_\mathrm{k}$ = \Ts, then the isothermal sound speed can be computed as $C_\mathrm{s} = \sqrt{k T_\mathrm{s}/\mu m_\mathrm{H}}$. Dividing $\Delta V$ and $V_\mathrm{turb,3D}$ by the sound speed $C_\mathrm{s}$ will respectively give the sonic Mach number ($M_\mathrm{s}$) and the turbulent Mach number ($M_\mathrm{t}$) of the CNM. Adopting a mean atomic weight of $\mu$ = 1.4 for the Galactic ISM, we may write:

\begin{equation}
M_\mathrm{t} = \frac{V_\mathrm{turb,3D}}{C_\mathrm{s}} = \sqrt{3\left( \frac{M^2_\mathrm{s}}{8~ln2} - \mu \right)}.
\label{eq:Ms}
\end{equation}

\begin{figure}[htbp]
\includegraphics[width=1.0\linewidth]{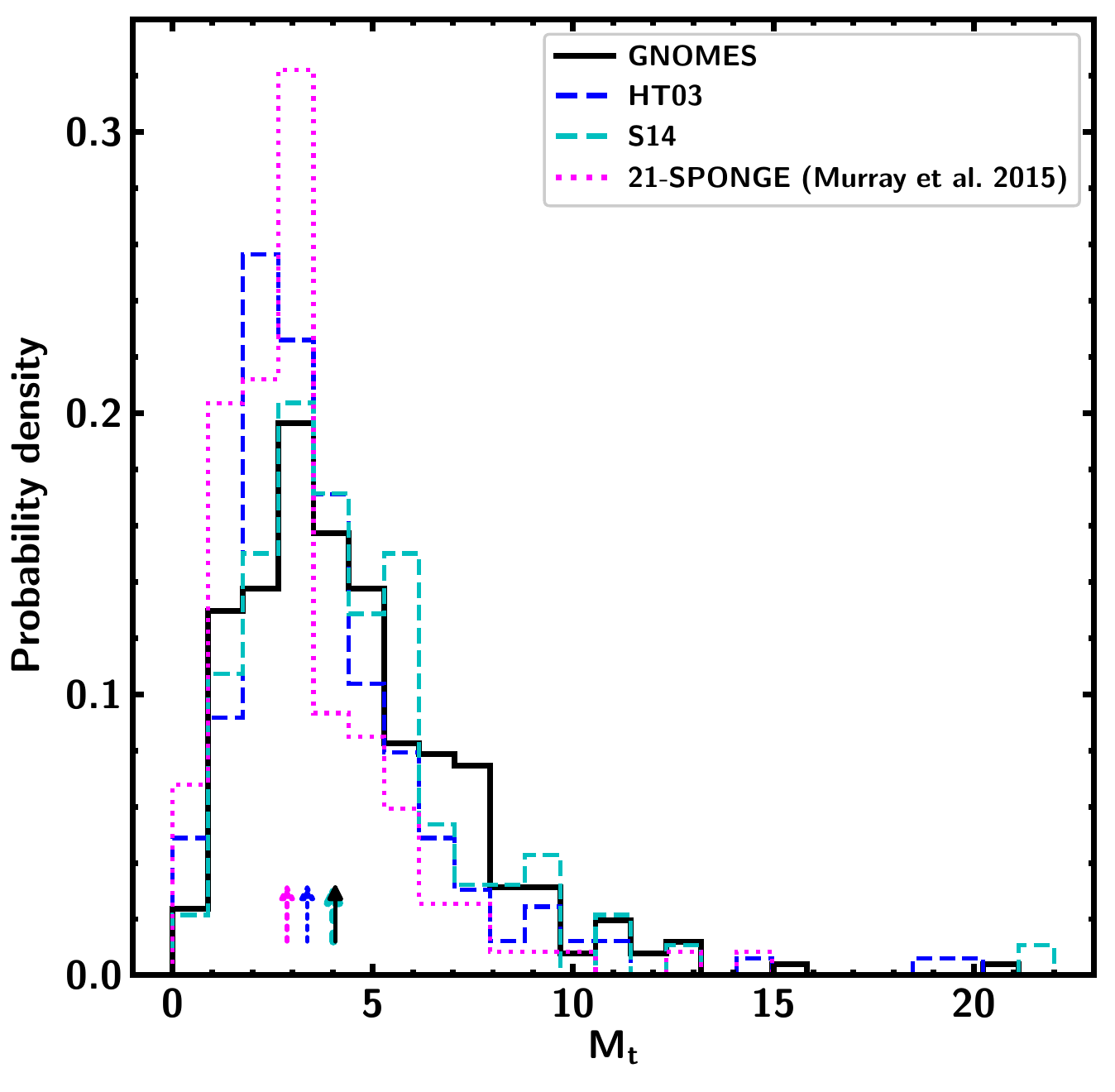}
\caption{Normalized histograms of turbulent Mach numbers for all CNM Gaussian components: GNOMES in black, HT03 in blue, S14 in green and M15 in pink.}
\label{fig:Ms_hist}
\end{figure}

We show a histogram of derived $M_\mathrm{t}$ values in Figure \ref{fig:Ms_hist}. The median sonic Mach number for the whole sample of 77 sources is $M_\mathrm{t}$ = 4.1$\pm$0.3, and the histogram peaks around  $M_\mathrm{t}$ = 3.5--4.0. This result is slightly higher than found by HT03, M15 and M18 (at $M_\mathrm{t}$ = 3.4, 2.9 and 3.1 respectively) and agrees best with S14 in the Perseus molecular cloud region ($M_\mathrm{t}$ = 4.0). We also note that \citet{Burkhart2010} find $M_\mathrm{t}$ $\sim$ 4.0, for CNM around the SMC bar even though the gas properties in the SMC are very different from those in the Milky Way. All of these results support a picture in which internal macroscopic motions in the CNM are highly supersonic. One possible explanation for this dynamical property of the CNM is that the cold \hi\ gas along the line of sight consists of a few individual long living CNM structures (clumps); while the individual clumps are subsonic, their relative velocities are supersonic with respect to the CNM sound speed as they were inherited from the velocity dispersion of the WNM (from which CNM was formed). This would imply that the observed Mach number along Galactic sightlines is due to the velocity dispersion from the relative motions of clumps rather than the internal velocity dispersion of the cold \hi\ medium \citep[e.g.][]{KoyamaInutsuka2002,Heitsch2005,HennebelleAudit2007,Hennebelle2007MAMD,Saury2014}. Alternatively, it has been suggested that the supersonic motions of the cold \hi\ along the edges of the SMC bar may be related to the shearing/turbulent flows and/or shocks between the bar and the surrounding gas \citep{Burkhart2010}; it is possible that similar processes may also be implicated in some Milky Way gas.

\subsection{\hi\ column density distributions}
\label{subsec:nhi_dist}

The \hi\ column density of the cold absorbing \hi\ along each sightline is calculated by:

\begin{equation}
N_\mathrm{HI,CNM} = C_0 \int T_\mathrm{s}\ \tau_{v} ~dv
\label{eq:nhi_cnm}
\end{equation}

\noindent where $C_0 = 1.823\times 10^{18}$ cm$^{-2}$ K$^{-1}$ (km s$^{-1}$)$^{-1}$, and for the non-absorbing emission components, we estimate the column density as

\begin{equation}
N_\mathrm{HI,WNM} = C_0 \int T_{B} ~dv
\label{eq:nhi_wnm}
\end{equation}

\noindent where \TB\ is the brightness temperature.

\begin{figure}
\includegraphics[width=1.0\linewidth]{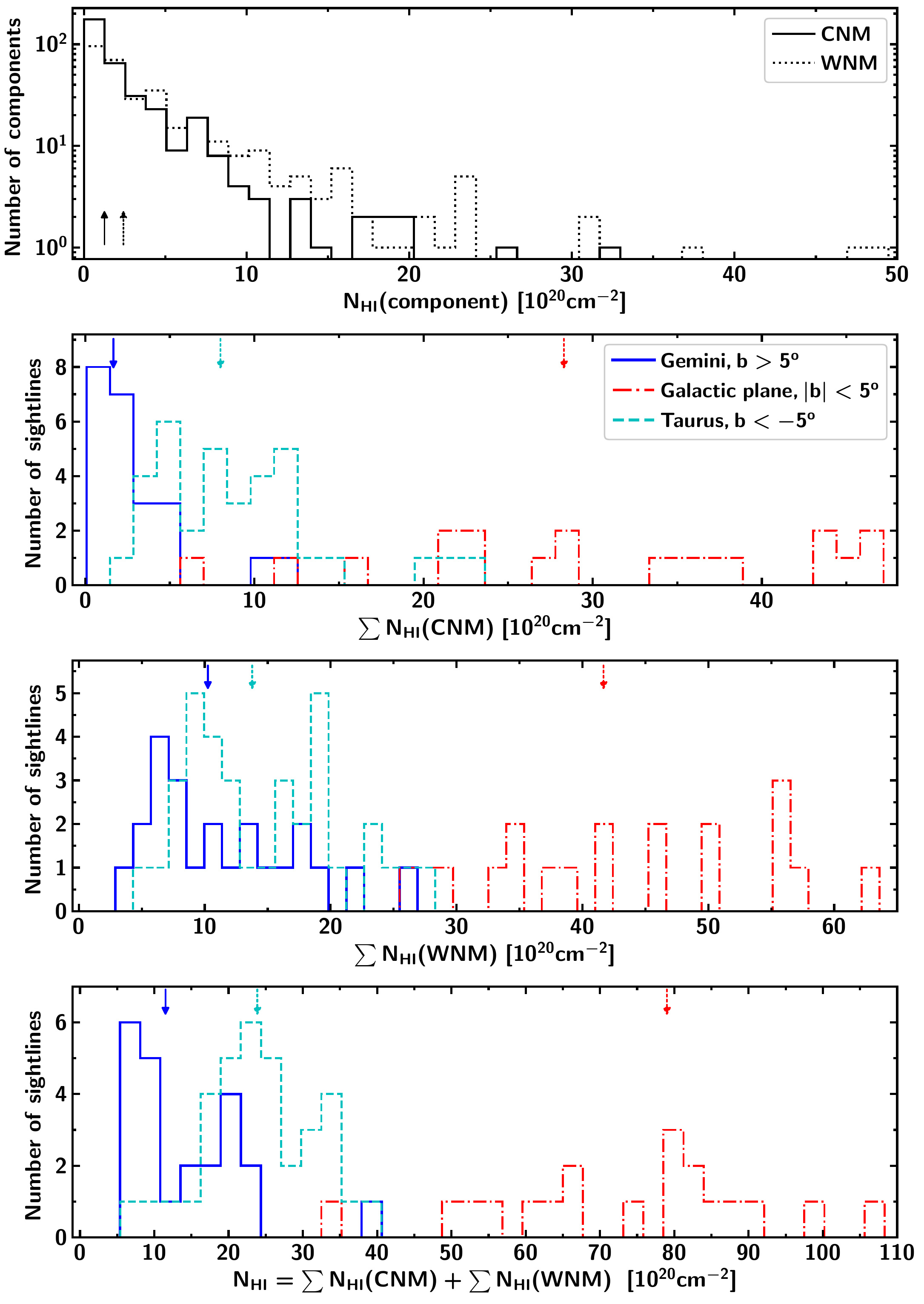}
\caption{Histograms of \NHI\ for all CNM (solid) and WNM (dotted) Gaussian components (top), $\sum N_\mathrm{HI,CNM}$ (second panel), $\sum N_\mathrm{HI,WNM}$ (third panel), and total \NHI\ = $\sum N_\mathrm{HI,CNM} + \sum N_\mathrm{HI,WNM}$ (bottom) for all sightlines in different regions: blue for Gemini, red for Galactic plane, cyan for Taurus. The arrows show the medians.}
\label{fig:hist_nhi_3regions}
\end{figure}

We summarize the column densities from the Gaussian fits in Figure \ref{fig:hist_nhi_3regions}. In the top panel we plot histograms of \NHI\ for all CNM and WNM Gaussian components. The two phases have similar column density distributions regardless of Galactic latitudes. They have relatively close medians (1.4 \nhiUnit\ for CNM and 2.3 \nhiUnit\ for WNM) even though the column density ranges are enormous ((0.04--32.0)\nhiUnit\ for CNM and (0.06--48.6)\nhiUnit\ for WNM). In the three remaining panels of Figure \ref{fig:hist_nhi_3regions}, we show histograms of $\sum N_\mathrm{HI,CNM}$, $\sum N_\mathrm{HI,WNM}$ and total \NHI\ along sightlines (color-coded by region). Table \ref{table:nhi_medians} lists the means and medians of the distributions (also indicated by arrows in the histograms). Now, when summing column densities of all CNM and WNM components along each line of sight, it is obvious that the column densities distributions in the three regions are well separated, except for the case of $\sum N_\mathrm{HI,WNM}$ in Gemini and Taurus. Gemini occupies the low \NHI\ regimes, the Galactic plane samples the highest \NHI\ with the sightlines toward the Perseus and Outer arms, and so Taurus is in the intermediate range. Our total CNM column density is about two thirds of the total WNM column density (956.1 \nhiUnit\ versus 1609.7 \nhiUnit). The median of total \NHI\ in Taurus is nearly double that of Gemini (23.8 \nhiUnit\ against 11.5 \nhiUnit), but only about one third of the \NHI\ median in Galactic plane (79.0 \nhiUnit). While the \NHI\ distribution in Galactic plane is widely spread, the histograms of other regions are concentrated in well-defined ranges. Most Galactic plane sightlines ($|b|<5^{\circ}$) with high column densities (\NHI\ $>$ 50 \nhiUnit) are close to NGC 2264 and Rosette molecular clouds where the CO contours peak; on the contrary, all other diffuse sightlines above the plane are found further away from the two molecular clouds ($b>5^{\circ}$), and apparently do not have significant CO detections. In most cases, the CNM contributes significantly to the total \hi\ column density.

\begin{table}
\begin{center}
\fontsize{8}{7}\selectfont
\caption{Means and medians of histograms in Figure \ref{fig:hist_nhi_3regions}}
\centering
\label{table:nhi_medians}
\begin{tabular}{lcc}
\noalign{\smallskip} \hline \hline \noalign{\smallskip}
\shortstack{Region\\ \ } & \shortstack{Mean \NHI \\ $\left[10^{20}\ \mathrm{cm}^{-2}\right]$} & \shortstack{Median \NHI \\$\left[10^{20}\ \mathrm{cm}^{-2}\right]$}\\
\hline
CNM, Gemini ($b > 5^{\circ}$) & 2.7 & 1.7\\
CNM, Galactic plane ($|b|<5^{\circ}$) & 30.3 & 28.3\\
CNM, Taurus ($b < -5^{\circ}$) & 9.1 & 8.0\\
\\
WNM, Gemini & 11.6 & 10.3\\
WNM, Galactic plane & 43.4 & 41.7\\
WNM, Taurus & 14.8 & 13.8\\
\\
CNM+WNM, Gemini & 14.3 & 11.5\\
CNM+WNM, Galactic plane & 73.7 & 79.0\\
CNM+WNM, Taurus & 23.9 & 23.8\\
\hline
\end{tabular}
\end{center}
\end{table}

\begin{figure}
\includegraphics[width=1.0\linewidth]{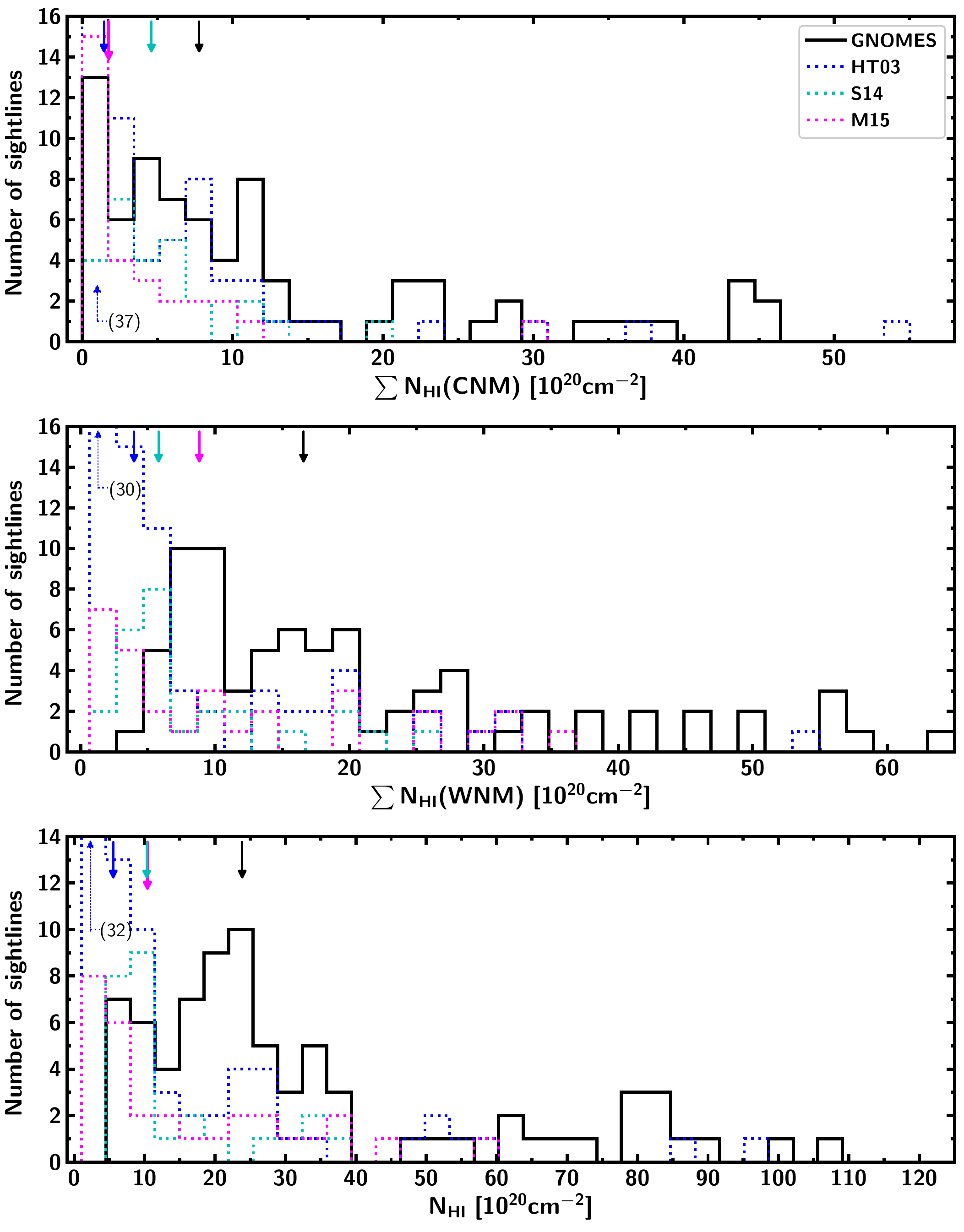}
\caption{Histograms of $\sum N_\mathrm{HI,CNM}$ (top), $\sum N_\mathrm{HI,WNM}$ (middle) and total \NHI\ = $\sum N_\mathrm{HI,CNM} + \sum N_\mathrm{HI,WNM}$ (bottom) for each sightline estimated from Gaussian decomposition fit: black for GNOMES, blue for HT03, cyan for S14 and pink for M15. The solid downward arrows show the medians. The upward arrows indicate the maximum values beyond the tops of the figure (HT03 only).}
\label{fig:hist_nhi}
\end{figure}

In Figure \ref{fig:hist_nhi}, we compare our results with the findings of previous surveys (here HT03, S14 and M15) that applied the same basic methodology in observation and data analysis, but for different regions of the sky. In all cases, our \NHI\ distributions are slightly shifted toward higher column densities. Our lowest \NHI\ median is found in the Gemini region, which is comparable with the median from S14 for Perseus molecular cloud (10.3 \nhiUnit) and M15 (10.4 \nhiUnit), but still higher than that obtained by HT03 (5.5 \nhiUnit). These differences likely reflect genuine differences in the sightlines probed by the different studies. The present work contains a high proportion of positions either in the Galactic Plane or close to molecular regions; meanwhile, HT03 and M15 mostly observed random sightlines in the neighborhood of the Solar system, and the S14 sightlines around Perseus mostly sampled material away from the main body of the molecular cloud.

\subsection{\NHIthin\ vs \NHI}
\label{subsec:nhi_scaling}
\begin{figure}
\includegraphics[width=1.0\linewidth]{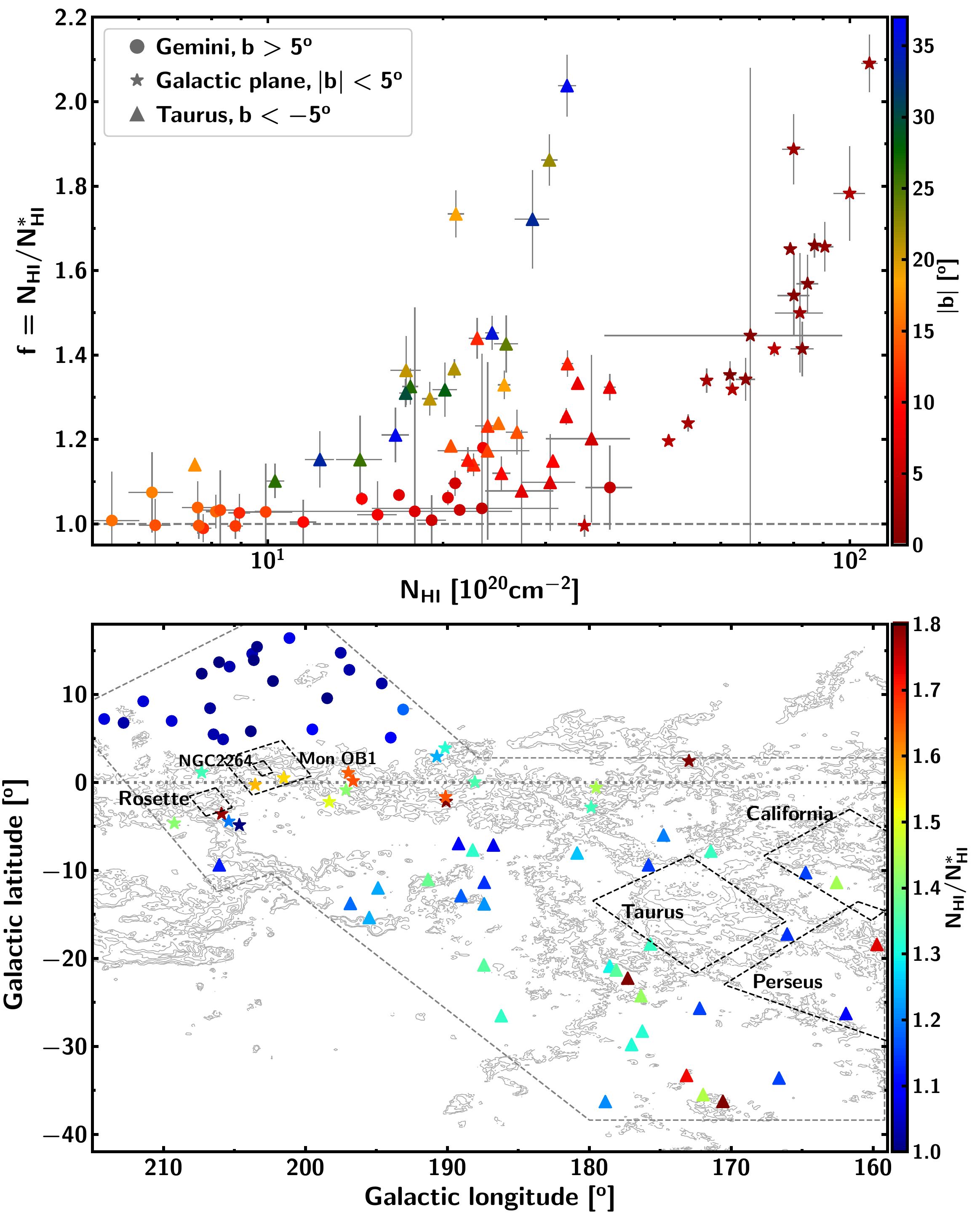}
\caption{Top panel: $f=$ \NHI/\NHIthin\ as a function of \NHI\ in Gemini (dots), Galactic plane (stars) and Taurus (triangles) regions. The colors represent the absolute values of Galactic latitude $|b|$; the dashed line marks where \NHIthin\ = \NHI. Bottom panel: The map of $f=$ \NHI/\NHIthin\ for our 77 sightlines toward the three regions; the horizontal dotted line shows the Galactic plane; the dashed rectangles roughly show the areas of molecular clouds; and the gray dashed lines roughly show the boundaries of the regions of interest. The colors represent the values of the ratio $f$.}
\label{fig:ratio_vs_nhi}
\end{figure}

The optically thin \hi\ column density, \NHIthin, is directly estimated from the expected emission profile, so is proportional to the profile area. On the contrary, to calculate the opacity-corrected \NHI\ one needs both spin temperature and optical depth, which can be derived from the combination of on-/off-source spectra. Based on our results, we can compare the two estimates. The ratio $f=$ \NHI/\NHIthin\ for all 77 sightlines is plotted as a function of total \NHI\ in Figure \ref{fig:ratio_vs_nhi}. For the full sample, $f$ has a mean and median of 1.28 and 1.21 respectively. At low column densities below $\sim$1$\times 10^{21}$ cm$^{-2}$, \NHIthin\ is comparable to the corrected \NHI\ ($f\sim 1.0$); the ratio then rises as column density increases, reaching $f\sim1.8$ by the time \NHI\ has increased by an order of magnitude ($1\times 10^{22}$ cm$^{-2}$). Interestingly, the ratio \NHI/\NHIthin\ for the Gemini region is almost flat and consistent with unity over the whole of its \NHI\ range (5--40) \nhiUnit, and as seen in the lower panel, no sightlines in this region fall within CO contours. Meanwhile, the ratio $f$ around Taurus stays mostly around the mean value of 1.3, whereas within the Galactic plane it exhibits a rising trend with \NHI. This indicates that significant corrections are needed to account for opacity effects, especially in denser gas regimes or at low latitudes \citep[see also][]{Dickey2003,Bihr2015}.

Nevertheless, the question of how best to correct for \hi\ opacity effects when emission/absorption pairs are not available is still open (see the discussion in Section \ref{sec:opacity_corrections}). A few methods have been proposed: e.g. a linear correlation between $f$ and log$_{10}(N^*_\mathrm{HI})$ \citep{Dickey2000,Lee2015} or a simple isothermal correction assuming the same \Ts\ along the lines-of-sight \citep[as employed by][]{Liszt2014b,Remy2017}. The general increase of the correction factor $f$ as a function of \NHI\ is obvious, but the relationship between the two appears to be environmentally dependent, as just discussed above for our regions. Literature values of $f$ include $\sim$1.1 in the range \NHI\ = (3.9--13.0) \nhiUnit\ for 26 sightlines within and around the Perseus molecular cloud  \citep{Lee2015}; $f\sim$ (1.1--1.3) for 79 random sightlines in Millennium survey (HT03); and $f<1.2$ for reddenings of $E(B-V) \lesssim 0.5$ mag, for sightlines at high Galactic latitudes \citep{Liszt2014}.

\subsection{The CNM fraction, \FCNM}
\label{subsec:cnm}
\begin{figure}
\includegraphics[width=1.0\linewidth]{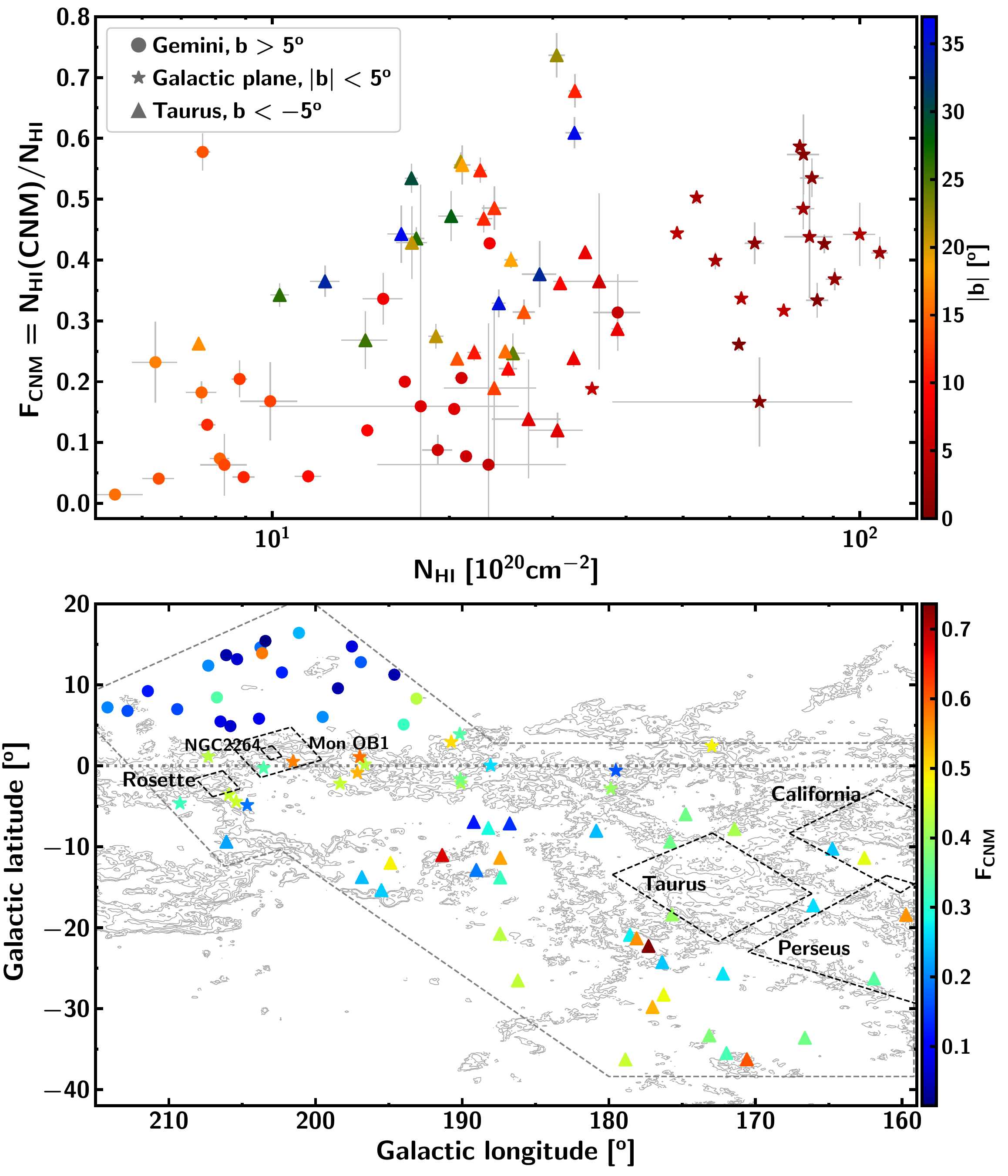}
\caption{Top panel: \FCNM\ = \NCNM/\NHI\ as a function of total \NHI\ toward Gemini (dots), Galactic plane (stars) and Taurus (triangles) regions. The colors represent the absolute values of Galactic latitude $|b|$. Bottom panel: The map of \FCNM\ for all 77 sightlines toward the three regions; the horizontal dotted line shows the Galactic plane; the dashed rectangles roughly show the areas of molecular clouds; and the gray dashed lines roughly show the boundaries of the regions of interest. The colors represent the values of \FCNM.}
\label{fig:fcnm}
\end{figure}

In the first panel of Figure \ref{fig:fcnm}, we present the variation of the CNM fraction (\FCNM\ = \NCNM/\NHI) as a function of \NHI. While the \FCNM\ does increase with \NHI, there is clearly a very large scatter. We measure median values of \FCNM\ = 0.43 and 0.37 respectively for the Galactic plane and Taurus regions -- more than double the value of 0.16 for Gemini. We find 9/77 sightlines with \FCNM\ $<$ 0.1, all of them in Gemini, above the Galactic Plane. The highest \FCNM\ values are found in the proximity of the giant molecular clouds (Rosette, NGC 2264 and in the south of Taurus and California) as shown in the map in the lower panel of the Figure. The Taurus \FCNM\ median is comparable to the value of 0.35 found in Perseus (S14) -- a similar region in and around a molecular cloud. Similarly, the median for the Gemini region is consistent with the value of 0.20 found by M15 and 0.23 by HT03, both of which sampled largely diffuse sightlines. Thus, the CNM fraction around molecular clouds seems to be higher than in diffuse regions. This agrees well with a scenario in which a high CNM fraction is required for molecule formation, and GMCs are built-up stage by stage -- from WNM-rich gas to CNM-rich gas to molecular clouds.

Overall the \FCNM\ from all GNOMES sightlines does not exceed 75\% -- a result that agrees with the \FCNM\ range observed in different regions of the Galaxy by HT03 and S14, also using the Arecibo telescope, and M15 and M18, using the VLA for absorption. Our CNM fraction is also close to the range of \FCNM\ $\sim 40$--70\% found in the numerical simulations of \citet{Kim2014}.

\begin{figure}
\includegraphics[width=1.0\linewidth]{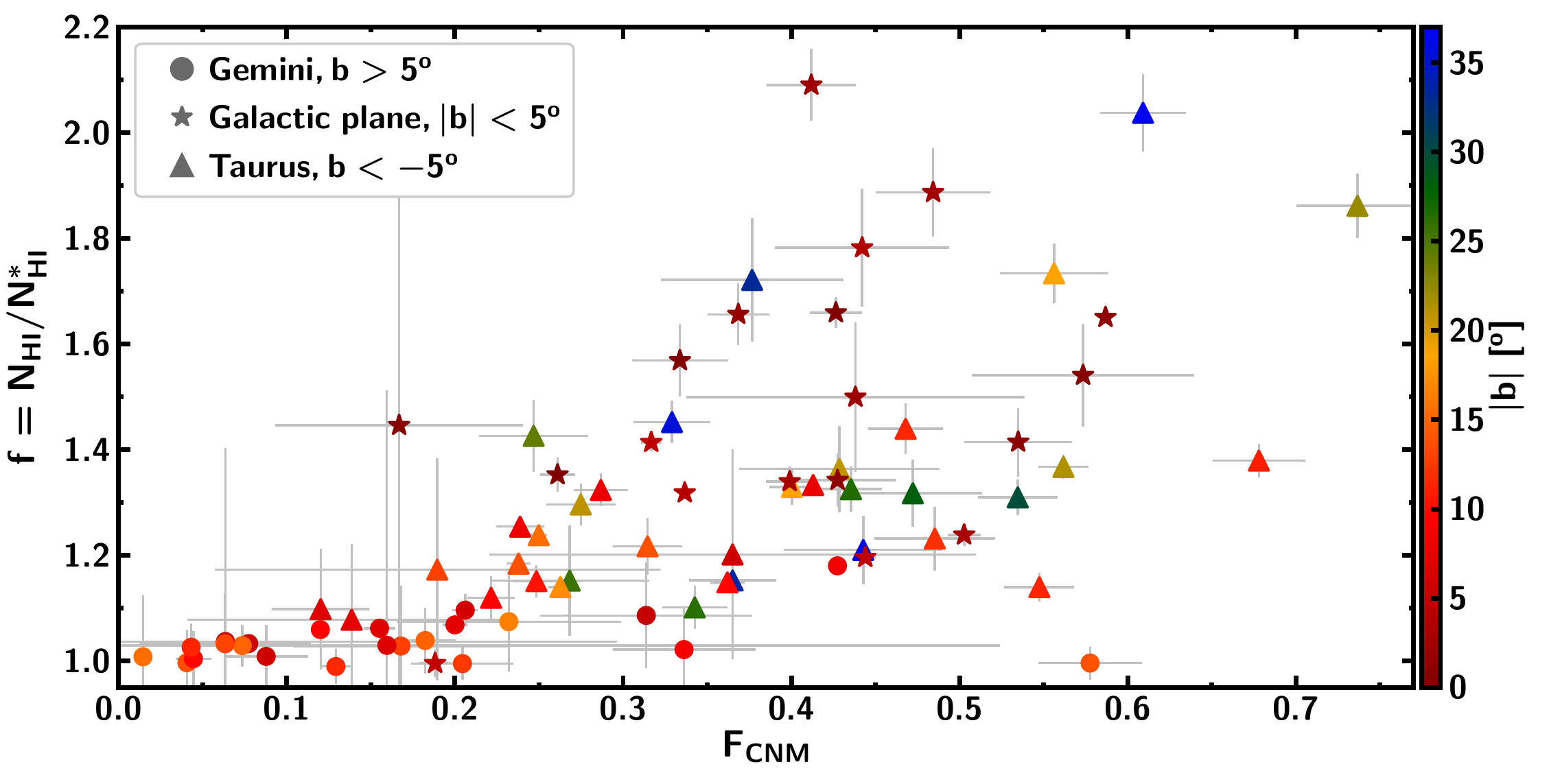}
\caption{$f=$ \NHI/\NHIthin\ versus \FCNM\ along 77 sightlines. The colors represent the absolute values of Galactic latitude $|b|$.}
\label{fig:f_vs_fcnm}
\end{figure}

In Figure \ref{fig:f_vs_fcnm}, we plot \FCNM\ against the \NHI/\NHIthin\ ratio. Obviously, \NHI/\NHIthin\ increases with the increase of \FCNM\ as expected. This simply demonstrates that the higher the CNM fraction is along the line-of-sight, the more important the opacity correction is. Nevertheless, we note that with the CNM fraction below 20\%, \NHIthin\ is consistent with \NHI, to within the errors. After this point, the \NHI/\NHIthin\ ratio rises up to a value of $\sim$2 as \FCNM\ increases from 20\% to 75\%.\\

\section{\hi\ opacity corrections}
\label{sec:opacity_corrections}

Pursuing the ultimate goal of deriving ``true'' \NHI\ maps toward our three regions around Taurus and Gemini, we will test two methods of opacity correction using \hi\ emission data from the GALFA-\hi\ survey. This is an important preparatory step for future work aiming to derive dark gas maps in these two regions using dust and $\gamma$-ray data as tracers of total proton column density.

\subsection{From the best fit of $f$ = \NHI/\NHIthin\ vs $\mathrm{log_{10}}$(\NHIthin)}
\label{subsection:nhi_corr_from_best_fit}
\begin{figure}
\includegraphics[width=1.0\linewidth]{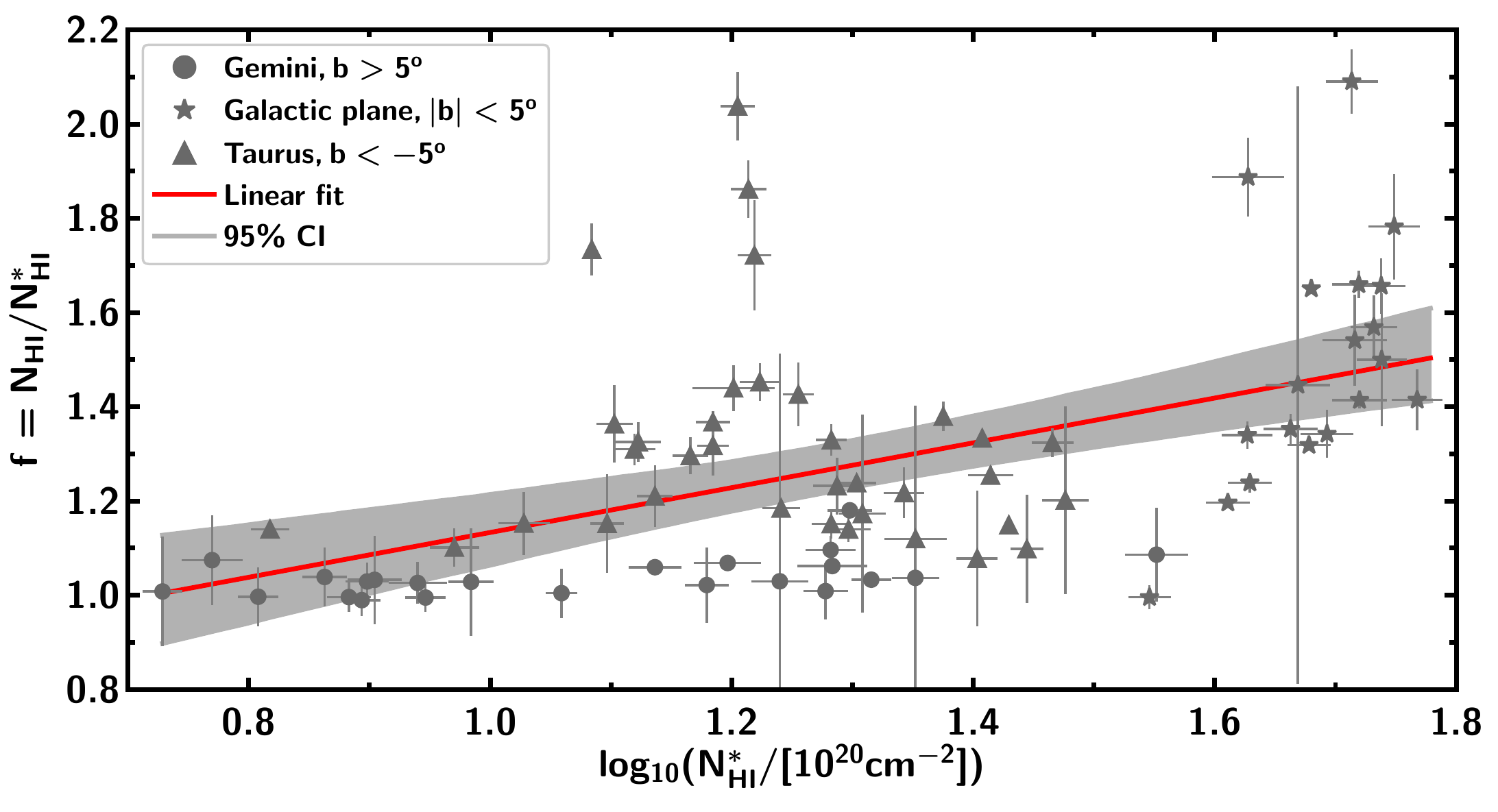}
\caption{Ratio $f$=\NHI/\NHIthin\ versus $\mathrm{log_{10}}(N^*_\mathrm{HI}/10^{20})$. The red line represents the best linear fit:  $f$=$(0.47\pm0.09) \mathrm{log_{10}}(N^*_\mathrm{HI}/10^{20}) + (0.66\pm0.12)$. The shaded region shows the 95\% confidence interval (CI) obtained from bootstrap re-sampling.}
\label{fig:ratio_vs_thin}
\end{figure}

The first method estimates an appropriate correction factor for each sightline from the best fit of $f$=\NHI/\NHIthin\ vs $\mathrm{log_{10}}(N^*_{HI}/10^{20})$, using the data from all 77 sightlines \citep[e.g.][]{Lee2015}. From our data, the ratio $f$ increases with increasing \NHIthin, with a Pearson linear coefficient of 0.5. To find an appropriate correlation, we fit the datapoints using four models: linear, quadratic, first-order exponential (in the form of $e^{ax}+b$) and second-order exponential ($e^{ax^2}+b$). We then compute the ``Bayesian factors'' (R) and ``Bayesian $p$-values'' ($p$) from the Bayesian model selection for each pair of models, finding R(linear/quadratic) = 72.08, $p$(linear) = 98.6\%; R(linear/first-order exponential) = 1.22, $p$(linear) = 54.9\%; R(linear/second-order exponential) = 1.02, $p$(linear) = 50.5\%. Thus the Bayesian factors favor the linear model (since their R values are greater than 1 and their $p$-values are greater than 50\%). It is worth noting that the linear model is definitely much more preferred than the quadratic, but the posterior probability in favor of the first-order exponential model is 46\% (compared to the linear model), which is too large to support rejecting the exponential model; meanwhile the posterior probabilities in favor of the second-order exponential and linear models are almost the same. Figure \ref{fig:ratio_vs_thin} shows the \textit{general} linear fit of $f$=\NHI/\NHIthin\ vs $\mathrm{log_{10}}(N^*_\mathrm{HI}/10^{20})$ with 95\% confidence intervals obtained from bootstrap re-sampling. The correction factor from the best fit is: $f=(0.47\pm0.09)\mathrm{log_{10}}(N^*_\mathrm{HI}/10^{20}) + (0.66\pm0.12)$ \citep[see also][for the gas around Perseus molecular cloud]{Lee2015}. The \hi\ column density obtained from this method is denoted as $N($H{\sc i}$)_\mathrm{fit}$. We note that our slope is quite different from \citet{Lee2015}'s. However, as mentioned before, it is clear that the correction factor and its variation with \NHI\ for each region are quite distinct, hence applying a single linear relationship for all regions may not be reliable. In fact, when fitting on a region-by-region basis we find that the best fit for the Gemini region is nearly flat with $f=(0.09\pm0.04)\mathrm{log_{10}}(N^*_\mathrm{HI}/10^{20}) + (0.94\pm0.04)$; for sightlines through the Galactic plane area the correction factor increases rapidly as \NHIthin\ increases: $f=(2.41\pm0.93)\mathrm{log_{10}}(N^*_\mathrm{HI}/10^{20}) - (2.57\pm1.57)$; but the best fit $f$ for Taurus region \textit{decreases} with increasing \NHIthin: $f=(-0.18\pm0.26)\mathrm{log_{10}}(N^*_\mathrm{HI}/10^{20}) + (1.54\pm0.32)$. A linear correlation between $f$ and $\mathrm{log_{10}}(N^*_\mathrm{HI})$ is therefore not convincing for every region.

\subsection{Using region-dependent uniform \Ts}
\label{subsection:nhi_corr_from_uniform_ts_maps}

The second approach is to apply a simple isothermal correction to GALFA-\hi\ using a single value of spin temperature for each sub-region. Here, an opacity-corrected \hi\ column density is obtained by integrating the following function of brightness temperature (\TB), and spin temperature \Ts\ (e.g. \citealt{Lockman1995,Wakker2011}):

\begin{equation}
N^{ISO}_\mathrm{HI} = 1.823 \times 10^{18} \int T_\mathrm{s}\ ln\left[\frac{T_\mathrm{s}}{T_\mathrm{s} - T_\mathrm{exp}}\right]\ dv.
\label{eq:nhi_lockman}
\end{equation}

\begin{figure}
\includegraphics[width=1.0\linewidth]{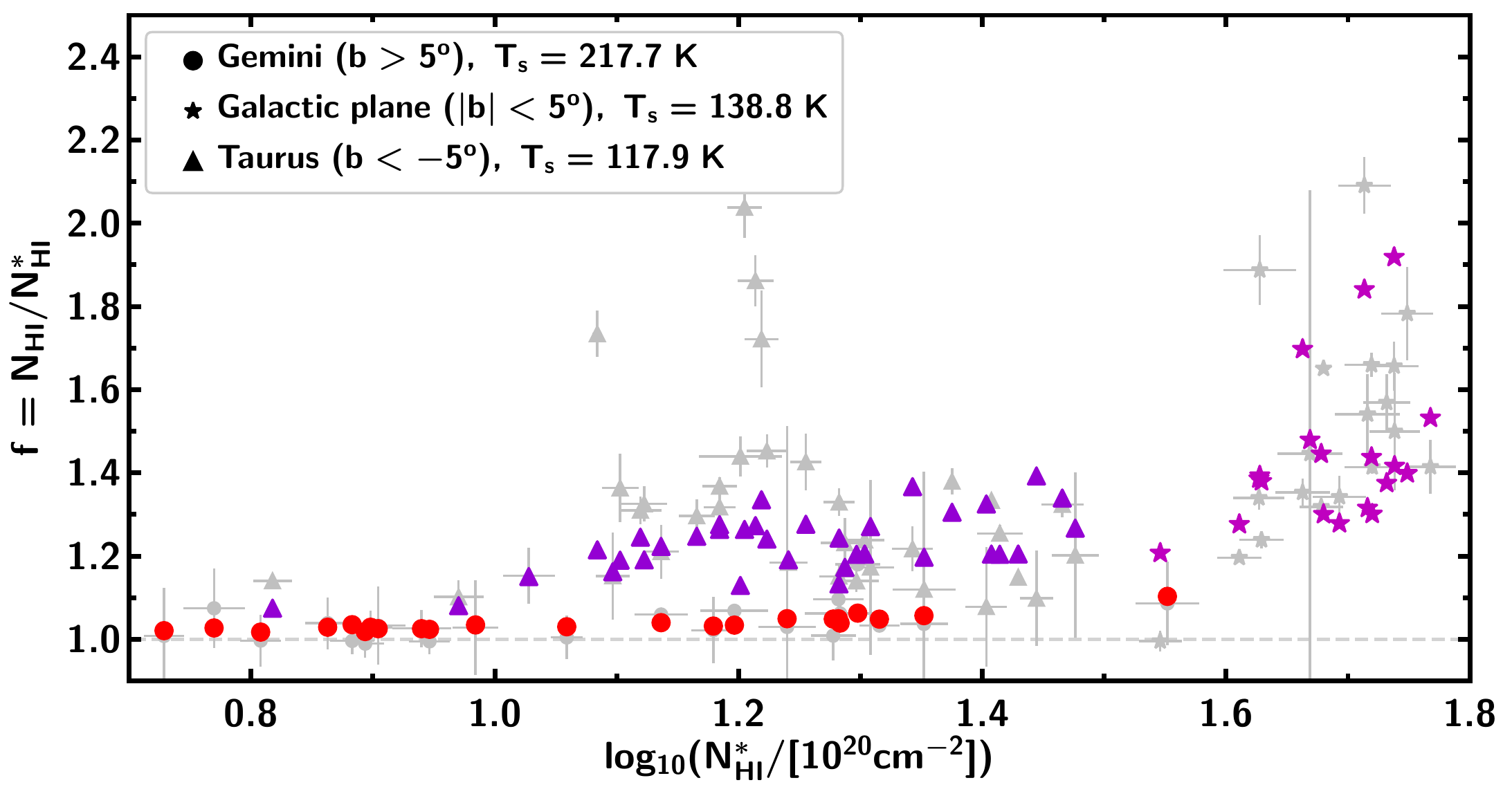}
\caption{Ratio $f=$ \NHI/\NHIthin\ versus $\mathrm{log_{10}}(N^*_{HI}/10^{20})$. The gray data points are results from Gaussian fit (same as for Figures \ref{fig:ratio_vs_nhi} and \ref{fig:ratio_vs_thin}). The color markers show the ratios estimated from corrected \NHI\ using single values of \Ts\ for different regions: stars for Galactic plane $|b|<5^{\circ}$, triangles for Taurus region $b<-5^{\circ}$, and dots for Gemini area $b>5^{\circ}$.}
\label{fig:ratio_vs_thin_same_ts}
\end{figure}

\noindent This approximation has the advantage that with an appropriate value of \Ts, it only requires \hi\ emission spectra to estimate \NHI, but it is not applicable at high optical depths where $T_\mathrm{s} \approx T_\mathrm{B}$ because the denominator in Equation \ref{eq:nhi_lockman} approaches zero. By conducting a least square fit of the \hi\ column densities obtained from this isothermal assumption and those from the Gaussian decomposition method (Section \ref{subsec:nhi_dist}) using \Ts\ as a parameter, we find best fit spin temperatures of \Ts\ = 217.7 K for the Gemini region ($b>5^{\circ}$), \Ts\ = 138.8 K for the Galactic Plane ($|b|<5^{\circ}$), and \Ts\ = 117.9 K for the Taurus region ($b<-5^{\circ}$). The \hi\ column density corrected by this method is denoted as $N($H{\sc i}$)_\mathrm{uniform}$, and its corresponding $f$ ratios are shown in Figure \ref{fig:ratio_vs_thin_same_ts}. It can be seen that the isothermal approximation with uniform spin temperatures generally reproduces quite well the correction factors derived from the original Gaussian decomposition.

\subsection{Comparing corrected \NHI\ maps}
\label{subsection:nhi_corr_comparison}
\begin{figure}
\includegraphics[width=1.0\linewidth]{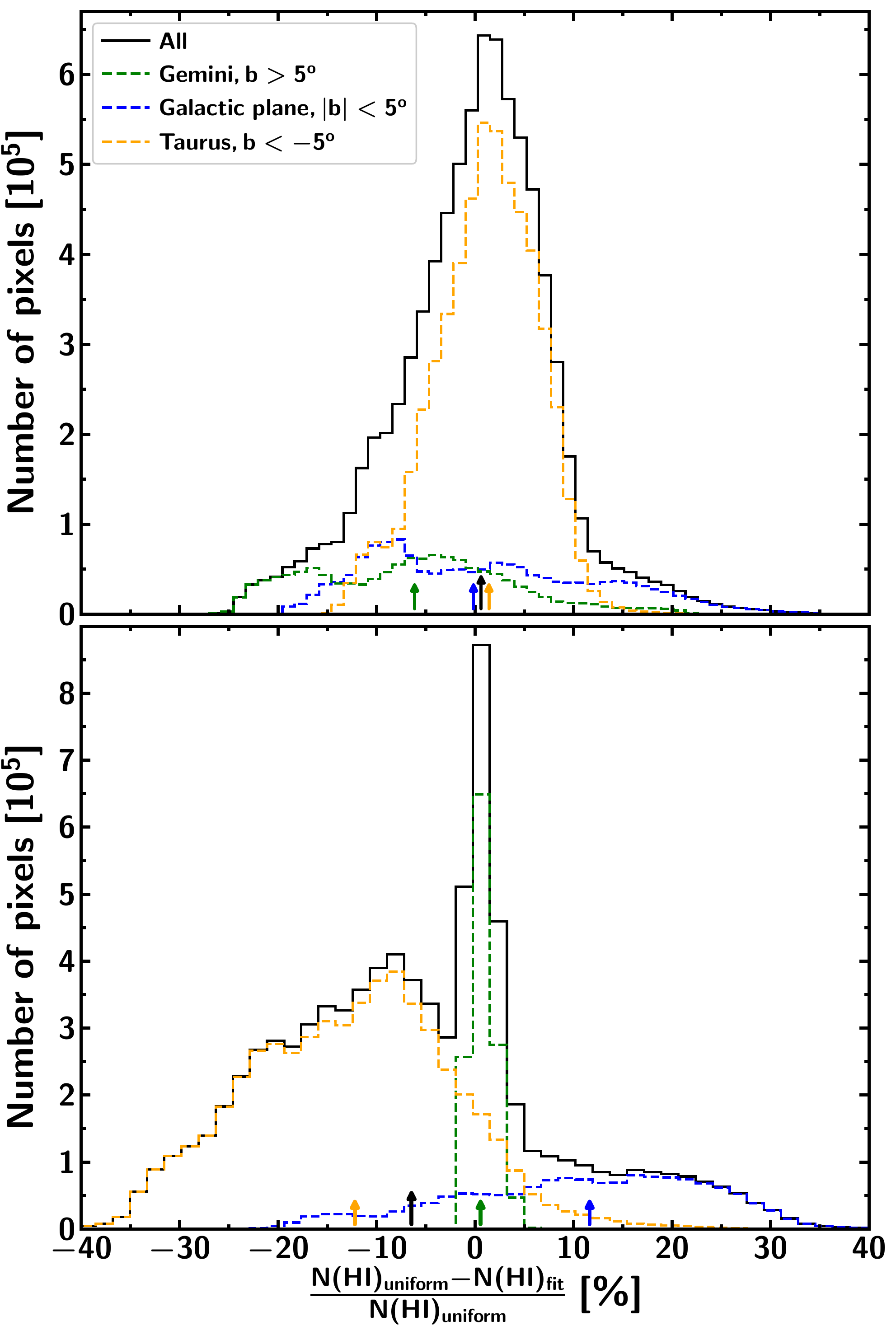}
\caption{Comparisons of \hi\ column densities obtained from two methods of opacity correction for GALFA-\hi\ data toward the regions around Taurus and Gemini: (1) from the best fit of $f$=\NHI/\NHIthin\ vs $log_{10}(N^*_{HI})$ (subscript ``fit'') and (2) from uniform spin temperature maps (subscript ``uniform''). Upper panel for the \textit{general} best linear fit, lower panel for three region-dependent linear fits. The blue histogram shows the \NHI\ relative difference for Galactic plane area ($|b|<5^{\circ}$), the green histogram for Gemini ($b>5^{\circ}$), the yellow one for Taurus region ($b<-5^{\circ}$) and black line for all three areas. The arrows show the medians.}
\label{fig:compare_nhi_corr}
\end{figure}

\begin{figure}
\includegraphics[width=1.0\linewidth]{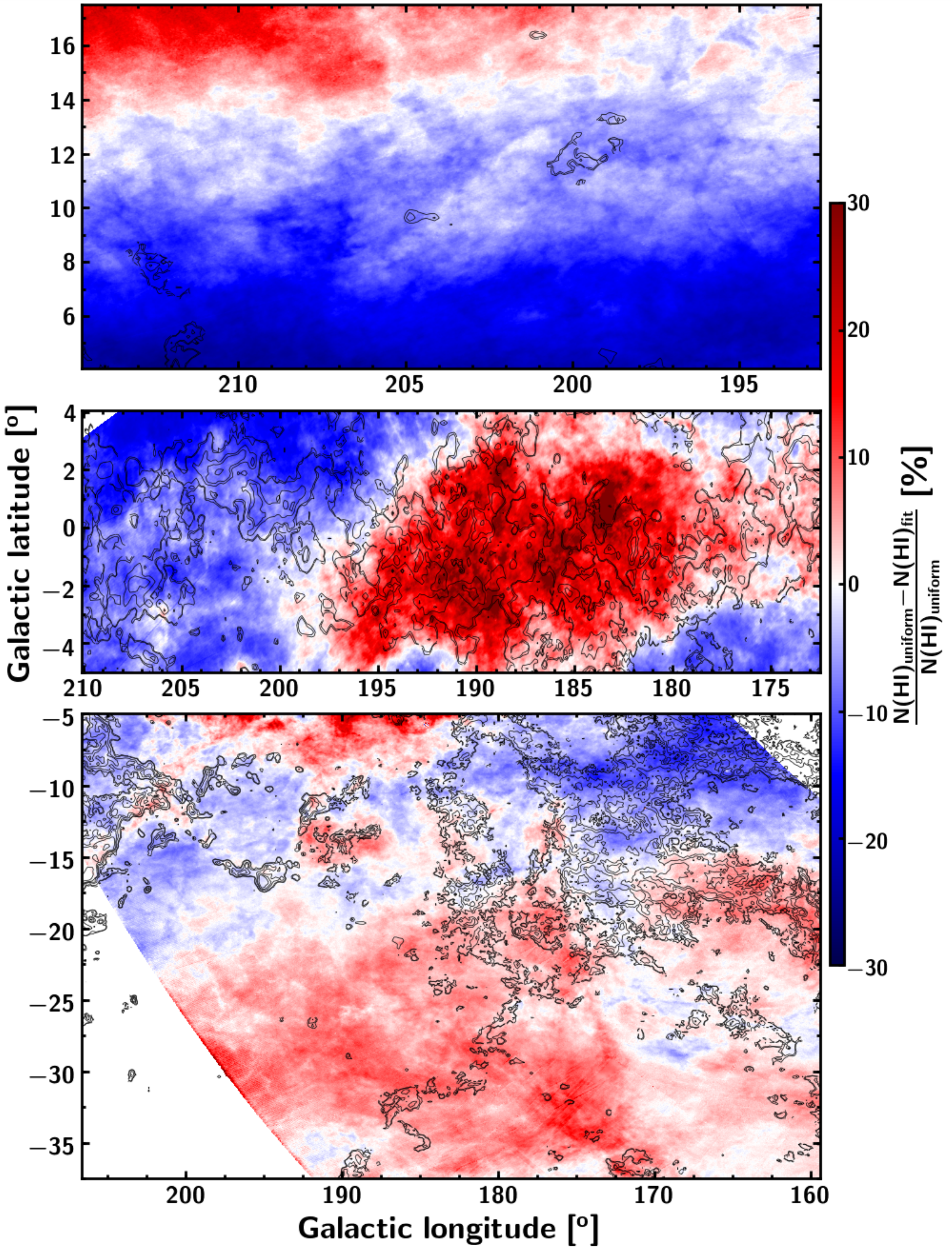}
\caption{Relative difference of opacity-corrected GALFA-\hi\ column densities obtained from the two methods discussed in the text (linear fit and region-dependent uniform spin temperature). The top panel shows the Gemini region ($b>5^{\circ}$), the middle panel for the Galactic plane region ($|b|<5^{\circ}$), and the bottom panel shows the Taurus region ($b<-5^{\circ}$). The colors indicate the relative difference $[N($H{\sc i}$)_\mathrm{uniform}$ $-$ $N($H{\sc i}$)_\mathrm{fit}]$/$N($H{\sc i}$)_\mathrm{uniform}$ $(\%)$. The contours are from $W_\mathrm{CO(1-0)}$ map as described in Figure \ref{fig:src_locations}.}
\label{fig:compare_nhi_corr_maps}
\end{figure}

We apply the two methods described above to correct for opacity effects pixel-by-pixel in the GALFA-\hi\ maps of the Gemini, Galactic plane and Taurus regions. In upper panel of Figure \ref{fig:compare_nhi_corr} we compare the corrected \NHI\ results from the \textit{general} linear fit and uniform \Ts\ maps by plotting the histograms of their relative differences in different regions, and Figure \ref{fig:compare_nhi_corr_maps} projects these relative differences onto the corresponding maps. Although we see individual pixel-by-pixel differences of up to $\sim$25\%, the \NHI\ corrected from the two methods are in general comparable with a median difference of 0.6\%, a mean of 0.1\%, and standard deviation of 7.9\%. The difference between the two corrected column densities is largest in the Gemini region ($b>5^{\circ}$) with (median, mean, standard deviation) of (6.1\%, 6.5\%, 10.0\%). In the Taurus region, the two correction methods agree well, with (median, mean, standard deviation) of (1.4\%, 1.1\%, 5.0\%). Within the Galactic plane, \NHI\ corrected with uniform \Ts\ is slightly lower than from the linear fit, the values of (median, mean, standard deviation) are (0.2\%, 1.2\%, 11.5\%). Generally, the largest differences (more than 10\%) are found in regions assumed to have higher CNM fractions: near the Galactic plane and in close vicinity to GMCs (see Figure \ref{fig:fcnm} and \ref{fig:compare_nhi_corr_maps}).

These differences may arise in part from the fact that the \textit{general} linear fit is made to the whole map for all three areas, whereas the different uniform spin temperatures were applied to each region. It appears that the two types of opacity correction are not consistent in Gemini and Galactic plane, but agree much better for the Taurus region, probably because our \textit{general} linear fit seems to reproduce well the correction factors for the Taurus area only (as seen in Figure \ref{fig:ratio_vs_thin}). By contrast, when region-by-region linear fitting is applied, the two  opacity correction methods agree very well for the Gemini region only, and depart further in the others, as seen in the lower panel of Figure \ref{fig:compare_nhi_corr}. This means that a regional linear correlation could be a good choice for Gemini, but is not suitable for our sightline samples toward the Galactic plane and Taurus regions. In addition, our region of interest spreads over a large sky area of different conditions, therefore, we prefer a region-based uniform spin temperature correction for \hi\ opacity throughout the Galactic ISM.

\section{summary and future work}
\label{sec:summary}
As a follow-up of the \hi\ study in the Perseus molecular cloud (\citealt{Stanimirovic2014,Lee2015}), we have presented a large-scale Arecibo survey of \hi\ emission/absorption observations to investigate the physical properties of cold and warm gas in the vicinity of five GMCs (Taurus, California, Rosette, Mon OB1, NGC 2264). This study is also a part of the GNOMES (Galactic Neutral Opacity and Molecular Excitation Survey) collaboration, which aims to explore the properties of neutral and molecular gas in/around molecular clouds. We detected strong \hi\ absorption in all directions toward 79 background radio continuum sources. By performing Gaussian decompositions (following the method of \citealt{Heiles2003a}) for all pairs of absorption/emission spectra, we directly determined \hi\ optical depths, temperatures and column densities for 349 CNM and 327 WNM Gaussian components. We used the distribution of CO integrated intensity $W_\mathrm{CO(1-0)}$ \citep{Dame2001} to separate our sample into three different interstellar environments -- diffuse ($b>5^{\circ}$), in-Plane ($|b|<5^{\circ}$), and around giant molecular clouds ($b<-5^{\circ}$). Comparing the \hi\ column density \NHI\ with that derived under the optically-thin assumption (\NHIthin) allows us to test two different methods for opacity-correction, which we apply to the GALFA-\hi\ data. Our key conclusions are as follows:

1. The peak optical depth of individual Gaussian components ranges from $\sim$0.01--16.2, with a median value of 0.35. We measure CNM spin temperatures between $\sim$10--480 K, finding that the distribution peaks at $\sim$50 K. The typical sonic Mach number for CNM is $\sim$4, which is consistent with previous studies and suggests that the turbulent motion of cold gas in the ISM is strongly supersonic. The median values of column densities for individual CNM and WNM components are 1.4 \nhiUnit\ and 2.3 \nhiUnit, respectively.

2. The properties of individual CNM components in our survey toward the three characteristic physical regions are consistent. More interestingly, they also agree well with those of previous observations along all-sky sightlines (HT03, M15, M18) and in/around the intermediate-mass Perseus molecular cloud (S14). This suggests that the properties of cold \hi\ gas in the Galactic ISM are fairly universal. The same conclusion was pointed out by S14.

3. The WNM constitutes $\sim$60\% of the total \hi\ gas, about 40\% of which is in thermally unstable regime with an upper limit on kinetic temperature of 500--5000 K. The implied fractions of cold, unstable and warm medium (by mass) are therefore 40\%, 24\% and 36\%, respectively. 

4. The fraction of cold gas along each sightline, \FCNM, increases with increasing \NHI\ to a maximum of 75\%. This is consistent with previous observations (HT03, M15, M18, S14) and also close to the 40--70\% \FCNM\ range found in the numerical simulations of \citet{Kim2014}. The \FCNM\ around molecular clouds is higher than in diffuse regions. This may support a staged build-up scenario for GMCs: from WNM-rich gas to CNM-rich gas to molecular clouds, with a high fraction of cold gas required for molecules to form.

5. The \hi\ opacity correction factor $f=$ \NHI/\NHIthin\ increases as total \NHI\ increases, with a median value of 1.21 for the full sample of 77 sightlines. However, the variation behaves differently in each individual region: while the ratio $f$ for Gemini is almost flat, it scatters around a mean value of 1.3 in Taurus and rises very steeply within the Galactic plane. Therefore, a linear relationship between the correction factor $f$ and $\mathrm{log_{10}}$(\NHIthin) is not convincing for every region.

6. We tested two methods of opacity correction: A linear fit of $f$=\NHI/\NHIthin\ vs $\mathrm{log_{10}}$(\NHIthin) made to the full sample, and the use of region-dependent uniform spin temperatures. We applied both methods to GALFA-\hi\ emission cubes, finding that the relative difference on a pixel-by-pixel basis is up to $\sim$25\%, however the mean offset between the derived \NHI\ distributions is small ($\sim$1\%). Because separate linear fits do not well describe the relation between $f$ and \NHIthin\ for each region, we prefer the uniform \Ts\ method.

Future work will analyse OH observations toward the same continuum source sample (A. Petzler in prep.), then combine these with CO data, and broad-band tracers of molecular hydrogen and total proton column density to begin constructing dark gas maps of the Taurus and Gemini regions.

\section*{Acknowledgments}
JRD is the recipient of an Australian Research Council (ARC) DECRA Fellowship (project number DE170101086). MYL was partially funded through the sub-project A6 of the Collaborative Research Council 956, funded by the Deutsche Forschungsgemeinschaft (DFG). CEM is supported by a National Science Foundation Astronomy and Astrophysics Postdoctoral Fellowship under award AST-1801471.

We are very grateful to Professor Isabelle Grenier and Professor Mark Wardle for a number of useful discusssions. We are indebted to T. Dame for generously providing the CO survey data. We thank the anonymous referee for constructive comments and criticisms which allowed us to improve this work. This publication utilizes data from Galactic ALFA \hi\ (GALFA \hi) survey data set obtained with the Arecibo L-band Feed Array (ALFA) on the Arecibo 305m telescope. The Arecibo Observatory is operated by SRI International under a cooperative agreement with the National Science Foundation (AST-1100968), and in alliance with Ana G. M{\'e}ndez -- Universidad Metropolitana, and the Universities Space Research Association. The GALFA \hi\ surveys have been funded by the NSF through grants to Columbia University, the University of Wisconsin, and the University of California.

\section*{Appendix A:\\Comparison of Gaussian and PSEUDO-VOIGT fittings}
\label{apd:compare-GnV}
\begin{figure}
\includegraphics[width=1.0\linewidth]{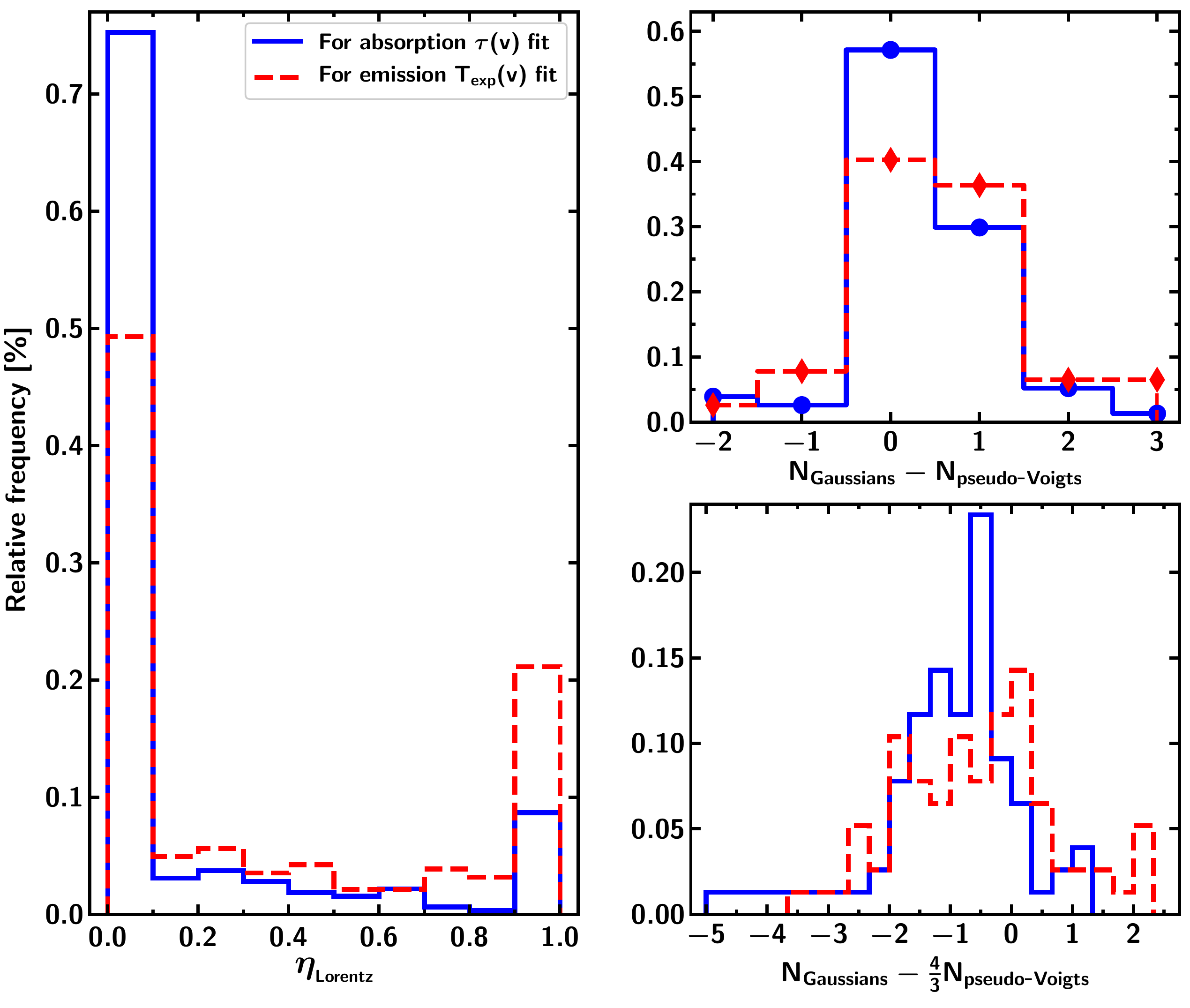}
\caption{Left: Histograms of $\eta_\mathrm{Lorentz}$, the fraction of Lorentzian in \psv\ function, for CNM and WNM components from absorption (blue) and emission (red) fits. Right: Comparison of the number of WNM/CNM components from Gaussian and \psv\ fittings without (top) and with (bottom) consideration of free parameter number for each mathematical function (3 for Gaussian and 4 for \psv).}
\label{fig:compare_GV_comp_number}
\end{figure}

In this section, we will carry out a detailed comparison of the parameters obtained from Gaussian and \psv\ decomposition fittings. The left panel of Figure \ref{fig:compare_GV_comp_number} shows a histogram of $\eta_{Lorentz}$, the fraction of the Lorentzian function in the \psv\ profile, for all CNM and WNM components obtained from \psv\ absorption and emission fits. For 327 CNM components, 67\% (217 out of 327) are pure Gaussian ($\eta_{Lorentz}=0$), 25\% (81/327) are a mixture of Gaussian and Lorentzian functions ($0<\eta_{Lorentz}<1$) (and basically in the mixture the components are more Gaussian than Lorentzian), and only 8\% (26/303) are pure Lorentzian ($\eta_{Lorentz}=1$). Of the 284 WNM components, 42\% (119/284) are pure Gaussian, 40\% (114/284) are mixed, and they in general distribute evenly in the range $0<\eta_{Lorentz}<1$, and 18\% (51/284) are pure Lorentzian. Our results show that, compared to WNM, the CNM in general appears to have a smaller portion of Lorentzian components.

In the upper right panel of Figure \ref{fig:compare_GV_comp_number} we compare the numbers of WNM and CNM components from Gaussian and \psv\ fits. In general, Gaussian fits require more components than \psv. For absorption, 44 out of 77 sightlines (57\%) require the same numbers of Gaussian and \psv\ CNM components, 28 sightlines (36\%) need more Gaussians (up to 3 components), 5 sightlines (7\%) take more \psv\ components. In the case of emission, most of the sightlines (28/77, or 36\%) demand one extra Gaussian component compared to \psv; the number of sightlines having the same number of Gaussian and \psv\ components is 31 (corresponding to 40\%); 10 sightlines (13\%) need more Gaussian than \psv\ from two to three extra components; 11\% of the remaining 8 sightlines take at maximum two more \psv\ components than Gaussian ones. However, when the numbers of free parameters for the two line-shape functions are taken into consideration, \psv\ fits turn out to require more components than Gaussian in a major part of sightlines, namely 65\% for emission and 86\% for absorption (as shown in the lower right panel).

\begin{figure}
\includegraphics[width=1.0\linewidth]{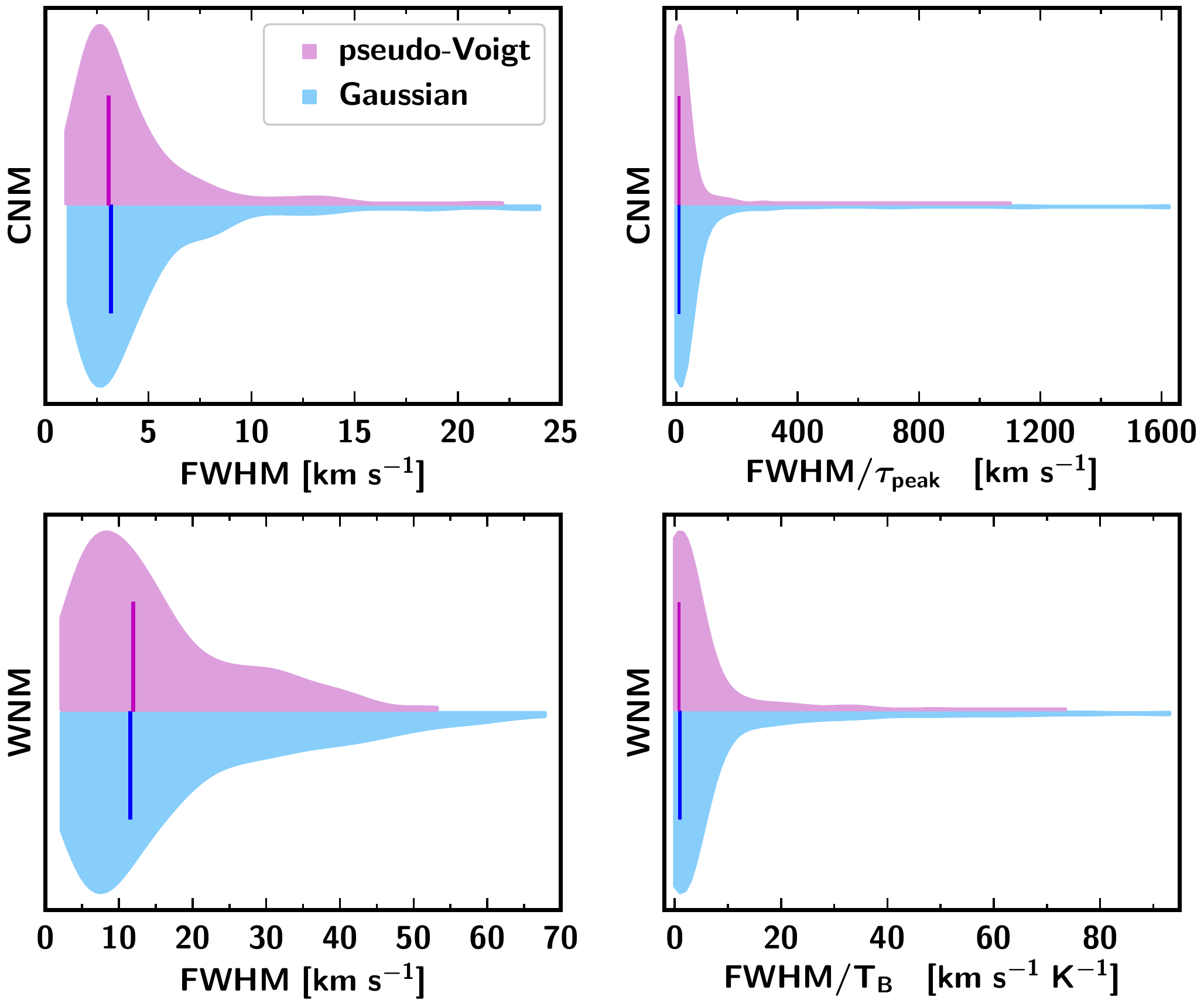}
\caption{Left column: Violin plots for comparing the widths (FWHM) of Gaussian (pink) and \psv\ (blue) components for CNM (top) and WNM (bottom). Right column: Violin plots of FWHM/\taupeak\ ratio for CNM and FWHM/$T_\mathrm{B}$ for WNM. The vertical lines show the medians.}
\label{fig:compare_GV_comp_width}
\end{figure}

The first column of Figure \ref{fig:compare_GV_comp_width} shows violin plots of the widths for both WNM and CNM components derived from Gaussian and \psv\ fits; the middle column displays violin plots of the FWHM/height ratios for each component. The medians of the widths and the FWHM/height ratios are almost identical, but compared to \psv, the Gaussian fitting needs slightly more wider WNM and CNM components; it also has more broad-but-weak components.

\begin{figure}
\includegraphics[width=1.0\linewidth]{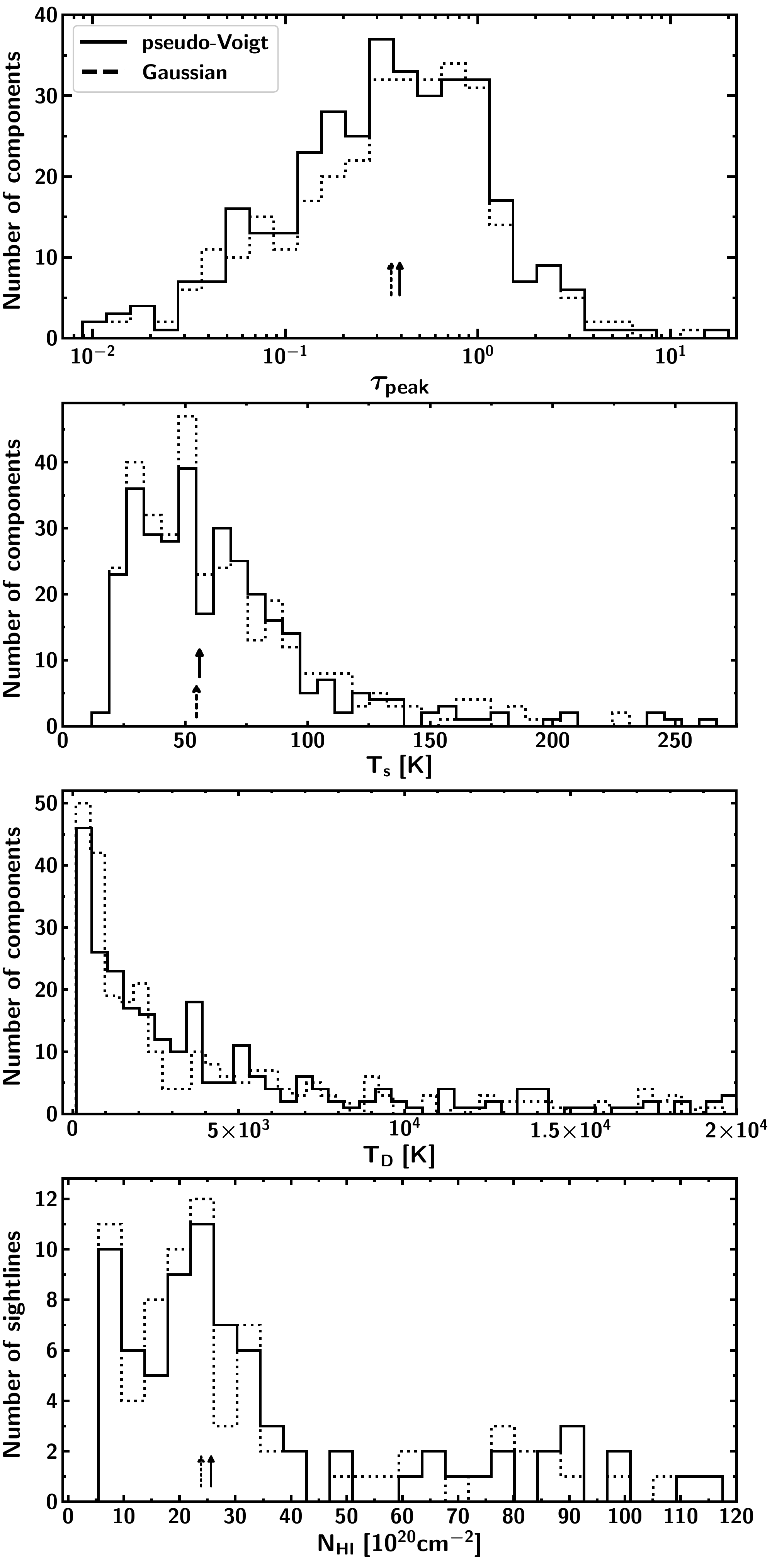}
\caption{Histograms of peak optical depth \taupeak, spin temperature \Ts, Doppler temperature $T_D$, and total column density \NHI\ derived from Gaussian (dotted) and \psv\ (solid) fits.}
\label{fig:compare_GnV_hist}
\end{figure}

\begin{figure}
\includegraphics[width=1.0\linewidth]{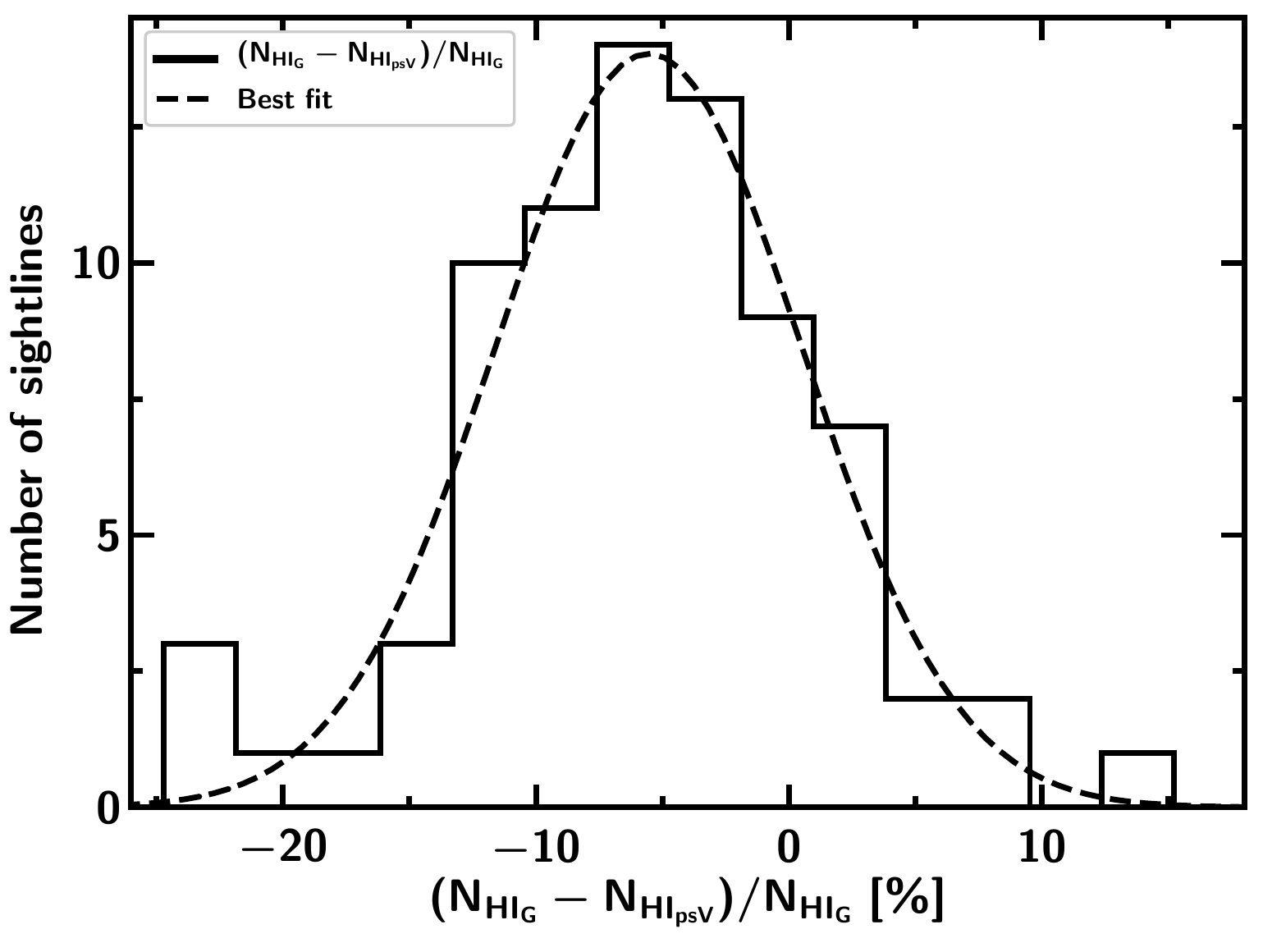}
\caption{Comparison of \NHI\ derived from Gaussian (subscript G) and \psv\ (subscript psV) fits.}
\label{fig:nhi_GnV}
\end{figure}

We plot the histograms of \taupeak, \Ts, $T_\mathrm{D}$ and total \NHI\ derived from \psv\ and Gaussian decompositions in Figure \ref{fig:compare_GnV_hist}. The fittings with the two different line-shapes appear to be consistent, since the distributions of their properties are almost equivalent. The mean and median values from each pair are close; namely the medians of the Gaussian and \psv\ fits are (0.35, 0.39) respectively for \taupeak, (54.6 K, 55.8 K) for \Ts, (2926 K, 3116 K) for \TD\ and (23.9, 25.6) \nhiUnit\ for total \NHI. Figure \ref{fig:nhi_GnV} shows the relative difference between \NHI\ as derived from Gaussian and \psv\ fittings. The distribution is well fit by a Gaussian with a peak at $-5.2$\% and a width of 6.1\%. This means the \NHI\ from Gaussian fitting is $\sim$5\% slightly lower than that obtained from the \psv\ fits. This leads us to a conclusion that the derived properties of Gaussian and \psv\ fits are compatible.

\label{compare_psv_gauss}

\section*{Appendix B:\\Comparison of \Ts\ and \NHI\ estimated from other methods}
\label{apd:ts-nhi-from-other-mehods}
Given pairs of emission (off-source) and absorption (on-source) observations, our goal is to combine them to find the spin temperature and column density of interstellar \hi. This is challenging because gas at different temperatures can be projected onto the same velocity channel along a line of sight. In a general sense, every sightline will contain a mixture of cold and warm gas: cold gas can be seen in the opacity spectrum; whereas warm gas is almost invisible in the opacity spectrum but is easily detected from the off-source (emission) measurements. Then the question is how best to use on-source and off-source spectra to distinguish these two phases. Beside the multi-component decomposition used in this work, there exist several independent methods to estimate \hi\ spin temperature and column density. In this Section, we will apply two alternative methods (one \Ts\ per velocity channel and single harmonic-mean \Ts\ along full sightline) to derive the spin temperature and column density, and compare their results with those from our spectral decompositions. Similar analyses have been performed by M15, they found that these alternative methods overestimate the spin temperature; and that the Gaussian fit method is the most successful in reproducing the \Ts\ range predicted by theoretical models. We repeat the analyses here for our three regions for completeness.

\setcounter{secnumdepth}{0}
\subsection{B1: Isothermal assumption: One \Ts\ per velocity channel}
\label{subsection: isothermal}
This method is established from the assumption that the gas in each velocity channel is isothermal with a single corresponding spin temperature (e.g. \citealt{Dickey1982,Chengalur2013}; M18). In this case the spin temperature spectrum, $T_\mathrm{s,chan}(v)$, can be computed as:

\begin{equation}
T_{s,chan}(v) = \frac{T_\mathrm{exp}(v)}{1-e^{-\tau(v)}}
\label{eq:TsC}
\end{equation}

\noindent where $T_\mathrm{exp}(v)$ is expected emission profile, $\tau(v)$ is opacity profile, and the effect of the $T_{\mathrm{bg}}$ term (see Equation \ref{eq:off_simp}) is assumed to be negligible. Here we use this method to calculate spin temperature for each velocity channel above our 5$\sigma$ detection limit in absorption, with a velocity resolution of 0.16 km s$^{-1}$. The total column density along the sightline is thus given by:

\begin{equation}
N_\mathrm{HI} = C_0 \int \frac{\tau(v)\ T_\mathrm{exp}(v)}{1-e^{-\tau(v)}} dv
\label{eq:nhi_iso}
\end{equation}

\subsection{B2: Single harmonic-mean \Ts\ along full sightline}
\label{subsection: single_ts_along_sightline}
This method assumes that there exists a uniform \Ts\ along the full sightline, also known as harmonic-mean spin temperature, $T_\mathrm{s,mean}$, which is computed from the emission and absorption profiles as:

\begin{equation}
T_{s,mean} = \frac{\int T_\mathrm{exp}(v)dv}{\int (1-e^{-\tau(v)}) dv }
\label{eq:TsS}
\end{equation}

\noindent where the quantities are the same as in Equation \ref{eq:TsC}. Then we calculate \NHI\ as:

\begin{equation}
\label{eq:nhi_tsmean}
\begin{split}
N_{HI} & = C_0\ T_{s,mean} \int\tau(v)dv \\
& = -C_0\ T_{s,mean} \int ln\left[1- \frac{T_\mathrm{exp}(v)}{T_{s,mean}} \right]dv
\end{split}
\end{equation}

\subsection{B3: Comparing \Ts\ and \NHI\ from these methods}
\label{subsec:compare_ts_nhi}
\begin{figure}
\includegraphics[width=1.0\linewidth]{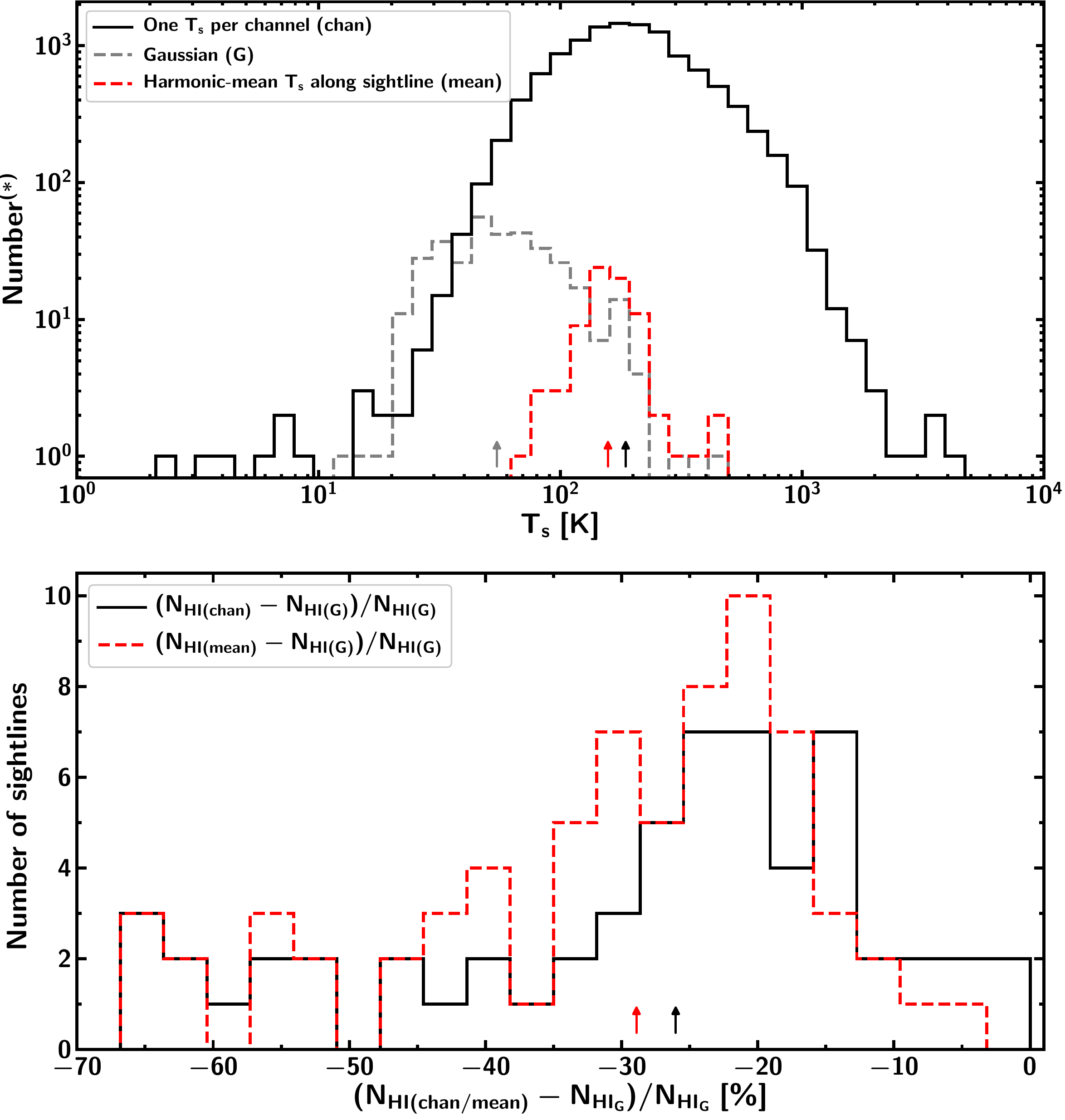}
\caption{Comparison of \Ts\ and \NHI\ derived from different methods: one-phase-per-channel (subscript ``(chan)'') and single harmonic mean \Ts\ along full sightline ``(mean)''. Top panel: Histograms of spin temperatures; black solid line for (chan) and red dashed line for (mean). Spin temperature from \hi\ Gaussian decomposition fit ``(G)'' is shown in gray for reference. Bottom panel: Relative difference of \NHI\ from isothermal channel (black) method and harmonic-mean assumption (red) compared with decomposition fit. The arrows show the medians. \newline $^{(*)}$Note: For (chan): y-axis is the number of velocity channels; for (G): number of components; and for (mean): the number of sightlines.}
\label{fig:compare_methods}
\end{figure}

We apply both the isothermal channel approximation and the harmonic-mean method to our GNOMES \hi\ data. We then compare the derived \Ts\ and \NHI\ with those from our Gaussian decompositions. The upper panel of Figure \ref{fig:compare_methods} shows histograms of the spin temperatures. The $T_\mathrm{s,chan}$ from the isothermal channel method spans a large range from 2 K to 4500 K with mean and median of (236 K, 186 K). Meanwhile, the harmonic-mean spin temperatures lie in a much narrower range, from 75 K to 467 K, with a mean and median of (173 K, 157 K). This is not surprising because these two methods both include warm and cold gas in their spin temperature estimates, thus their median temperatures are comparable. For the isothermal channel method, $T_\mathrm{s,chan}(v)$ is the mean temperature of the various clouds that contribute to each velocity channel. For the harmonic-mean assumption, $T_\mathrm{s,mean}$ is the column density weighted mean spin temperature of all warm and cold clouds along the line of sight. Thus, compared to spin temperatures from Gaussian decomposition fit, both $T_\mathrm{s,mean}$ and $T_\mathrm{s,chan}$ tend to be higher because the WNM accounts for a significant fraction of the emission along the sightline.

Compared to these methods, the advantage of the spectral decomposition approach is that it separates the two thermal phases within channels, and along the line of sight. As mentioned earlier, our CNM spin temperatures from Gaussian fitting range from 10 K to 465 K, peak at $\sim$50 K and have a median of 54.6 K, which is much lower than the estimates from the two single-phase approximations (which include both CNM and WNM). Theoretical models of a thermally bistable ISM \citep{Field1969,McKee1977,Wolfire2003} predict that the neutral CNM has typical temperatures of $30\lesssim T \lesssim 200$ K. So it is obvious that the spectral decomposition method in the current study successfully reproduces this \Ts\ range. The isothermal channel method gives a large fraction of thermally unstable gas with 500 $<$ T $<$ 5000 K; by contrast, the \Ts\ distribution from the harmonic-mean method overlaps with the CNM spin temperatures from our Gaussian fits at the high \Ts\ end, but cannot return colder temperatures (since the WNM always makes some contribution along a sightline).

In the bottom panel, we compare \NHI\ derived from the three methods. The column densities from the isothermal channel and harmonic-mean methods are relatively consistent, with differences mostly smaller than 8\%. However both underestimate the total \NHI\ by between $\sim$10--70\% (median $\sim$27\%) when compared to the Gaussian fits. These differences arise from the fact that the \hi\ gas in the ISM is a complex structure of multiple phases at different temperatures, where cold components contribute significantly to the total gas column density, but that both of the single-phase methods can only return a mean \Ts\ from the mixture of CNM and WNM within a sightline or channel. Based on a Monte Carlo simulation of multi-phase gas, \citet{Chengalur2013} examined how well the per-channel method is able to recover the true column density for gas with a complex spatial and temperature distribution along the sightline. They find that for \NHI\ $>10^{21}$ cm$^{-2}$, the ratio [\NHI/$N_\mathrm{HI,chan}$] = $-112.88+10.144\times \mathrm{log}_{10}N_\mathrm{HI}-0.2248\times [\mathrm{log}_{10}N_\mathrm{HI}]^{2}$. In our column density range (5--120) \nhiUnit, this translates to a true \NHI\ that is 20--50\% higher than $N_\mathrm{HI,chan}$. Our findings are thus in excellent agreement with their predictions.

\clearpage
\bibliographystyle{apj}
\bibliography{references}
\clearpage

\cleardoublepage

 


\begin{sidewaystable}[h!]
\fontsize{4}{3}\selectfont
\centering
\caption{Parameters for \psv\ and Gaussian fits \\ (See Section \ref{sec:results} for detailed descriptions)}
\label{table:parameters0}
\renewcommand{\arraystretch}{1.5}
\begin{tabular}{ lllllllllclllllll  }
\hline
\hline
\multirow{2}{*}{$l/b$} & \multicolumn{8}{c}{pseudo-Voigt} & \multicolumn{7}{c}{Gaussian}\\ 
\cline{2-9}
\cline{11-17}
& \multicolumn{1}{c}{$T_{B}$} & \multicolumn{1}{c}{$\tau$} & \multicolumn{1}{c}{$V_{lsr}$} & \multicolumn{1}{c}{$\Delta V$} & \multicolumn{1}{c}{$\eta$} & \multicolumn{1}{c}{$T_{s}$} & \multicolumn{1}{c}{\NHI} & \multicolumn{1}{c}{$F\ or\ O$} & &  \multicolumn{1}{c}{$T_{B}$} & \multicolumn{1}{c}{$\tau$} & \multicolumn{1}{c}{$V_{lsr}$} & \multicolumn{1}{c}{$\Delta V$} & \multicolumn{1}{c}{$T_{s}$} & \multicolumn{1}{c}{\NHI} & \multicolumn{1}{c}{$F\ or\ O$} \\ 
($^{\circ}$) & \multicolumn{1}{c}{$(K)$} & \multicolumn{1}{c}{$(10^{-2})$} & \multicolumn{1}{c}{$(km s^{-1})$} & \multicolumn{1}{c}{$(km s^{-1})$} & \multicolumn{1}{c}{$\ $} & \multicolumn{1}{c}{$(K)$} & \multicolumn{1}{c}{$(10^{20}cm^{-2})$} & \multicolumn{1}{c}{$\ $} & & \multicolumn{1}{c}{$(K)$} & \multicolumn{1}{c}{$(10^{-2})$} & \multicolumn{1}{c}{$(km s^{-1})$} & \multicolumn{1}{c}{$(km s^{-1})$} & \multicolumn{1}{c}{$(K)$} & \multicolumn{1}{c}{$(10^{20}cm^{-2})$} & \multicolumn{1}{c}{} \\ 
\hline

188.07/0.04&24.81&1.84$\pm$0.1&-2.29$\pm$0.03&1.53$\pm$0.06&0.0$\pm$0.0&29.49$\pm$2.21&1.61$\pm$0.16&2& & 29.13&1.87$\pm$0.01&-2.29$\pm$0.0&1.5$\pm$0.01&34.44$\pm$0.74&1.87$\pm$0.04&2\\
188.07/0.04&37.12&0.63$\pm$0.02&0.98$\pm$0.32&5.99$\pm$0.37&0.0$\pm$0.0&79.42$\pm$17.9&5.86$\pm$1.39&1& & 27.11&0.63$\pm$0.0&1.02$\pm$0.04&6.14$\pm$0.05&58.0$\pm$3.46&4.37$\pm$0.26&4\\
188.07/0.04&59.76&1.1$\pm$0.13&4.42$\pm$0.13&1.99$\pm$0.28&0.0$\pm$0.0&89.58$\pm$9.69&3.8$\pm$0.8&0& & 33.89&1.1$\pm$0.01&4.47$\pm$0.01&2.06$\pm$0.02&50.8$\pm$2.77&2.23$\pm$0.13&3\\
\hline
\end{tabular}
\end{sidewaystable}

\begin{sidewaystable}[h!]
\fontsize{4}{3}\selectfont
\renewcommand\thetable{\ref{table:parameters0} (cont)}
\centering
\caption{}
\label{}
\renewcommand{\arraystretch}{1.5}
\begin{tabular}{ lllllllllclllllll  }
\hline
\hline
\multirow{2}{*}{$l/b$} & \multicolumn{8}{c}{pseudo-Voigt} & \multicolumn{7}{c}{Gaussian}\\ 
\cline{2-9}
\cline{11-17}
& \multicolumn{1}{c}{$T_{B}$} & \multicolumn{1}{c}{$\tau$} & \multicolumn{1}{c}{$V_{lsr}$} & \multicolumn{1}{c}{$\Delta V$} & \multicolumn{1}{c}{$\eta$} & \multicolumn{1}{c}{$T_{s}$} & \multicolumn{1}{c}{\NHI} & \multicolumn{1}{c}{$F\ or\ O$} & &  \multicolumn{1}{c}{$T_{B}$} & \multicolumn{1}{c}{$\tau$} & \multicolumn{1}{c}{$V_{lsr}$} & \multicolumn{1}{c}{$\Delta V$} & \multicolumn{1}{c}{$T_{s}$} & \multicolumn{1}{c}{\NHI} & \multicolumn{1}{c}{$F\ or\ O$} \\ 
($^{\circ}$) & \multicolumn{1}{c}{$(K)$} & \multicolumn{1}{c}{$(10^{-2})$} & \multicolumn{1}{c}{$(km s^{-1})$} & \multicolumn{1}{c}{$(km s^{-1})$} & \multicolumn{1}{c}{$\ $} & \multicolumn{1}{c}{$(K)$} & \multicolumn{1}{c}{$(10^{20}cm^{-2})$} & \multicolumn{1}{c}{$\ $} & & \multicolumn{1}{c}{$(K)$} & \multicolumn{1}{c}{$(10^{-2})$} & \multicolumn{1}{c}{$(km s^{-1})$} & \multicolumn{1}{c}{$(km s^{-1})$} & \multicolumn{1}{c}{$(K)$} & \multicolumn{1}{c}{$(10^{20}cm^{-2})$} & \multicolumn{1}{c}{} \\ 
\hline

203.75/14.63&5.2&0.06$\pm$0.01&-3.46$\pm$0.28&3.29$\pm$0.67&0.0$\pm$0.0&89.34$\pm$6.49&0.33$\pm$0.09&1& & -&-&-&-&-&-&-\\
203.75/14.63&-&-&-&-&-&-&-&-& & 6.59&0.08$\pm$0.0&-2.9$\pm$0.08&4.62$\pm$0.18&85.65$\pm$6.15&0.6$\pm$0.05&1\\
203.75/14.63&9.51&0.15$\pm$0.01&0.66$\pm$0.14&3.01$\pm$0.51&1.0$\pm$0.0&68.3$\pm$3.52&0.87$\pm$0.17&0& & 8.66&0.13$\pm$0.0&0.67$\pm$0.03&2.24$\pm$0.08&71.0$\pm$3.89&0.39$\pm$0.03&2\\
203.75/14.63&-&-&-&-&-&-&-&-& & 2.88&0.06$\pm$0.0&4.95$\pm$0.25&5.91$\pm$0.41&49.4$\pm$13.8&0.31$\pm$0.09&3\\
203.75/14.63&3.89&0.12$\pm$0.01&6.51$\pm$0.11&2.23$\pm$0.32&0.49$\pm$0.38&34.36$\pm$3.1&0.23$\pm$0.06&2& & 2.19&0.09$\pm$0.0&6.66$\pm$0.03&1.76$\pm$0.09&25.43$\pm$4.59&0.08$\pm$0.01&0\\
\hline
\end{tabular}
\end{sidewaystable}

\begin{sidewaystable}[h!]
\fontsize{4}{3}\selectfont
\renewcommand\thetable{\ref{table:parameters0} (cont)}
\centering
\caption{}
\label{}
\renewcommand{\arraystretch}{1.5}
\begin{tabular}{ lllllllllclllllll  }
\hline
\hline
\multirow{2}{*}{$l/b$} & \multicolumn{8}{c}{pseudo-Voigt} & \multicolumn{7}{c}{Gaussian}\\ 
\cline{2-9}
\cline{11-17}
& \multicolumn{1}{c}{$T_{B}$} & \multicolumn{1}{c}{$\tau$} & \multicolumn{1}{c}{$V_{lsr}$} & \multicolumn{1}{c}{$\Delta V$} & \multicolumn{1}{c}{$\eta$} & \multicolumn{1}{c}{$T_{s}$} & \multicolumn{1}{c}{\NHI} & \multicolumn{1}{c}{$F\ or\ O$} & &  \multicolumn{1}{c}{$T_{B}$} & \multicolumn{1}{c}{$\tau$} & \multicolumn{1}{c}{$V_{lsr}$} & \multicolumn{1}{c}{$\Delta V$} & \multicolumn{1}{c}{$T_{s}$} & \multicolumn{1}{c}{\NHI} & \multicolumn{1}{c}{$F\ or\ O$} \\ 
($^{\circ}$) & \multicolumn{1}{c}{$(K)$} & \multicolumn{1}{c}{$(10^{-2})$} & \multicolumn{1}{c}{$(km s^{-1})$} & \multicolumn{1}{c}{$(km s^{-1})$} & \multicolumn{1}{c}{$\ $} & \multicolumn{1}{c}{$(K)$} & \multicolumn{1}{c}{$(10^{20}cm^{-2})$} & \multicolumn{1}{c}{$\ $} & & \multicolumn{1}{c}{$(K)$} & \multicolumn{1}{c}{$(10^{-2})$} & \multicolumn{1}{c}{$(km s^{-1})$} & \multicolumn{1}{c}{$(km s^{-1})$} & \multicolumn{1}{c}{$(K)$} & \multicolumn{1}{c}{$(10^{20}cm^{-2})$} & \multicolumn{1}{c}{} \\ 
\hline

202.29/11.53&2.94&0.12$\pm$0.01&-3.11$\pm$0.09&1.82$\pm$0.22&0.0$\pm$0.0&25.96$\pm$2.29&0.11$\pm$0.02&1& & 3.16&0.12$\pm$0.0&-3.12$\pm$0.02&1.81$\pm$0.05&27.96$\pm$2.41&0.12$\pm$0.01&1\\
202.29/11.53&2.49&0.04$\pm$0.01&0.03$\pm$0.68&7.67$\pm$1.33&1.0$\pm$0.0&63.41$\pm$19.62&0.61$\pm$0.25&0& & 4.9&0.04$\pm$0.0&-0.28$\pm$0.14&8.31$\pm$0.24&124.92$\pm$7.0&0.89$\pm$0.06&0\\
202.29/11.53&-&-&-&-&-&-&-&-& & 0.52$\pm$0.12& $<$0.015&-28.29$\pm$5.77&28.03$\pm$7.93& $>$33.59&0.28$\pm$0.1&0.5$\pm$0.29\\
202.29/11.53&1.13$\pm$0.3& $<$0.015&-18.05$\pm$7.71&32.09$\pm$10.49&0.92$\pm$0.54& $>$77.18&1.01$\pm$0.46&0.5$\pm$0.29& & -&-&-&-&-&-&-\\
\hline
\end{tabular}
\end{sidewaystable}

\begin{sidewaystable}[h!]
\fontsize{4}{3}\selectfont
\renewcommand\thetable{\ref{table:parameters0} (cont)}
\centering
\caption{}
\label{}
\renewcommand{\arraystretch}{1.5}
\begin{tabular}{ lllllllllclllllll  }
\hline
\hline
\multirow{2}{*}{$l/b$} & \multicolumn{8}{c}{pseudo-Voigt} & \multicolumn{7}{c}{Gaussian}\\ 
\cline{2-9}
\cline{11-17}
& \multicolumn{1}{c}{$T_{B}$} & \multicolumn{1}{c}{$\tau$} & \multicolumn{1}{c}{$V_{lsr}$} & \multicolumn{1}{c}{$\Delta V$} & \multicolumn{1}{c}{$\eta$} & \multicolumn{1}{c}{$T_{s}$} & \multicolumn{1}{c}{\NHI} & \multicolumn{1}{c}{$F\ or\ O$} & &  \multicolumn{1}{c}{$T_{B}$} & \multicolumn{1}{c}{$\tau$} & \multicolumn{1}{c}{$V_{lsr}$} & \multicolumn{1}{c}{$\Delta V$} & \multicolumn{1}{c}{$T_{s}$} & \multicolumn{1}{c}{\NHI} & \multicolumn{1}{c}{$F\ or\ O$} \\ 
($^{\circ}$) & \multicolumn{1}{c}{$(K)$} & \multicolumn{1}{c}{$(10^{-2})$} & \multicolumn{1}{c}{$(km s^{-1})$} & \multicolumn{1}{c}{$(km s^{-1})$} & \multicolumn{1}{c}{$\ $} & \multicolumn{1}{c}{$(K)$} & \multicolumn{1}{c}{$(10^{20}cm^{-2})$} & \multicolumn{1}{c}{$\ $} & & \multicolumn{1}{c}{$(K)$} & \multicolumn{1}{c}{$(10^{-2})$} & \multicolumn{1}{c}{$(km s^{-1})$} & \multicolumn{1}{c}{$(km s^{-1})$} & \multicolumn{1}{c}{$(K)$} & \multicolumn{1}{c}{$(10^{20}cm^{-2})$} & \multicolumn{1}{c}{} \\ 
\hline

207.31/12.37&5.04&0.04$\pm$0.01&-0.84$\pm$0.62&5.19$\pm$1.42&0.0$\pm$0.0&128.59$\pm$39.6&0.54$\pm$0.26&0& & 5.05&0.04$\pm$0.0&-0.81$\pm$0.18&5.43$\pm$0.45&128.67$\pm$5.07&0.55$\pm$0.06&1\\
207.31/12.37&-&-&-&-&-&-&-&-& & 10.6&0.1$\pm$0.01&6.5$\pm$0.59&4.29$\pm$0.81&111.35$\pm$2.81&0.95$\pm$0.22&0\\
207.31/12.37&11.37&0.12$\pm$0.02&7.02$\pm$0.33&4.72$\pm$0.65&0.0$\pm$0.0&100.52$\pm$22.82&1.1$\pm$0.34&1& & -&-&-&-&-&-&-\\
207.31/12.37&-&-&-&-&-&-&-&-& & 5.37&0.06$\pm$0.02&8.89$\pm$0.25&2.6$\pm$0.64&92.25$\pm$1.74&0.26$\pm$0.12&2\\
207.31/12.37&-&-&-&-&-&-&-&-& & 0.89&0.04$\pm$0.0&21.04$\pm$0.08&2.13$\pm$0.19&22.75$\pm$1.75&0.04$\pm$0.01&3\\
\hline
\end{tabular}
\end{sidewaystable}

\begin{sidewaystable}[h!]
\fontsize{4}{3}\selectfont
\renewcommand\thetable{\ref{table:parameters0} (cont)}
\centering
\caption{}
\label{}
\renewcommand{\arraystretch}{1.5}
\begin{tabular}{ lllllllllclllllll  }
\hline
\hline
\multirow{2}{*}{$l/b$} & \multicolumn{8}{c}{pseudo-Voigt} & \multicolumn{7}{c}{Gaussian}\\ 
\cline{2-9}
\cline{11-17}
& \multicolumn{1}{c}{$T_{B}$} & \multicolumn{1}{c}{$\tau$} & \multicolumn{1}{c}{$V_{lsr}$} & \multicolumn{1}{c}{$\Delta V$} & \multicolumn{1}{c}{$\eta$} & \multicolumn{1}{c}{$T_{s}$} & \multicolumn{1}{c}{\NHI} & \multicolumn{1}{c}{$F\ or\ O$} & &  \multicolumn{1}{c}{$T_{B}$} & \multicolumn{1}{c}{$\tau$} & \multicolumn{1}{c}{$V_{lsr}$} & \multicolumn{1}{c}{$\Delta V$} & \multicolumn{1}{c}{$T_{s}$} & \multicolumn{1}{c}{\NHI} & \multicolumn{1}{c}{$F\ or\ O$} \\ 
($^{\circ}$) & \multicolumn{1}{c}{$(K)$} & \multicolumn{1}{c}{$(10^{-2})$} & \multicolumn{1}{c}{$(km s^{-1})$} & \multicolumn{1}{c}{$(km s^{-1})$} & \multicolumn{1}{c}{$\ $} & \multicolumn{1}{c}{$(K)$} & \multicolumn{1}{c}{$(10^{20}cm^{-2})$} & \multicolumn{1}{c}{$\ $} & & \multicolumn{1}{c}{$(K)$} & \multicolumn{1}{c}{$(10^{-2})$} & \multicolumn{1}{c}{$(km s^{-1})$} & \multicolumn{1}{c}{$(km s^{-1})$} & \multicolumn{1}{c}{$(K)$} & \multicolumn{1}{c}{$(10^{20}cm^{-2})$} & \multicolumn{1}{c}{} \\ 
\hline

171.98/-35.48&13.89&0.23$\pm$0.03&-3.5$\pm$0.53&4.32$\pm$0.9&0.0$\pm$0.0&67.58$\pm$2.24&1.29$\pm$0.33&2& & 7.75&0.26$\pm$0.01&-3.51$\pm$0.11&4.76$\pm$0.17&33.85$\pm$2.56&0.82$\pm$0.07&5\\
171.98/-35.48&16.26&0.56$\pm$0.06&1.44$\pm$0.28&3.93$\pm$0.91&0.64$\pm$0.34&37.92$\pm$3.92&2.1$\pm$0.63&1& & 14.69&0.56$\pm$0.01&1.73$\pm$0.04&4.36$\pm$0.18&34.26$\pm$2.37&1.64$\pm$0.13&1\\
171.98/-35.48&-&-&-&-&-&-&-&-& & 15.42&0.33$\pm$0.03&4.91$\pm$0.08&1.75$\pm$0.19&54.86$\pm$3.61&0.61$\pm$0.09&2\\
171.98/-35.48&6.9&0.3$\pm$0.12&6.69$\pm$0.2&1.06$\pm$0.51&0.0$\pm$0.0&26.61$\pm$2.14&0.16$\pm$0.1&3& & 18.32&0.61$\pm$0.12&6.56$\pm$0.04&1.44$\pm$0.14&40.11$\pm$1.23&0.68$\pm$0.15&3\\
\hline
\end{tabular}
\end{sidewaystable}

\begin{sidewaystable}[h!]
\fontsize{4}{3}\selectfont
\renewcommand\thetable{\ref{table:parameters0} (cont)}
\centering
\caption{}
\label{}
\renewcommand{\arraystretch}{1.5}
\begin{tabular}{ lllllllllclllllll  }
\hline
\hline
\multirow{2}{*}{$l/b$} & \multicolumn{8}{c}{pseudo-Voigt} & \multicolumn{7}{c}{Gaussian}\\ 
\cline{2-9}
\cline{11-17}
& \multicolumn{1}{c}{$T_{B}$} & \multicolumn{1}{c}{$\tau$} & \multicolumn{1}{c}{$V_{lsr}$} & \multicolumn{1}{c}{$\Delta V$} & \multicolumn{1}{c}{$\eta$} & \multicolumn{1}{c}{$T_{s}$} & \multicolumn{1}{c}{\NHI} & \multicolumn{1}{c}{$F\ or\ O$} & &  \multicolumn{1}{c}{$T_{B}$} & \multicolumn{1}{c}{$\tau$} & \multicolumn{1}{c}{$V_{lsr}$} & \multicolumn{1}{c}{$\Delta V$} & \multicolumn{1}{c}{$T_{s}$} & \multicolumn{1}{c}{\NHI} & \multicolumn{1}{c}{$F\ or\ O$} \\ 
($^{\circ}$) & \multicolumn{1}{c}{$(K)$} & \multicolumn{1}{c}{$(10^{-2})$} & \multicolumn{1}{c}{$(km s^{-1})$} & \multicolumn{1}{c}{$(km s^{-1})$} & \multicolumn{1}{c}{$\ $} & \multicolumn{1}{c}{$(K)$} & \multicolumn{1}{c}{$(10^{20}cm^{-2})$} & \multicolumn{1}{c}{$\ $} & & \multicolumn{1}{c}{$(K)$} & \multicolumn{1}{c}{$(10^{-2})$} & \multicolumn{1}{c}{$(km s^{-1})$} & \multicolumn{1}{c}{$(km s^{-1})$} & \multicolumn{1}{c}{$(K)$} & \multicolumn{1}{c}{$(10^{20}cm^{-2})$} & \multicolumn{1}{c}{} \\ 
\hline

199.52/6.04&8.29&0.25$\pm$0.08&0.24$\pm$0.19&2.35$\pm$0.73&0.63$\pm$0.51&37.48$\pm$2.73&0.56$\pm$0.27&0& & 6.41&0.25$\pm$0.0&0.45$\pm$0.02&2.59$\pm$0.05&28.99$\pm$2.17&0.36$\pm$0.03&0\\
199.52/6.04&15.73&0.38$\pm$0.05&8.29$\pm$0.39&3.07$\pm$0.66&0.0$\pm$0.0&49.77$\pm$2.39&1.12$\pm$0.29&1& & 23.44&0.38$\pm$0.01&8.43$\pm$0.07&3.5$\pm$0.11&74.13$\pm$1.12&1.93$\pm$0.08&1\\
199.52/6.04&17.67&0.53$\pm$0.06&11.31$\pm$0.25&2.57$\pm$0.44&0.06$\pm$0.27&42.96$\pm$1.55&1.16$\pm$0.28&2& & 21.16&0.5$\pm$0.01&11.42$\pm$0.04&2.49$\pm$0.06&53.78$\pm$1.1&1.3$\pm$0.05&2\\
\hline
\end{tabular}
\end{sidewaystable}

\begin{sidewaystable}[h!]
\fontsize{4}{3}\selectfont
\renewcommand\thetable{\ref{table:parameters0} (cont)}
\centering
\caption{}
\label{}
\renewcommand{\arraystretch}{1.5}
\begin{tabular}{ lllllllllclllllll  }
\hline
\hline
\multirow{2}{*}{$l/b$} & \multicolumn{8}{c}{pseudo-Voigt} & \multicolumn{7}{c}{Gaussian}\\ 
\cline{2-9}
\cline{11-17}
& \multicolumn{1}{c}{$T_{B}$} & \multicolumn{1}{c}{$\tau$} & \multicolumn{1}{c}{$V_{lsr}$} & \multicolumn{1}{c}{$\Delta V$} & \multicolumn{1}{c}{$\eta$} & \multicolumn{1}{c}{$T_{s}$} & \multicolumn{1}{c}{\NHI} & \multicolumn{1}{c}{$F\ or\ O$} & &  \multicolumn{1}{c}{$T_{B}$} & \multicolumn{1}{c}{$\tau$} & \multicolumn{1}{c}{$V_{lsr}$} & \multicolumn{1}{c}{$\Delta V$} & \multicolumn{1}{c}{$T_{s}$} & \multicolumn{1}{c}{\NHI} & \multicolumn{1}{c}{$F\ or\ O$} \\ 
($^{\circ}$) & \multicolumn{1}{c}{$(K)$} & \multicolumn{1}{c}{$(10^{-2})$} & \multicolumn{1}{c}{$(km s^{-1})$} & \multicolumn{1}{c}{$(km s^{-1})$} & \multicolumn{1}{c}{$\ $} & \multicolumn{1}{c}{$(K)$} & \multicolumn{1}{c}{$(10^{20}cm^{-2})$} & \multicolumn{1}{c}{$\ $} & & \multicolumn{1}{c}{$(K)$} & \multicolumn{1}{c}{$(10^{-2})$} & \multicolumn{1}{c}{$(km s^{-1})$} & \multicolumn{1}{c}{$(km s^{-1})$} & \multicolumn{1}{c}{$(K)$} & \multicolumn{1}{c}{$(10^{20}cm^{-2})$} & \multicolumn{1}{c}{} \\ 
\hline

209.24/-4.64&-&-&-&-&-&-&-&-& & 65.45&0.98$\pm$0.01&7.45$\pm$0.04&5.09$\pm$0.06&104.78$\pm$1.42&10.16$\pm$0.2&1\\
209.24/-4.64&44.35&1.13$\pm$0.12&8.89$\pm$0.23&6.92$\pm$0.43&0.0$\pm$0.0&65.52$\pm$4.98&9.9$\pm$1.46&1& & -&-&-&-&-&-&-\\
209.24/-4.64&72.72&2.21$\pm$1.56&10.81$\pm$0.18&1.25$\pm$0.42&0.0$\pm$0.0&81.68$\pm$1.51&4.35$\pm$3.41&0& & 75.94&2.32$\pm$0.04&10.78$\pm$0.01&1.86$\pm$0.03&84.22$\pm$0.77&7.06$\pm$0.18&4\\
209.24/-4.64&-&-&-&-&-&-&-&-& & 20.04&0.32$\pm$0.01&14.01$\pm$0.04&3.17$\pm$0.09&73.17$\pm$6.19&1.43$\pm$0.13&0\\
209.24/-4.64&27.06&1.22$\pm$0.16&22.6$\pm$0.11&3.16$\pm$0.32&0.32$\pm$0.15&38.4$\pm$2.02&3.3$\pm$0.6&2& & 26.89&1.15$\pm$0.01&22.58$\pm$0.01&3.46$\pm$0.02&39.35$\pm$2.63&3.03$\pm$0.21&3\\
\hline
\end{tabular}
\end{sidewaystable}

\begin{sidewaystable}[h!]
\fontsize{4}{3}\selectfont
\renewcommand\thetable{\ref{table:parameters0} (cont)}
\centering
\caption{}
\label{}
\renewcommand{\arraystretch}{1.5}
\begin{tabular}{ lllllllllclllllll  }
\hline
\hline
\multirow{2}{*}{$l/b$} & \multicolumn{8}{c}{pseudo-Voigt} & \multicolumn{7}{c}{Gaussian}\\ 
\cline{2-9}
\cline{11-17}
& \multicolumn{1}{c}{$T_{B}$} & \multicolumn{1}{c}{$\tau$} & \multicolumn{1}{c}{$V_{lsr}$} & \multicolumn{1}{c}{$\Delta V$} & \multicolumn{1}{c}{$\eta$} & \multicolumn{1}{c}{$T_{s}$} & \multicolumn{1}{c}{\NHI} & \multicolumn{1}{c}{$F\ or\ O$} & &  \multicolumn{1}{c}{$T_{B}$} & \multicolumn{1}{c}{$\tau$} & \multicolumn{1}{c}{$V_{lsr}$} & \multicolumn{1}{c}{$\Delta V$} & \multicolumn{1}{c}{$T_{s}$} & \multicolumn{1}{c}{\NHI} & \multicolumn{1}{c}{$F\ or\ O$} \\ 
($^{\circ}$) & \multicolumn{1}{c}{$(K)$} & \multicolumn{1}{c}{$(10^{-2})$} & \multicolumn{1}{c}{$(km s^{-1})$} & \multicolumn{1}{c}{$(km s^{-1})$} & \multicolumn{1}{c}{$\ $} & \multicolumn{1}{c}{$(K)$} & \multicolumn{1}{c}{$(10^{20}cm^{-2})$} & \multicolumn{1}{c}{$\ $} & & \multicolumn{1}{c}{$(K)$} & \multicolumn{1}{c}{$(10^{-2})$} & \multicolumn{1}{c}{$(km s^{-1})$} & \multicolumn{1}{c}{$(km s^{-1})$} & \multicolumn{1}{c}{$(K)$} & \multicolumn{1}{c}{$(10^{20}cm^{-2})$} & \multicolumn{1}{c}{} \\ 
\hline

172.98/2.44&54.15&1.66$\pm$0.22&-17.96$\pm$0.38&3.56$\pm$0.49&0.0$\pm$0.0&66.86$\pm$0.75&7.67$\pm$1.47&1& & 47.21&1.67$\pm$0.02&-17.93$\pm$0.03&3.58$\pm$0.04&58.16$\pm$2.27&6.74$\pm$0.28&5\\
172.98/2.44&54.06&1.33$\pm$0.86&-15.23$\pm$0.66&1.64$\pm$1.31&0.0$\pm$0.0&73.5$\pm$1.62&3.12$\pm$3.2&0& & 44.01&1.36$\pm$0.04&-15.18$\pm$0.03&1.65$\pm$0.07&59.21$\pm$5.31&2.59$\pm$0.27&0\\
172.98/2.44&88.32&5.98$\pm$7.62&-12.35$\pm$0.39&2.37$\pm$1.44&0.0$\pm$0.0&88.54$\pm$0.9&24.33$\pm$34.37&2& & 59.47&6.79$\pm$0.42&-12.39$\pm$0.02&2.16$\pm$0.07&59.54$\pm$5.63&16.97$\pm$2.0&4\\
\hline
\end{tabular}
\end{sidewaystable}

\clearpage

\end{document}